\documentclass[11pt]{article}
\pdfoutput=1

\usepackage{jheppub}
\usepackage{amsmath}
\usepackage{bbm}
\usepackage{url}

\newcommand{\beq}{\begin{eqnarray}}
\newcommand{\eeq}{\end{eqnarray}}

\newcommand{\beqa}{\begin{eqnarray}}
\newcommand{\eeqa}{\end{eqnarray}}

\def\centeron#1#2{{\setbox0=\hbox{#1}\setbox1=\hbox{#2}\ifdim
\wd1>\wd0\kern.5\wd1\kern-.5\wd0\fi
\copy0\kern-.5\wd0\kern-.5\wd1\copy1\ifdim\wd0>\wd1
\kern.5\wd0\kern-.5\wd1\fi}}
\def\ltap{\;\centeron{\raise.35ex\hbox{$<$}}{\lower.65ex\hbox{$\sim$}}\;}
\def\gtap{\;\centeron{\raise.35ex\hbox{$>$}}{\lower.65ex\hbox{$\sim$}}\;}

\def\CO{{\cal O}}

\def\chii0{\chi_i^0}
\def\chij0{\chi_j^0}

\def\foursqr#1#2{{\vcenter{\vbox{
 \hrule height.#2pt
 \hbox{\vrule width.#2pt height#1pt \kern#1pt
 \vrule width.#2pt}
 \hrule height.#2pt
 \hrule height.#2pt
 \hbox{\vrule width.#2pt height#1pt \kern#1pt
 \vrule width.#2pt}
 \hrule height.#2pt
     \hrule height.#2pt
 \hbox{\vrule width.#2pt height#1pt \kern#1pt
 \vrule width.#2pt}
 \hrule height.#2pt
     \hrule height.#2pt
 \hbox{\vrule width.#2pt height#1pt \kern#1pt
 \vrule width.#2pt}
 \hrule height.#2pt}}}}
\def\psqr#1#2{{\vcenter{\vbox{\hrule height.#2pt
 \hbox{\vrule width.#2pt height#1pt \kern#1pt
 \vrule width.#2pt}
 \hrule height.#2pt \hrule height.#2pt
 \hbox{\vrule width.#2pt height#1pt \kern#1pt
 \vrule width.#2pt}
 \hrule height.#2pt}}}}
\def\sqr#1#2{{\vcenter{\vbox{\hrule height.#2pt
 \hbox{\vrule width.#2pt height#1pt \kern#1pt
 \vrule width.#2pt}
 \hrule height.#2pt}}}}

\def\figin{\epsfcheck\figin}\def\figins{\epsfcheck\figins}
\def\epsfcheck{\ifx\epsfbox\UnDeFiNeD
\message{(NO epsf.tex, FIGURES WILL BE IGNORED)}
\gdef\figin##1{\vskip2in}\gdef\figins##1{\hskip.5in}
\else\message{(FIGURES WILL BE INCLUDED)}%
\gdef\figin##1{##1}\gdef\figins##1{##1}\fi}
\def\DefWarn#1{}
\def\figinsert{\goodbreak\midinsert}
\def\ifig#1#2#3{\DefWarn#1\xdef#1{fig.~\the\figno}
\writedef{#1\leftbracket fig.\noexpand~\the\figno}%
\figinsert\figin{\centerline{#3}}\medskip\centerline{\vbox{\baselineskip12pt
\advance\hsize by -1truein\noindent\footnotefont{\bf
Fig.~\the\figno:\ } \it#2}}
\bigskip\endinsert\global\advance\figno by1}

\def\fig#1#2#3#4{\vskip 0.5cm \begingroup \midinsert \centerline{
\psfig{file=#1,width=#2}} \vskip 0.4cm
\global\advance\figno by 1
\centerline{\vbox{\baselineskip=12pt \noindent Figure \the\figno: #3}}
\endinsert \endgroup {\xdef#4{\the\figno}} }

\def\figcrop#1#2#3#4#5#6#7#8{\vskip 0.5cm \begingroup \midinsert \centerline{
\psfig{file=#1,width=#2,bbllx=#3,bblly=#4,bburx=#5,bbury=#6}} \vskip 0.4cm
\global\advance\figno by 1
\centerline{\vbox{\baselineskip=12pt \noindent Figure \the\figno: #7}}
\endinsert \endgroup {\xdef#8{\the\figno}} \vskip .5cm}

\def\figlabel#1{\xdef#1{\the\figno}}
\def\encadremath#1{\vbox{\hrule\hbox{\vrule\kern8pt\vbox{\kern8pt
\hbox{$\displaystyle #1$}\kern8pt}
\kern8pt\vrule}\hrule}}
\def\underarrow#1{\vbox{\ialign{##\crcr$\hfil\displaystyle
 {#1}\hfil$\crcr\noalign{\kern1pt\nointerlineskip}$\longrightarrow$\crcr}}}

\title{Searching for Signs of the Second Higgs Doublet}

\author[a,b]{Nathaniel Craig,}

\author[c]{Jamison Galloway,}

\author[a]{Scott Thomas}

\affiliation[a]{Department of Physics, Rutgers University \\
Piscataway, NJ 08854 }
\affiliation[b]{ School of Natural Sciences, Institute for Advanced Study \\
Princeton, NJ 08540}
\affiliation[c]{Dipartimento di Fisica, Universit\`{a} di Roma ``La Sapienza''
and INFN Sezione di Roma \\
 I-00185 Rome, Italy}
 
 \emailAdd{ncraig@ias.edu}
 \emailAdd{jamison.galloway@roma1.infn.it}
\emailAdd{scthomas@physics.rutgers.edu}

\preprint{RU-NHETC-2013-07}

\abstract{The search for evidence of extended electroweak symmetry breaking has entered a new phase with the discovery of a Standard Model (SM)-like Higgs at the LHC. The measurement of Higgs couplings and direct searches for additional scalars provide complementary avenues for the discovery of new degrees of freedom. This complementarity is particularly sharp in two Higgs doublet models (2HDMs) where the couplings of the SM-like Higgs may be directly related to the LHC signals of additional scalars. In this work we develop a strategy for searching for the second Higgs doublet given the LHC signals of the recently discovered SM-like Higgs.  We focus on a motivated parameter space of flavor- and CP-conserving 2HDMs in which the couplings of all scalars to SM states are controlled by two parameters. We construct fits in this parameter space to the signals of the SM-like Higgs and translate these fits into signal expectations for future measurements of both the SM-like Higgs and additional scalars, identifying the most promising search channels for discovery or exclusion of new physics. When kinematically accessible, decays of the heavy neutral scalar Higgs to two light Higgs scalars, $H \to hh$, and decays of the pseudoscalar Higgs to a light Higgs scalar and $Z$ boson, $A \to Zh$, provide promising avenues for discovery even when the couplings of the light Higgs are within a few percent of SM predictions.  When the couplings of the light Higgs are exceptionally close to those of the SM, decays of heavier neutral scalars to $\gamma \gamma$ and $\tau^+ \tau^-$ become particularly important for discovery. 
}

\begin{document}

\maketitle


\section{Introduction}

The discovery of a Higgs-like boson at the LHC \cite{Aad:2012tfa, Chatrchyan:2012ufa} provides an unprecedented opportunity in the search for physics beyond the Standard Model (BSM). Any significant deviations from SM predictions for Higgs couplings would provide an immediate indication of new physics, while the Higgs may also serve as a portal into an extended electroweak symmetry breaking (EWSB) sector with additional scalar degrees of freedom. These two possibilities are closely related, as mixing among scalars in an extended EWSB sector leads to deviations from Standard Model Higgs couplings. Given the discovery of one Higgs boson, the search for additional Higgs scalars is a crucial objective for current and future searches at the LHC.

There are three principal avenues to searching for a second Higgs. The first is to study the couplings of the SM-like Higgs $h$ itself, as the couplings of $h$ are altered from their Standard Model values by mixing between scalars. The second is to search for new states in SM Higgs channels \cite{Craig:2012vn}, since additional scalars share many of the same production and decay modes as the SM-like Higgs. The third is to search for new states in additional channels, such as those in which heavier scalars decay to final states involving the SM-like Higgs \cite{Craig:2012pu}. There is extensive interplay between these three avenues, since the couplings of the SM-like Higgs are correlated with the variation in production and decay modes of additional scalars. Indeed, many types of extended EWSB sectors are already tightly constrained by measurements of the SM-like Higgs couplings alone.

In order to optimally exploit the interplay between these search channels, it is instrumental to develop a map between the couplings of the observed SM-like Higgs and the couplings of additional Higgs scalars. Typically this map depends on the details of the extended EWSB sector. In this work we will focus on EWSB sectors whose low-energy effective theory is described by two Higgs doublets \cite{Lee:1973iz, Lee:1974jb, Fayet:1974fj, Flores:1982pr}. Theories with two Higgs doublets provide a motivated parameter space for probing extended electroweak symmetry breaking.  Additional Higgs doublets arise in many models of natural BSM physics, including the minimal supersymmetric extension of the SM \cite{Dimopoulos:1981zb}, twin Higgs models \cite{Chacko:2005vw}, and certain composite Higgs models \cite{Mrazek:2011iu}. More generally, two Higgs doublet models (2HDMs) provide a simple parameterization of extensions of the Higgs sector that captures the most important features; other extended EWSB sectors are qualitatively similar. 

In this paper we classify the correlations between search channels, focusing on the most important production and decay topologies for the discovery of additional states. Two Higgs doublet models exhibit a vast signal space involving the five physical scalar particles that remain after EWSB: two neutral CP-even scalars, $h$, $H$; one neutral CP-odd pseudoscalar, $A$; and two charged scalars, $H^+$ and $H^-$.  
The parameter space of these 2HDMs can accommodate a wide range of variations in the production and decay modes of the SM-like Higgs boson, as well as discoverable rates for the production and decay of additional scalars \cite{Gunion:1986nh, Spira:1993bb, Carena:1998gk,Carena:1999bh, Carena:2011dm, Chang:2011kr, Craig:2012pu}.\footnote{For reviews of 2HDM phenomenology, see \cite{Gunion:1989we, Branco:2011iw}. For recent work on 2HDM at the LHC in light of the Higgs discovery, see e.g. \cite{Ferreira:2011aa, Ferreira:2012my, Craig:2012vn, Alves:2012ez, Belanger:2012sd, Craig:2012pu, Altmannshofer:2012ar, Bai:2012ex, Drozd:2012vf, 
Chang:2012zf, Chang:2012bq, Belanger:2012gc, Chen:2013kt, Celis:2013rcs, Chiang:2013ixa, Grinstein:2013npa}.} Although the available parameter space of completely general 2HDMs is quite broad, there is a strongly motivated subspace that provides a predictive map between the couplings of the SM-like Higgs and the potential signals of additional scalars. In particular, tight constraints on flavor-changing neutral currents disfavor 2HDM with tree-level flavor violation. Such prohibitive flavor violation may be avoided by four discrete choices of tree-level Yukawa couplings between the Higgs doublets and SM fermions. Similarly, limits on additional sources of CP violation favor 2HDM with a CP-conserving potential. The avoidance of tree-level FCNCs and explicit CP violation limits the available parametric freedom to the extent that the couplings of all states to SM fermions and gauge bosons may be described solely in terms of two mixing angles.\footnote{Note that even if tree-level flavor violation is forbidden, significant flavor violation may still arise at one loop and contribute to precision flavor observables such as $B \to X_s \gamma$. In this work we will not consider additional constraints coming from precision flavor measurements, since loop-induced contributions to these processes from additional Higgs scalars may be reduced by destructive interference.} 
  
 Within this framework, we will study the interplay between current fits to the SM-like Higgs couplings driven by LHC data and the range of couplings available to additional scalars. For definiteness, we focus on the scenario in which the observed SM-like Higgs is the lightest CP-even neutral scalar $h$, although the alternate case remains interesting. First, we construct fits to current Higgs signals at the LHC and Tevatron in terms of the 2HDM parameter space following the approach of \cite{ Azatov:2012bz, Craig:2012vn}. At present these fits have the greatest impact on the 2HDM parameter space among current measurements. We find all four discrete 2HDM types are constrained to lie close to the alignment limit where the couplings of $h$ are SM-like. Two of these types (Type 2 and Type 4) must lie particularly close to the alignment limit, with the coupling of the SM-like Higgs to vector bosons constrained to lie within 10\% of the Standard Model value at 95\% CL over the entire parameter space.
 
Given the constraints imposed by the coupling fits of the SM-like Higgs, we identify the most promising search channels for discovering additional scalars at the LHC. There is a natural ordering of these channels when the additional scalars are similar in mass, as is expected to be the case when the observed CP-even Higgs is mostly SM-like. In this case the kinematically available decay modes include the usual decays to SM gauge bosons and fermions, as well as decays involving one or more SM-like Higgs bosons. For example, the heavy CP-even Higgs may decay to two light CP-even Higgses, $H \to hh$; the CP-odd pseudoscalar Higgs may decay to a light CP-even Higgs and a $Z$ boson, $A \to Zh$; and the charged Higgses $H^\pm$ may decay to an SM-like Higgs and a $W^\pm$ boson, $H^\pm \to W^\pm h$. These modes are often complementary to direct decays to SM final states. 

    \begin{table}[h]

\begin{center}
\begin{tabular}{|crclccc|} \hline    
&   & & & & &\\
& \multicolumn{3}{c}{Second Higgs Doublet} & & Alignment &\\
& \multicolumn{3}{c}{   Decay Topology}  & &  Limit &\\
&   & & & & &\\
 \hline
&   & & & & &\\
& $H$ & $ \! \! \! \to $ & $\! \! \! WW, ZZ $ & & $-$ &\\
& $H,A$ & $\! \! \! \to$ & $\! \! \! \gamma \gamma $  &  & \checkmark & \\
& $H,A$ & $\! \! \! \to$ & $\! \! \! \tau \tau, \mu \mu $  &  & \checkmark & \\
& & & & & &\\
& $H,A$ & $\! \! \! \to$ & $\! \! \! tt $  &  & \checkmark & \\
& & & & & &\\
& $A$ &  $\! \! \! \to$ & $\! \! \! Zh$ & & $-$ &\\
& $H$ & $\! \! \! \to$ & $\! \! \! hh$  & & $-$ &\\
& & & & & &\\
 & $ t$ &  $\! \! \! \to$ & $\! \! \! H^{\pm} b $  &  & \checkmark &\\
& & & & & &\\
\hline
\end{tabular}
\caption{Leading 2HDM decay topologies with unsuppressed production cross 
sections near the alignment limit for neutral scalars
with mass spectra $m_H \sim m_A > 2 m_h$,
and top quark 
with $m_{H^{\pm}} < m_t+m_b$.
A checkmark (dash) indicates that the partial
decay width approaches a constant (vanishes) in the
$\cos(\beta-\alpha)=0$ alignment limit. 
The first five 
topologies give additional non-Standard Model 
contributions to Standard Model Higgs search channels. 
The remaining topologies are specific Second Higgs Doublet search channels.  
Altogether these topologies form a basis for 
Second Higgs Doublet searches 
with non-SM-like boson masses greater than the SM-like Higgs boson. 
 \label{tab:channels} }
\end{center}
\end{table}

The consistency of Higgs coupling fits with SM predictions suggests that the 2HDM Higgs sector is close to the alignment limit where the couplings of one CP-even Higgs scalar are SM-like. As illustrated in Table \ref{tab:channels}, in the exact alignment limit certain decay modes of heavy scalars vanish, including $H \to VV$, $H \to hh$, and $A \to Zh$. However, even when close to the alignment limit, these modes may still dominate the decay products of the heavy scalars if they are kinematically accessible. This is particularly likely when $m_{H}, m_A < 2m_t$, in which case the decays $H \to VV$, $H \to hh$, and $A \to Zh$ only compete with decays into bottom quarks. If the partial widths for $H \to VV$, $H \to hh$, and $A \to Zh$ exceed the partial widths for decays into fermions, the resulting branching ratios only weakly depend on proximity to the alignment limit. However, proximity to the alignment limit {\it does} imply that associated production modes for $H$ involving vector couplings -- such as vector boson fusion or production in association with a $W$ or $Z$ boson -- are likely to be suppressed. Thus the production of both $H$ and the pseudoscalar $A$ are likely to be dominated by gluon fusion, in which case discovering $H$ and $A$ at the LHC requires focusing on distinctive final states.

The most promising final states for the discovery of additional Higgs scalars change as the couplings are varied. For the heavy neutral Higgs $H$, the decays to $hh, VV$ and $\gamma \gamma, \tau^+ \tau^-$ play complementary roles depending on proximity to the alignment limit. When kinematically available, the decay of $H \to hh$ is significant even for small deviations from the alignment limit; it dominates the decay products of $H$ for $2 m_h < m_H < 2 m_t$ and may continue to dominate even when $m_H > 2 m_t$. However, this partial width falls  as the alignment limit is approached, much like the rate for $H$ to decay into two vectors. This alone does not guarantee that the branching ratio is small, since $\Gamma(H \to hh) \propto m_H^3 / v^2$ and typically dominates the total width until $g_{HVV}/g_{H_{SM} VV} \lesssim \mathcal{O}(m_b / m_H)$ unless the coupling of $H$ to bottom quarks is parametrically enhanced. Since the coupling of the SM-like Higgs to vector bosons can be measured at the 14 TeV LHC to only $\mathcal{O}(8\%)$ accuracy with 3000 fb$^{-1}$ of data \cite{Klute:2012pu}, this suggests that $H \to hh$ may constitute a substantial fraction of $H$ decays even if asymptotic measurements of Higgs couplings at the LHC remain consistent with SM expectations. Of course, it is also possible that the 2HDM Higgs sector lies very close to the alignment limit. For $g_{HVV}/g_{H_{SM} VV} \lesssim \mathcal{O}(m_b/m_H)$, the branching ratio for $H \to hh$ is diminished, but at the same time decays to $\gamma \gamma$ and $\tau^+ \tau^-$ become increasingly important because they do not decouple in the alignment limit. 

The same is true of the pseudoscalar Higgs $A$ with respect to the decays to $Zh$ and $\gamma \gamma, \tau^+ \tau^-$, respectively. The partial width $\Gamma(A \to Zh)$ likewise scales as $\propto m_A^3 / v^2$ and vanishes in the alignment limit. Unless the coupling of $A$ to fermions is parametrically enhanced, $A\to Zh$ often dominates the total width until very close to the alignment limit. In the exact alignment limit, decays to $\gamma \gamma$ and $\tau^+ \tau^-$ become important. 

Our paper is organized as follows: In Section \ref{sec2} we discuss the relevant aspects of 2HDMs, focusing on the alignment limit in which one CP-even neutral scalar is approximately SM-like, and define the restricted parameter space used in our study. In Section \ref{sec3}
we construct fits to the couplings of the SM-like Higgs in the context of various  2HDM types, using all available data from the LHC and Tevatron. In Section \ref{sec4} we present our procedure for parameterizing the production cross sections and branching ratios of the additional Higgs scalars. We use the coupling fits in Section \ref{sec5} to explore the range of signals available in future measurements of the SM-like Higgs in production and decay modes that are currently poorly constrained. In Section \ref{sec6} we use the coupling fits to the SM-like Higgs to probe the range of signals available to the additional physical Higgs bosons in a 2HDM. We conclude in Section \ref{sec7} with various suggestions for LHC search strategies motivated by the range of possible 2HDM signals. Details of the 2HDM scalar potential and the Higgs coupling fit procedure are reserved for appendices.


\section{The Second Higgs Doublet \label{sec2}}

The general parameter space of 2HDMs is large, but it can be efficiently reduced using a small set of motivated assumptions. Specifically, in this work we focus on CP conservation in the potential and the absence of tree-level contributions to flavor-changing neutral currents; these are reasonable assumptions since additional sources of CP violation and FCNCs are tightly constrained. Although the mass spectrum of 2HDM is in principle arbitrary, here we wish to focus on a spectrum with two approximate scales: a light SM-like Higgs $h$ at $m_h \sim 126$ GeV, and the remaining physical Higgs scalars $H, A, H^\pm$ clustered together at an equal or higher scale with $m_H \sim m_A \sim m_{H^\pm}.$ In this case the only available decay modes of the SM-like Higgs are those involving SM states. This allows us to describe the couplings of the SM-like Higgs in terms of two free parameters, and gives a  natural ordering for the additional scalar decay modes that are kinematically accessible. With these assumptions, it is straightforward to map the current fits to the signals of the SM-like Higgs to the production and decay rates of the remaining scalars.\footnote{Note that although we focus on the case where the light CP-even scalar $h$ is the SM-like Higgs, the coupling fits are identical (under $\alpha \to \alpha + \pi/2$) when the heavy CP-even scalar $H$ is the SM-like Higgs, up to the possible effects of new decay modes involving additional scalars.}

The absence of tree-level flavor-changing neutral currents in theories with multiple Higgs doublets is guaranteed by the Glashow-Weinberg condition \cite{Glashow:1976nt} that all fermions of a given representation receive their masses through renormalizable Yukawa couplings to a single Higgs doublet, in which case the tree-level couplings of neutral Higgs bosons are diagonal in the mass eigenbasis. This restriction may be enforced by a discrete symmetry acting on the doublets.  In theories with only two Higgs doublets, the Yukawa couplings are
\begin{eqnarray}
V_{yukawa} = - \sum_{i = 1,2} \left(Q \tilde \Phi_i y_i^u \bar u + Q \Phi_i y_i^d \bar d + L \Phi_i y_i^e \bar e + {\rm h.c.} \right) 
\end{eqnarray}
and the Glashow-Weinberg condition is satisfied by four discrete assignments, where by convention up-type quarks are always taken to couple to $\Phi_2$:
\begin{itemize}
\item Type 1, in which $y_1^{u,d,e} = 0$; all fermions couple to one doublet.
\item Type 2, in which $y_1^u = y_2^d = y_2^e = 0$; the up-type quarks couple to one doublet and the down-type quarks and leptons couple to the other.
\item Type 3, in which $y_1^u = y_1^d = y_2^e = 0$; quarks couple to one doublet and leptons to the other.
\item Type 4, in which $y_1^u = y_1^e = y_2^d = 0$; up-type quarks and leptons couple to one doublet and down-type quarks couple to the other.
\end{itemize}
The signals of Type 3 and Type 4 2HDM typically resemble those of Type 1 and Type 2 2HDM, respectively, since these pairings share the same quark assignments and thus the same parametric scaling for dominant production and decay modes.\footnote{Note that some other 2HDM studies switch Type 3 and Type 4 assignments.} The primary exception is for signals involving leptonic final states, for which the branching ratios are parametrically enhanced (suppressed) in Type 3 (4) 2HDM compared to their Type 1 (2) counterparts. In what follows we will largely focus on 2HDM of Type 1 and 2, though we will discuss distinctive features of Type 3 and 4 where appropriate. 

The most general scalar potential for a CP-conserving 2HDM allowed by gauge invariance is given in Appendix \ref{app:a}. Including the vacuum expectation values, there are 12 real degrees of freedom in the potential; 9 remain free after minimizing the potential and fixing the electroweak symmetry breaking vev $v^2 = v_1^2 + v_2^2 = (246 \, {\rm GeV})^2$. A convenient basis for the remaining free parameters consists of the ratio of vacuum expectation values of $\Phi_{1,2}^0$, parameterized by
\beq
\tan \beta \equiv | \langle \Phi_2^0 \rangle /  \langle \Phi_1^0 \rangle | ~;
\eeq
the mixing angle $\alpha$ that diagonalizes the $2 \times 2$ neutral scalar $h-H$ mass squared matrix,
\beq
 \left( 
\begin{matrix} \sqrt{2}~ {\rm Re} (\Phi_2^0) - v_2  \\  
     \sqrt{2}~ {\rm Re} (\Phi_1^0) - v_1 \end{matrix} \right) 
= 
\left( \begin{array}{rr} \cos \alpha  &\sin \alpha  \\ 
                              - \!  \sin \alpha & \cos \alpha  \end{array} \right) 
\left( \begin{matrix} h \\ H \end{matrix} \right) ~;
\label{hHmixing} 
\eeq
the four physical masses $m_h, m_H, m_A, m_{H^\pm}$; and the couplings $\lambda_5, \lambda_6,$ and $\lambda_7$, shown explicitly in Appendix \ref{app:a}. The discrete symmetry that ensures the Glashow-Weinberg condition also requires $\lambda_6 = \lambda_7 = 0$, though in what follows we will consider the effects of nonzero $\lambda_{6,7}$ where appropriate.


The angles $\alpha$ and $\beta$ fully determine the couplings between a single physical Higgs boson and two gauge bosons or two fermions, as well as the coupling between two Higgses and a single gauge boson.  Only renormalizable couplings involving three and four physical Higgs bosons depend on the additional parameters of the potential. Therefore if we identify the lightest CP-even neutral Higgs scalar $h$ with the observed SM-like Higgs at 126 GeV, with the remaining scalars $H, A, H^\pm$ equal in mass or heavier, then deviations in the production and decay rates of the SM-like Higgs from the SM prediction may be parameterized entirely in terms of $\alpha$ and $\beta$.

Thus far, the signals of the Higgs boson measured by ATLAS and CMS have remained largely consistent with SM predictions. This consistency suggests that if the EWSB sector is described by a 2HDM, it is likely to lie near the alignment limit where $\sin(\beta - \alpha) = 1$ and the coupling of $h$ to vector bosons is SM-like  \cite{Gunion:2002zf}.\footnote{Note that we distinguish the alignment limit $\sin(\beta - \alpha) = 1$ from the decoupling limit $m_A^2 \gg |\lambda_i| v^2$. When $m_{H,A,H^\pm} \gg m_h$ these limits coincide, but in general we also wish to consider the case where all scalars are relatively light but the couplings of $h$ are entirely SM-like, perhaps due to accidental cancellations in the 2HDM potential.} Given this preference, it is useful to express the couplings of various Higgs scalars to SM fermions and gauge bosons in terms of deviations from the alignment limit. In particular, the couplings to fermions in various types of 2HDM depend on four trigonometric functions of $\alpha, \beta$ that may be expanded  near the alignment limit. The couplings of the CP-even scalar $h$ depend entirely on the combinations
\begin{eqnarray}
& \sin(\beta \!-\! \alpha) - \tan \! \beta \cos (\beta \!-\! \alpha) \! &\simeq 1 - \tan \! \beta \cos (\beta \!-\! \alpha) \!-\!{1 \over 2}\! \cos^2 (\beta \!-\! \alpha) 
 + {\cal O}( \cos^4 (\beta \!-\! \alpha)) \\
& \sin (\beta \!-\! \alpha) + \cot \! \beta \cos (\beta \!-\! \alpha)  \! &\simeq 1  + \cot \! \beta \cos (\beta \!-\! \alpha) \!-\!{1 \over 2}\! \cos^2 (\beta \!-\! \alpha)  + {\cal O}( \cos^4 (\beta \!-\! \alpha)) 
\end{eqnarray}
while the couplings of the remaining scalars depend on the combinations
\begin{eqnarray}
& \tan \! \beta \sin (\beta \!-\! \alpha) \!+\! \cos(\beta \!-\! \alpha) \! &\simeq  \tan \! \beta \Big[ 1 \!+\! \cot \! \beta \cos(\beta \!-\! \alpha)  \!-\!{1 \over 2}\! \cos^2 (\beta \!-\! \alpha)  + {\cal O}( \cos^4 (\beta \!-\! \alpha)) \Big] \\
& \cot \! \beta \sin (\beta \!-\! \alpha) \!-\! \cos (\beta \!-\! \alpha)   \! &\simeq   \cot \! \beta \Big[ 1  \!-\! \tan \! \beta \cos (\beta \!-\! \alpha)  \!-\!{1 \over 2}\! \cos^2 (\beta \!-\! \alpha)  + {\cal O}( \cos^4 (\beta \!-\! \alpha)) \Big]
  ~~~~~~~~
 \end{eqnarray}
In the second equality we have expanded around $\sin(\beta - \alpha) = 1.$ In Table \ref{tab:couplings} we use these trigonometric identities to express the fermion and vector couplings of all scalars in the four discrete types of flavor-preserving 2HDM as a function of $\tan \beta$ and $\beta - \alpha$.

\begin{table}[h]
\begin{center}
\begin{tabular}{|c|c|c|c|c|} \hline
$y_{\rm 2HDM} / y_{\rm SM}$ & 2HDM 1& 2HDM 2 & 2HDM 3 & 2HDM 4 \\ \hline
$hVV$ & $s_{\beta - \alpha}$ &  $s_{\beta - \alpha}$ &  $s_{\beta - \alpha}$ &  $s_{\beta - \alpha}$ \\
$h Q u $ & $s_{\beta - \alpha} +  c_{\beta - \alpha} / t_\beta$ & $s_{\beta - \alpha} + c_{\beta - \alpha}/ t_\beta$ & $s_{\beta - \alpha} +  c_{\beta - \alpha}/ t_\beta$& $s_{\beta - \alpha} +  c_{\beta - \alpha}/ t_\beta$  \\
$h Q d$ & $s_{\beta - \alpha} +  c_{\beta - \alpha}/ t_\beta$ & $s_{\beta - \alpha} - t_\beta c_{\beta - \alpha}$ & $s_{\beta - \alpha} +  c_{\beta - \alpha}/ t_\beta$& $s_{\beta - \alpha} - t_\beta c_{\beta - \alpha}$  \\
$h L e$ & $s_{\beta - \alpha} + c_{\beta - \alpha}/ t_\beta$ & $s_{\beta - \alpha} - t_\beta c_{\beta - \alpha}$ & $s_{\beta - \alpha} - t_\beta c_{\beta - \alpha}$& $s_{\beta - \alpha} +  c_{\beta - \alpha}/ t_\beta$  \\ \hline
$HVV$ & $c_{\beta - \alpha}$ & $c_{\beta - \alpha}$ & $c_{\beta - \alpha}$& $c_{\beta - \alpha}$   \\
$H Q u$ & $c_{\beta - \alpha} -  s_{\beta - \alpha}/ t_\beta$ & $c_{\beta - \alpha} -  s_{\beta - \alpha}/ t_\beta$ & $c_{\beta - \alpha} - s_{\beta - \alpha}/ t_\beta$& $c_{\beta - \alpha} -  s_{\beta - \alpha}/ t_\beta$  \\
$H Q d$ & $c_{\beta - \alpha} -  s_{\beta - \alpha}/ t_\beta$ & $c_{\beta - \alpha} + t_\beta s_{\beta - \alpha}$ & $c_{\beta - \alpha} -  s_{\beta - \alpha}/ t_\beta$&$c_{\beta - \alpha} + t_\beta s_{\beta - \alpha}$  \\
$H L e$ & $c_{\beta - \alpha} - s_{\beta - \alpha}/ t_\beta$ & $c_{\beta - \alpha} + t_\beta s_{\beta - \alpha}$ &$c_{\beta - \alpha} + t_\beta s_{\beta - \alpha}$& $c_{\beta - \alpha} - s_{\beta - \alpha}/ t_\beta$  \\\hline
$AVV$ & 0 & 0 & 0 & 0 \\
$AQu$ & $1/t_\beta$ & $1/t_\beta$ & $1/t_\beta$& $1/t_\beta$   \\
$AQd$ & $-1/t_\beta$ & $t_\beta$& $- 1/t_\beta$ & $t_\beta$   \\
$ALe$ & $- 1/t_\beta$ & $t_\beta$  & $t_\beta$& $- 1/t_\beta$  \\ \hline
\end{tabular}
\caption{The tree-level couplings of the neutral Higgs bosons $h, H,$ and $A$ to up- and down-type quarks, leptons, and massive gauge bosons relative to the SM Higgs boson couplings as functions of $\alpha$ and $\beta$ in the four types of 2HDM models satisfying the Glashow-Weinberg condition. The coefficients of the couplings of the charged scalars $H^{\pm}$ are
the same as those of the pseudo-scalar $A$. \label{tab:couplings}}
\end{center}
\end{table}%

\subsection{Decays to the SM-like Higgs}

In addition to the couplings involving one scalar, we will be interested in three couplings involving two or more scalars: the coupling of $h$ to the pseudoscalar $A$ and a $Z$ boson, $g_{hZA}$; the coupling of $h$ to the charged Higgs $H^\pm$ and a $W$ boson, $g_{h W^\mp H^\pm}$; and the coupling of the heavy Higgs scalar $H$ to two SM-like scalars $h$: $g_{Hhh}$. These control the rates of the three processes $A \to Zh, H^\pm \to W^\pm h,$ and $H \to hh$ that may be kinematically available when $m_h < m_A \sim m_H \sim m_{H^\pm}$. 

The couplings of two scalars to a SM vector are transparently written in terms of departure from the alignment limit. In particular, we have
\begin{eqnarray}
g_{hZA}=& \frac 12 \sqrt{g^2 + g'^2} \cos(\beta-\alpha) \hspace{12mm} 
g_{hW^\mp H^\pm}=& \mp \frac i2 g \cos(\beta-\alpha) \hspace{9mm} 
\label{eqn:hhv}
\end{eqnarray}
The form of these couplings is guaranteed by unitarity constraints. In a 2HDM the couplings $g_{hZA}, g_{HZA}$ and $g_{hW^\mp H^\pm}, g_{H W^\mp H^\pm}$ obey a unitarity constraint akin to that satisfied by $g_{hVV}, g_{HVV}$, namely
\begin{eqnarray}
g_{hZA}^2 + g_{HZA}^2 &=& \frac{1}{4 m_Z^2} g_{h_{SM} ZZ}^2 \\
g_{h W^\mp H^\pm}^2 + g_{H W^\mp H^\pm}^2 &=& g_{A W^\mp H^\pm}^2 = \frac{1}{4 m_W^2} g_{h_{SM} WW}^2
\end{eqnarray}

Unlike the other couplings involving SM vectors or fermions, the triple Higgs coupling $g_{Hhh}$ depends on additional parameters beyond the physical masses and mixing angles. As we show in detail in Appendix \ref{app:a}, this coupling is conveniently expressed as a function of the physical masses, the mixing angles, and the quartic couplings $\lambda_5, \lambda_6, \lambda_7$; in terms of $\tan \beta$ and $\beta - \alpha$ it is given by
\begin{eqnarray} 
\label{gHhheq} 
g_{Hhh}  =   
\frac{\cos(\beta \! - \! \alpha)}{v} && \hspace{-0.5cm}
\left[ \left(3 m_A^2 + 3 \lambda_5 v^2 - 2 m_h^2 - m_H^2 \right) 
\left( \cos(2 \beta - 2 \alpha) - \frac{ \sin(2 \beta - 2 \alpha)}{\tan(2 \beta)} \right)  \right. \nonumber \\
& -&   m_A^2 - \lambda_5 v^2 + \frac{\lambda_6 v^2}{2}  
\left( - \cot \beta + 3 \sin(2\beta - 2 \alpha) + 3 \cot \beta \cos (2 \beta - 2 \alpha) \right) \nonumber \\ 
& +&  \left. \frac{\lambda_7 v^2}{2}  
 \left( - \tan \beta - 3 \sin(2 \beta - 2 \alpha) + 3 \tan \beta \cos(2 \beta - 2 \alpha) \right) \right] 
\end{eqnarray}
Various simplifying limits are presented in Appendix \ref{app:a}.

There are a number of key points worth emphasizing about the coupling $g_{Hhh}$ and the corresponding partial width for $H \to hh$. Note that $g_{Hhh} \propto \cos(\beta - \alpha)$ so that in the exact alignment limit $g_{Hhh} \to 0$. Indeed, $g_{Hhh}$ approaches zero in the alignment limit at the same rate as the vector coupling $g_{HVV}$, so that neither $H \to hh$ nor $H \to VV$ is available when $h$ is exactly SM-like. However, these processes may be important even if deviations from the alignment limit are small, and in fact may dominate the total width of $H$ since both partial widths grow as $\Gamma(H \to VV, hh) \propto m_H^3 / v^2$. Which of the two processes dominates is then a matter of numerical coefficients. Significantly, it  is often the case that $\Gamma(H\to hh) \gtrsim \Gamma(H \to VV)$ when kinematically open, in which case $H \to hh$ dominates over $H \to VV$. Note that $g_{Hhh}$ grows at large and small $\tan \beta$, but only well away from the alignment limit, as in the limit of small $\cos(\beta - \alpha)$ the leading $\tan \beta$- and $\cot \beta$-enhanced terms in $g_{Hhh}$ scale as $\cos^2(\beta - \alpha)$.  Away from the exact alignment limit, at large $\tan \beta$ we have $\Gamma(H \to hh) \gg \Gamma(H \to VV)$ due to the $\tan \beta$ enhancement of $\Gamma(H \to hh)$. This is important in determining the optimal search channels for the heavier Higgs $H$ as $m_H$ is varied, since it implies that ${\rm Br}(H \to VV)$ may be small even when the partial width is appreciable. Of course, there are some exceptions; $g_{Hhh}$ approaches zero as $\sin(\beta - \alpha) \to 1$ and also when $\cos(2 \beta - 2 \alpha) - \frac{\sin(2 \beta - 2 \alpha)}{\tan(2 \beta)} \approx \frac{1}{2} \frac{1}{1 - m_h^2/m_H^2}$, where the decay $H\to VV$ may dominate instead.

It is an oft-repeated truism that $H \to hh$ is unimportant at large $\tan \beta$ in a Type 2 2HDM because the $\tan \beta$-enhanced coupling of $H$ to bottom quarks rapidly leads to $\Gamma(H \to b \bar b)$ taking over the total width. This is true to a certain extent when $\lambda_6 = \lambda_7 = 0$, in which case the leading contributions to $g_{Hhh}$ near the alignment limit are not enhanced by $\tan \beta$ and $\tan \beta$-enhanced terms first arise at $\mathcal{O}(\cos^2(\beta - \alpha))$. The bottom coupling $y_{Hbb}$ is $\tan \beta$-enhanced, however, and so rapidly takes over the total width as $\tan \beta$ is increased, suppressing the $H \to hh$ branching ratio. (Note that this is not an issue in Type 1 2HDM, where there are no $\tan \beta$-enhanced couplings and $H \to hh$ continues to dominate the total width at large $\tan \beta$.) However, even in Type 2 2HDM this ceases to be the case when $\lambda_6$ and $\lambda_7$ are nonzero. Near the alignment limit, the leading $\tan \beta$-enhanced contribution to $g_{Hhh}$ scales as $\sim 2  \lambda_7  v \tan \beta \cos(\beta - \alpha) $. This may easily compete with the $\tan \beta$ enhancement of $y_{Hbb}$, in which case $H \to hh$ continues to govern the total width even when $\tan \beta$ is large. While we will not study the consequences of nonzero $\lambda_{6,7}$ extensively in what follows, it bears emphasizing that when $\lambda_{6,7}$ are non-vanishing $H \to hh$ may dominate the total width even in cases where fermion couplings are parametrically enhanced.

Given this simplified parameter space for theories with two Higgs doublets, we can fit the current signals of the SM-like Higgs $h$ in terms of $\alpha$ and $\beta$ and then map this fit onto the range of production and decay rates for additional scalars.

\section{Couplings of the SM-like Higgs \label{sec3}}

In order to ascertain which production and decay modes may be promising in future LHC searches, we first construct fits to the signals of the SM-like Higgs in terms of the 2HDM parameter space.\footnote{For a partial list of recent work on Higgs coupling fits at the LHC, see e.g. \cite{Carmi:2012yp, Azatov:2012bz, Espinosa:2012ir, Li:2012ku, Ellis:2012rx, Azatov:2012rd, Klute:2012pu, Azatov:2012wq, Low:2012rj,  Corbett:2012dm, Montull:2012ik, Espinosa:2012im, Carmi:2012in, Giardino:2013bm, Ellis:2013lra, Djouadi:2013qya}.} The fit to SM-like Higgs couplings is constructed with all of the most recent available data from the LHC and Tevatron.  We use fully exclusive channel breakdowns, as required for correct identification of the couplings controlling a given mode's production cross-section.  The profile of the signal strength modifier for each channel is fit with a two-sided Gaussian when best fit information is provided, otherwise we rely on likelihood reconstruction techniques reviewed in \cite{AGreview}.

\begin{figure}[htb]
\begin{center}
\includegraphics[width=7.5cm]{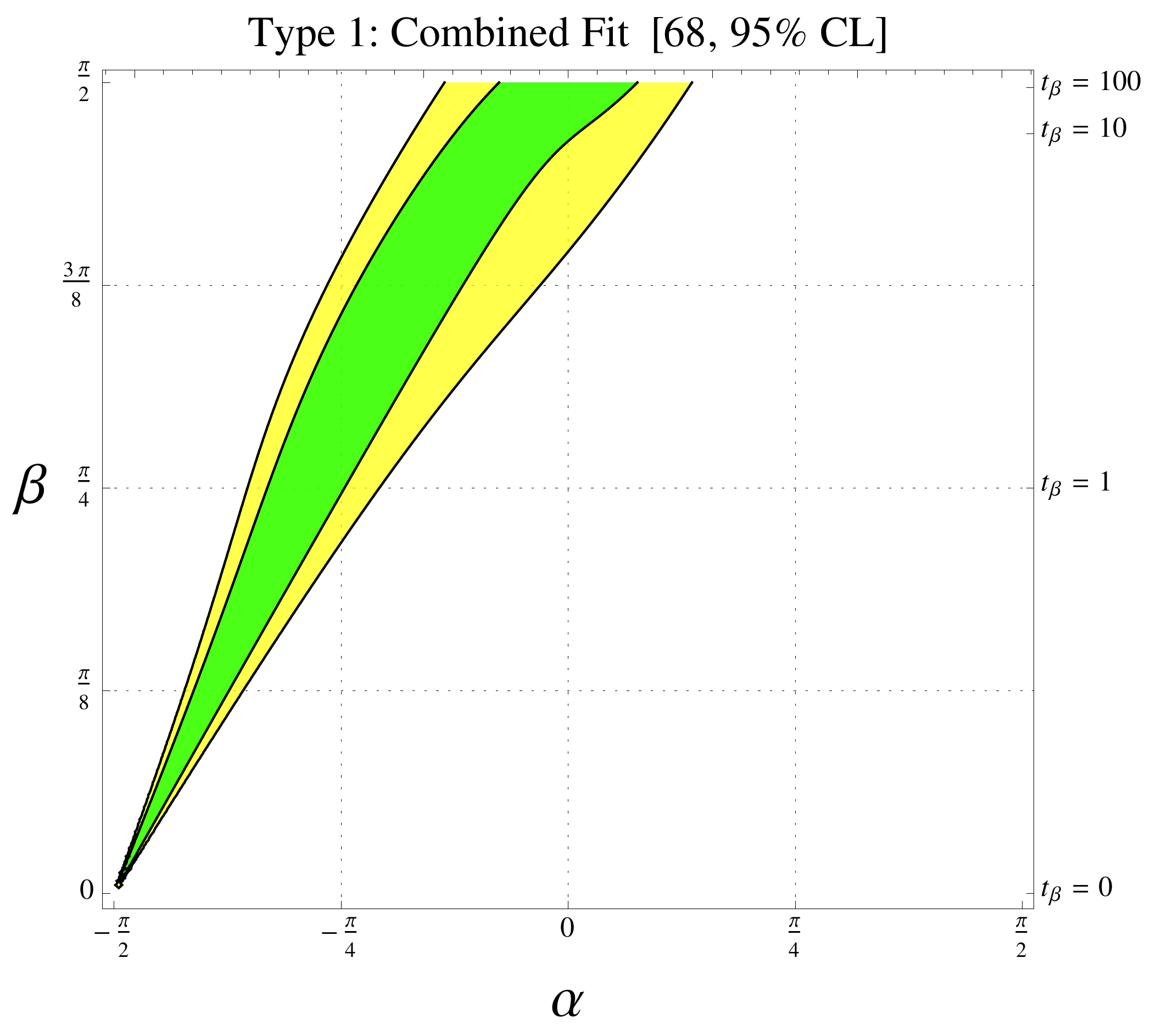} 
\includegraphics[width=7.5cm]{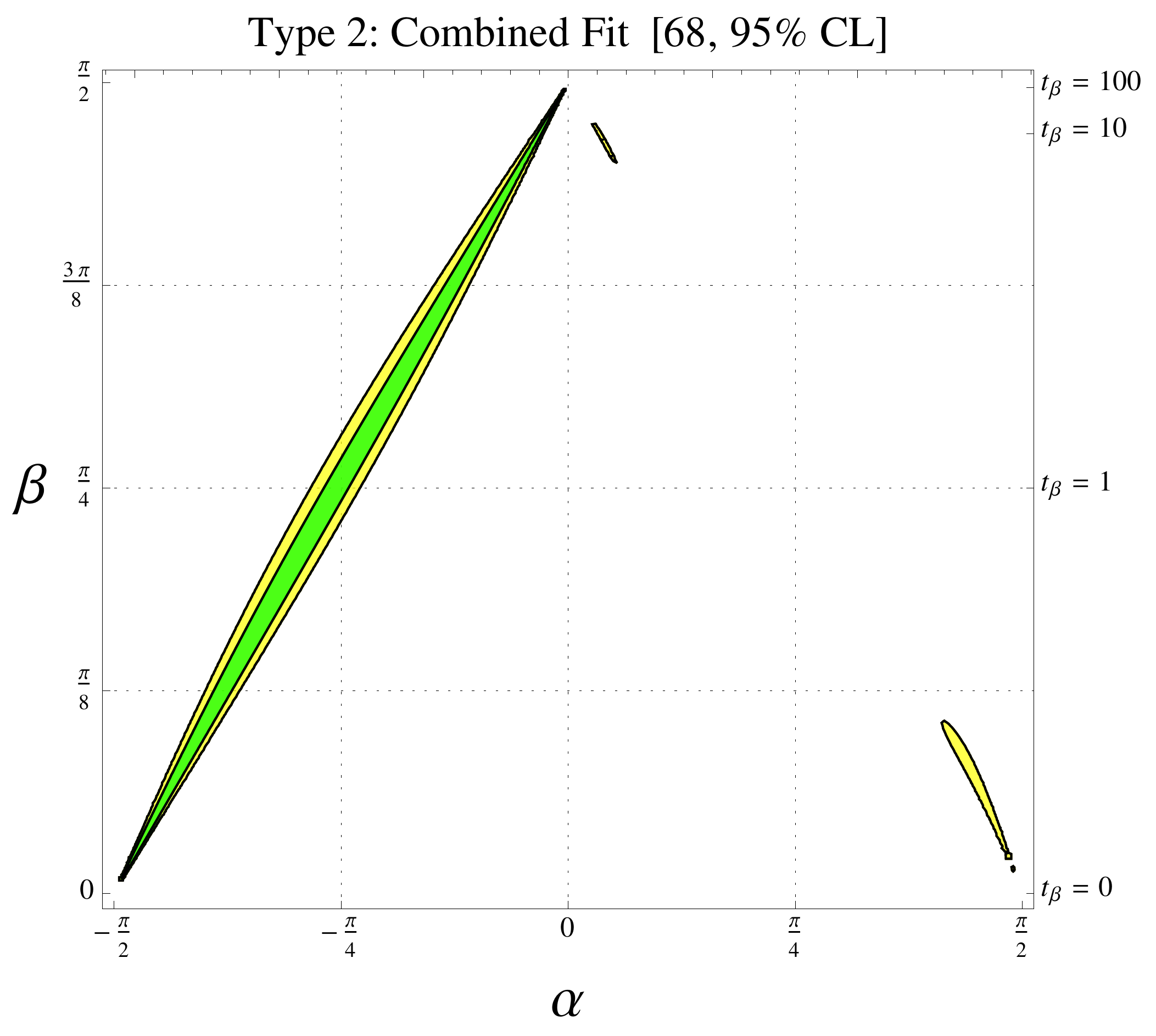} 
\includegraphics[width=7.5cm]{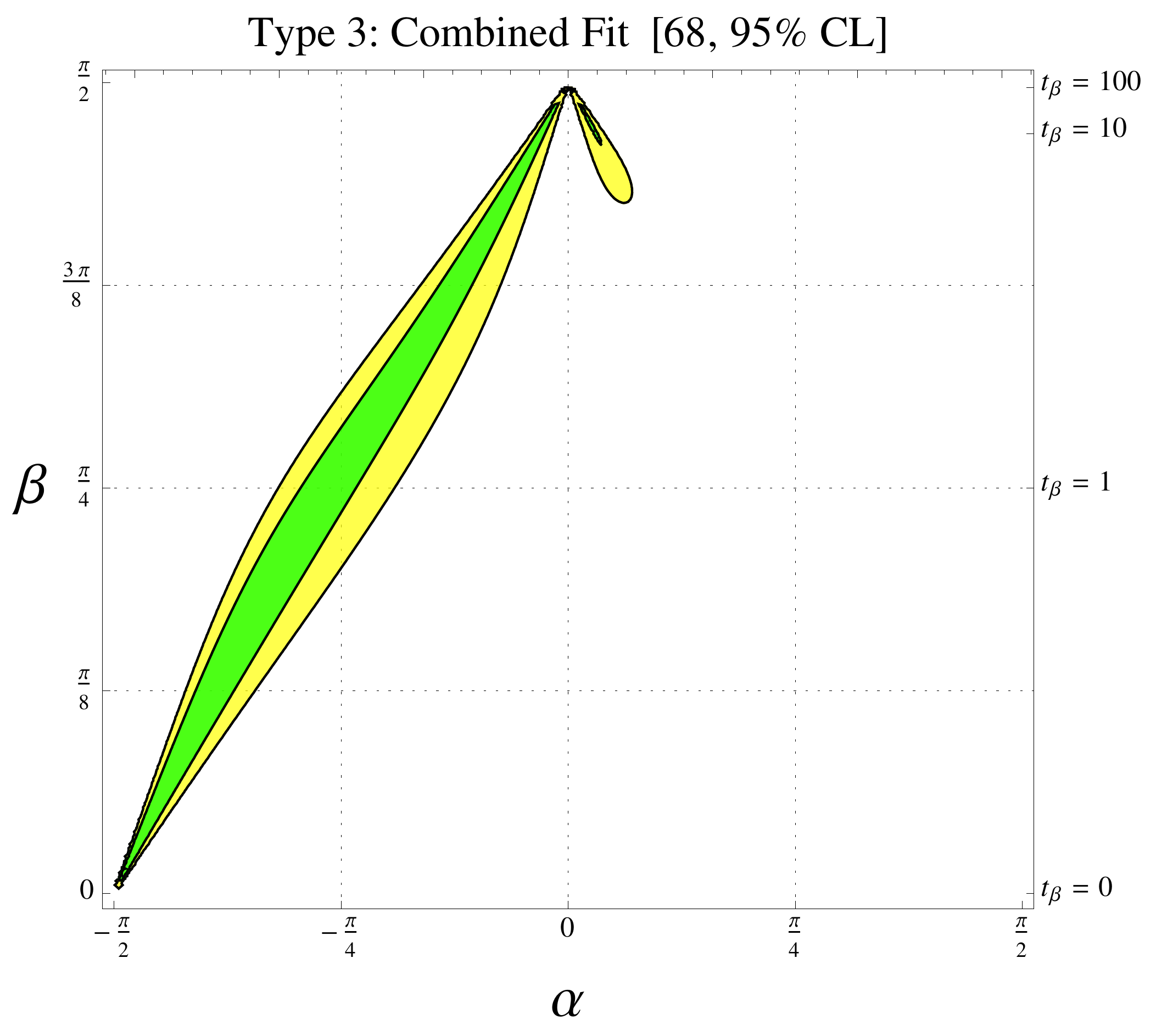} 
\includegraphics[width=7.5cm]{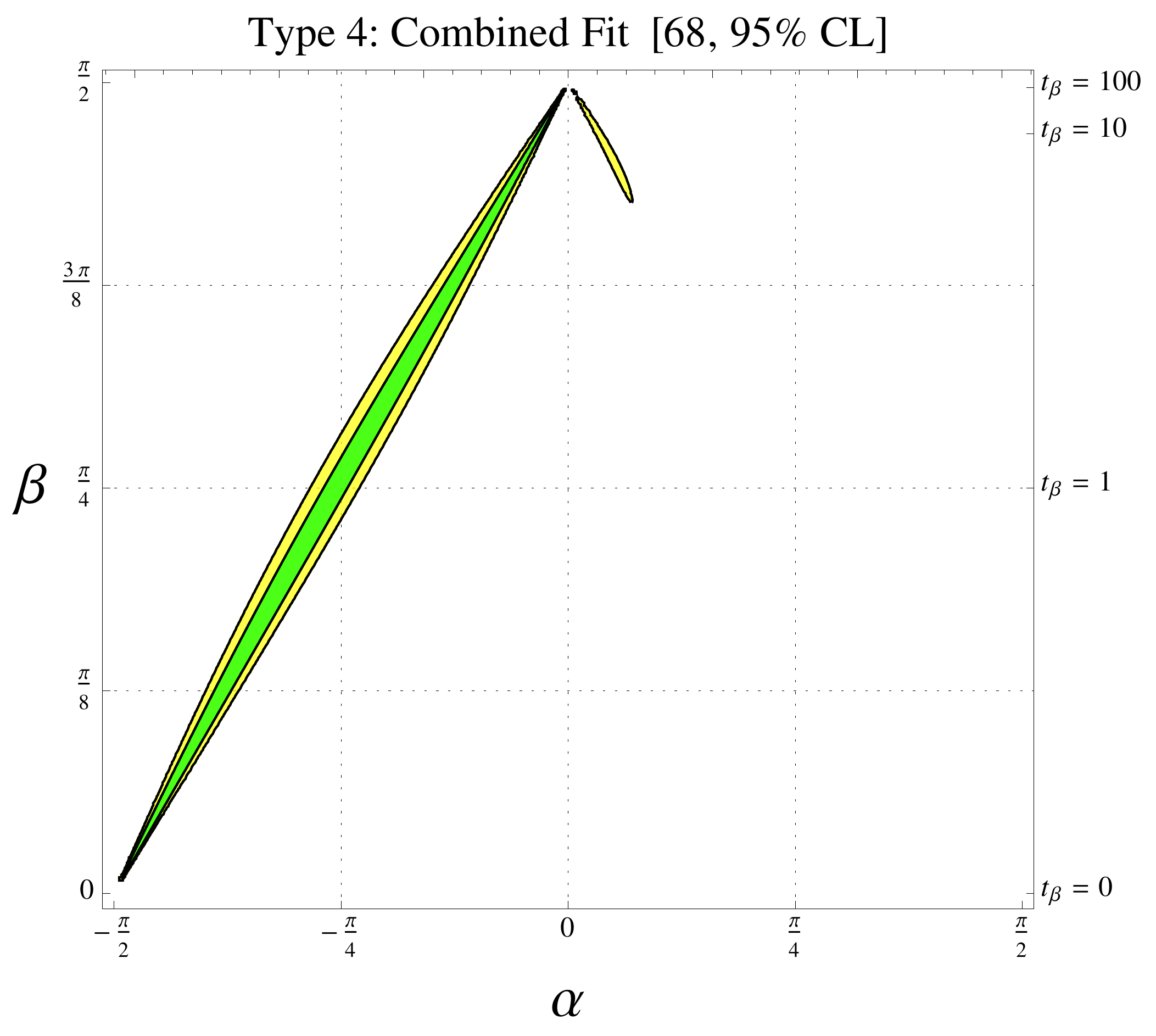} 
\caption{\small Upper left: Global fit of SM-like Higgs couplings in 2HDM of Type 1.  Upper right: Global fit of SM-like Higgs couplings in 2HDM of Type 2. Lower left: Global fit of SM-like Higgs couplings in 2HDM of Type 3.  Lower right: Global fit of SM-like Higgs couplings in 2HDM of Type 4. }
\label{fig:1234fit}
\end{center}
\end{figure}

In Fig.~\ref{fig:1234fit} we show the total global fits for all four discrete 2HDM types as a function of $\alpha$ and $\beta$. From these fits it is strongly apparent that all four types favor the alignment limit $\beta - \alpha = \pi/2$. The fits in Type 2 and Type 4 2HDM models are particularly tight around the alignment limit due to the variation in the bottom quark coupling, as we will discuss in detail below. It is apparent that Types 2-4 all feature a second lobe of the fit away from the alignment limit with $\alpha > 0,$ corresponding to changing the sign of of $g_{hVV}$ relative to the fermion couplings. This region was favored by previous fits to 2HDM couplings due to enhanced values of $h \to \gamma \gamma$ observed by both ATLAS and CMS in early Higgs data. However, the most recent CMS measurement of $h \to \gamma \gamma$ falls below the SM rate, so that this region is no longer favored by the combined fit. Note that the region with $\alpha > 0$ is not allowed in the specific Type 2 2HDM corresponding to the MSSM Higgs sector.

\begin{figure}[htb]
\begin{center}
\includegraphics[width=7.5cm]{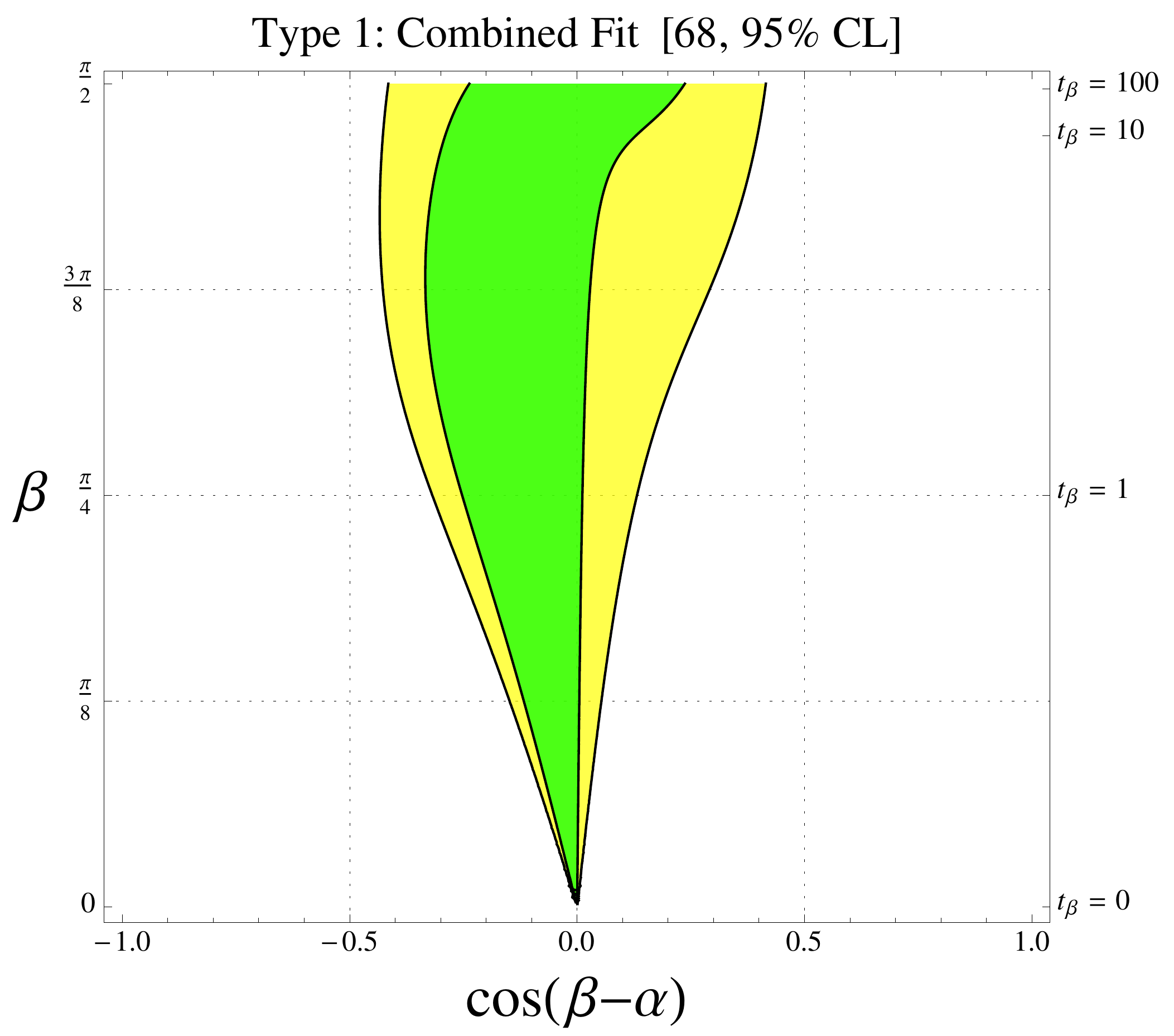} 
\includegraphics[width=7.5cm]{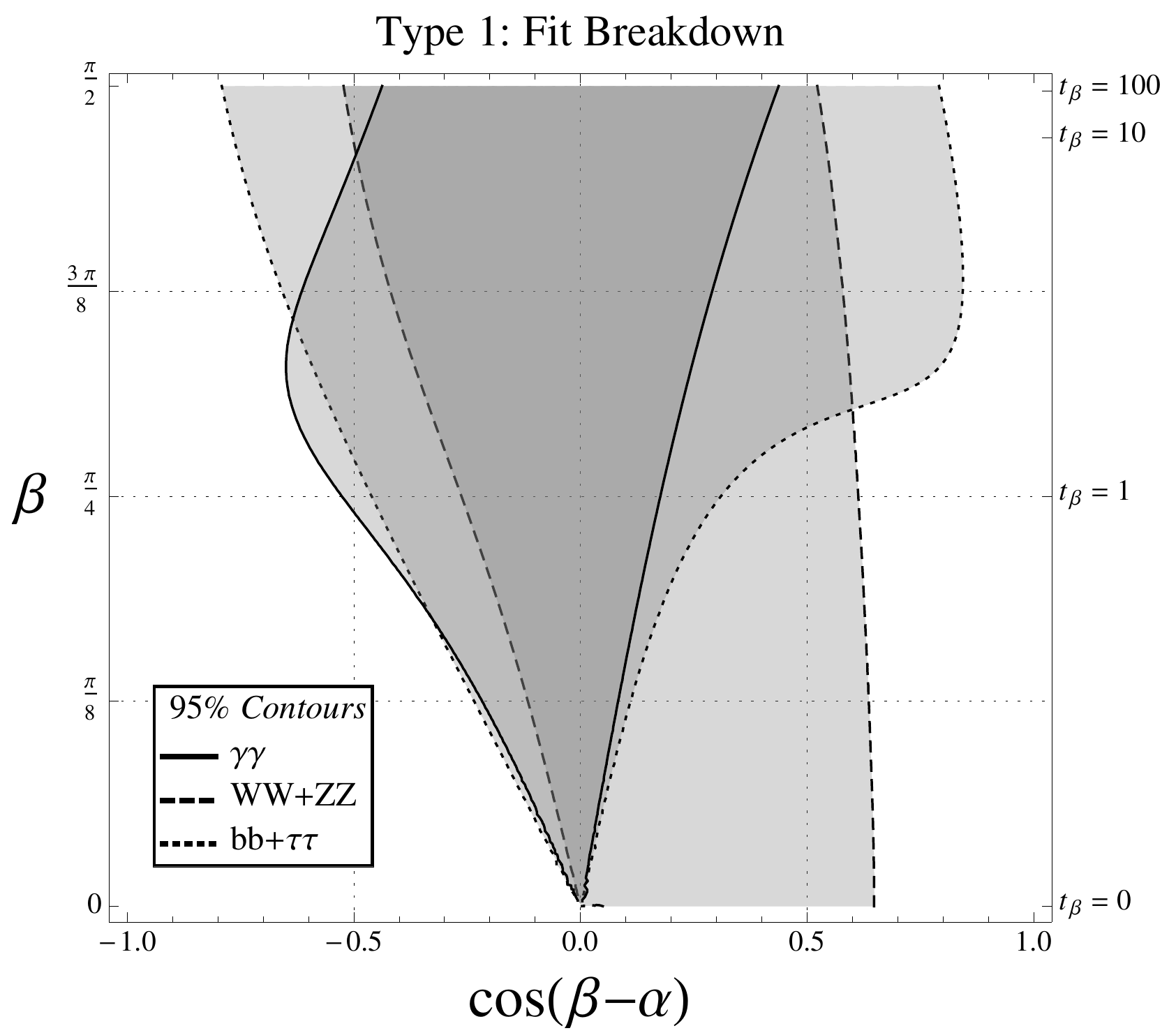} 
\caption{\small Left: Global fit of SM-like Higgs couplings in 2HDM of Type 1.  Right: Contributions to the fit coming separately from $h \to \gamma \gamma$, $h\to VV$, and $h \to bb, \tau \tau$.}
\label{fig:1fit}
\end{center}
\end{figure}

\begin{figure}[htb]
\begin{center}
\includegraphics[width=7.5cm]{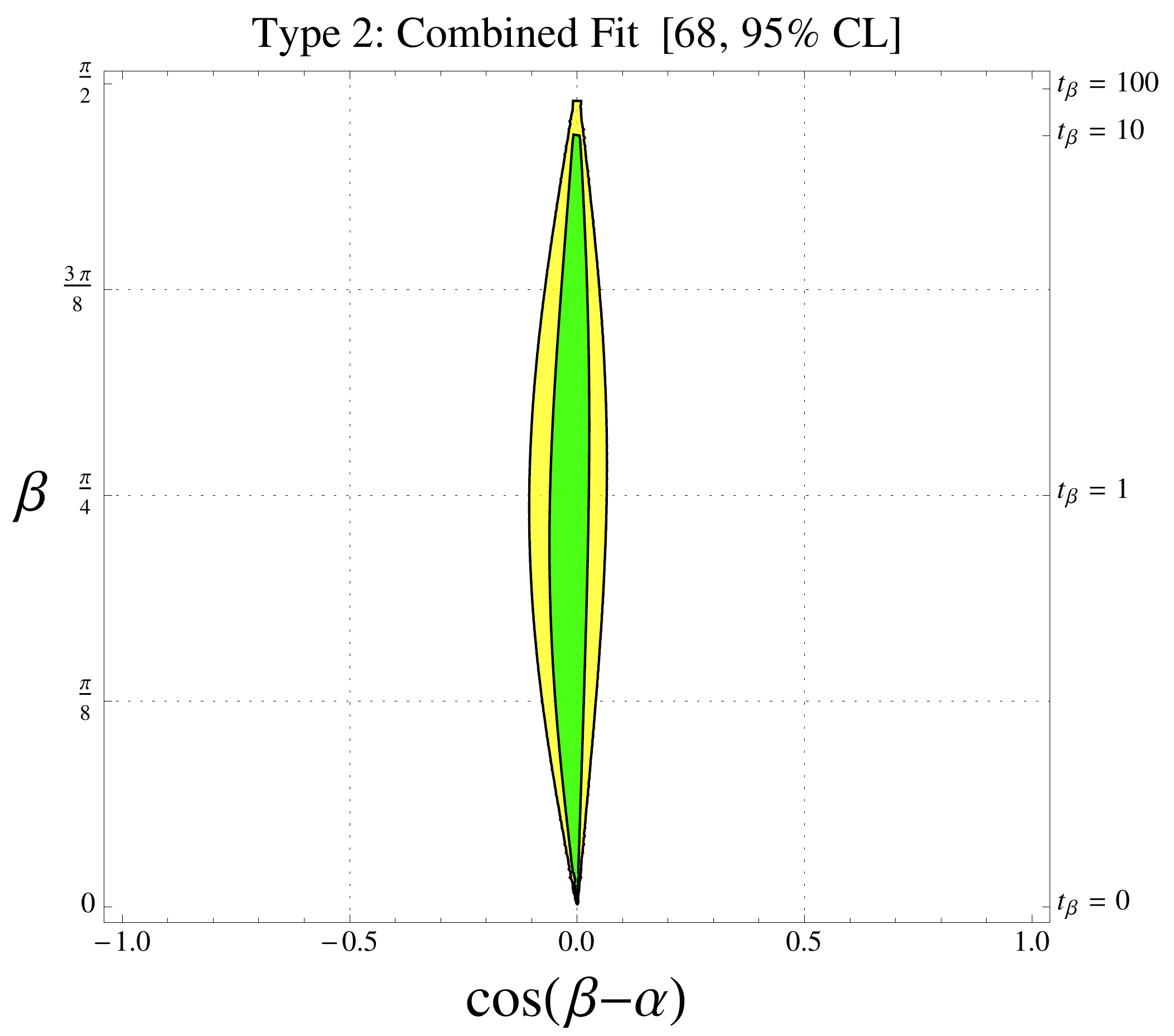} 
\includegraphics[width=7.5cm]{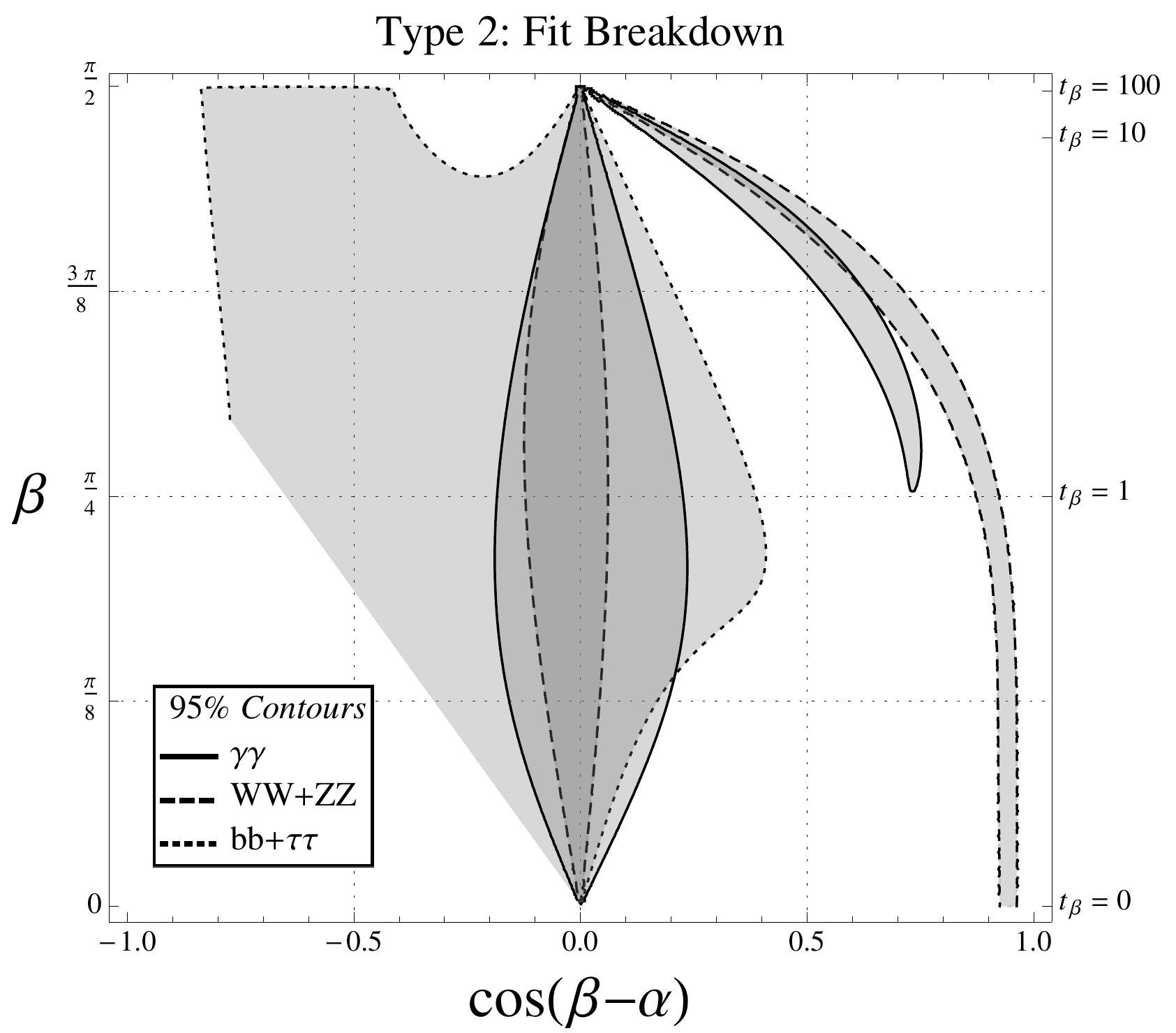} 
\caption{\small Left: Global fit of SM-like Higgs couplings in 2HDM of Type 2. Right: Contributions to the fit coming separately from $h \to \gamma \gamma$, $h\to VV$, and $h \to bb, \tau \tau$.  The peninsulas of the $\gamma \gamma$ and $VV$ contours correspond to regions where the bottom coupling has changed sign (i.e. $\alpha >0$) and has a magnitude that is respectively suppressed/unsuppressed relative to the decoupling limit (see Type 2 SM-like Higgs decay rates  in Figs.~\ref{fig:htautau},~\ref{fig:VBFhtautau},~\ref{fig:VBFhgammagamma} for comparison). Note also that the apparent discontinuity in the $bb+\tau \tau$ contour arises from the requirement $\alpha > -\pi/2$.}
\label{fig:2fit}
\end{center}
\end{figure}

In light of the evident preference for the alignment limit, in Figs.~\ref{fig:1fit} and \ref{fig:2fit} we show respectively the total global fits for 2HDMs of Type 1 and 2 around the alignment limit as a function of $\beta$ and $\cos(\beta- \alpha)$.  In addition to showing the combined fit,  we also indicate how the different final states $\gamma \gamma, WW, ZZ, \tau \tau$, and $bb$ contribute individually to the combined fit in order to illustrate the interplay between different 2HDM couplings governing the shape of the fits.  For reference, in Table~\ref{tab:95bounds} we quote values of the quantity $\cos (\beta - \alpha)$ at the $95\%$ CL contour for three benchmark values of $\tan \beta$.
\begin{table}[hhh]
\footnotesize
\centering
\renewcommand{\arraystretch}{1.1}
\begin{tabular}{| c | c | c | c | }
\hline
Model & $\tan \beta$ & $(c_{\beta-\alpha})^{95\%}_-$  & $(c_{\beta-\alpha})^{95\%}_+$  \\
\hline \hline
 & $1$ & $-0.32$ & $0.42$ \\
Type 1 & $10$ & $-0.43$ & $0.40$  \\
 & $100$ & $-0.42$ & $0.13$  \\
\hline
 & $1$ & $-0.11$ & $0.06$  \\
Type 2 & $10$ & $-0.02$ & $0.01$  \\
 & $100$ & --- &---\\
\hline
\end{tabular}
\caption{\small Values of $\cos (\beta -\alpha)$ on positive and negative sides of the alignment limit, denoted by subscripts $\pm$, at the 95\% CL contour for various values of $\tan \beta$.  For Type 2 2HDM, at very large values of $\tan \beta$ the width of the region in $\cos(\beta - \alpha)$ allowed by measured coupling values of the SM-like Higgs is narrower than the resolution of our fit procedure; see Fig.~\ref{fig:2fit}. \label{tab:95bounds}}

\end{table}

We can gain some intuition for the shape of the coupling fits as a function of $\alpha$ and $\beta$ by considering how couplings and rates scale in the limits of small and large values of $\beta$ in each case.

For Type 1 2HDM, we have the following important limits to consider:
\begin{itemize}
\item At large $\tan \beta$, the couplings scale very simply, namely
\beq
c_V, c_t, c_b, c_\tau  \to \cos \alpha
\eeq
where $c_V \equiv y_{hVV}/y_{h_{SM}VV}, c_t \equiv y_{hQu} /y_{h_{SM} Qu}, c_b \equiv y_{hQd}/y_{h_{SM} Qd},$ and $c_\tau \equiv y_{hLe}/ y_{h_{SM} Le}$. Thus all rates are rescaled equally, such that along the $\tan \beta = \infty$ line the fit region corresponds to a best fit on a simple one-dimensional signal strength modifier.  
\item At small $\tan \beta$, the couplings scale instead as
\beq
c_V &\to& -\sin \alpha \\
c_t, c_b, c_\tau &\to& \beta^{-1}  \cos \alpha.
\eeq
Assuming the fermion couplings dominate the total width, the rates of the important individual channels therefore get rescaled in simple ways:
\beq
R_{\gamma \gamma}\ \approx \ \frac{\sigma(gg \to h) \times {\rm BR} (h \to \gamma \gamma)}{\sigma(gg \to h_{SM}) \times {\rm BR} (h_{SM} \to \gamma \gamma)}\ \sim \ c_t^2 \times  \frac{c_V^2}{c_b^2}  & \to & \sin^2 \alpha \\
R_{VV} \ \approx \ \frac{\sigma(gg \to h) \times {\rm BR} (h \to VV^*)}{\sigma(gg \to h_{SM}) \times {\rm BR} (h_{SM} \to VV^*)} \ \sim \ c_t^2 \times \frac{c_V^2}{c_b^2} & \to & \sin^2 \alpha \\
R_{bb}^{ (Vh)} \ = \ \frac{\sigma(p p \to Vh) \times {\rm BR} (h \to bb)}{\sigma(p p \to Vh_{SM}) \times {\rm BR} (h_{SM} \to bb)} \ \sim \ c_V^2 \times \frac{c_b^2}{c_b^2} & \to & \sin^2 \alpha \\
R_{bb}^{ (ttH)} \ = \ \frac{\sigma(pp \to t \bar t h)  \times {\rm BR} (h \to bb)}{\sigma(pp \to t \bar t h_{SM})  \times {\rm BR} (h_{SM} \to bb)} \ \sim \ c_t^2 \times \frac{c_b^2}{c_b^2} & \to & \beta^{-2} \cos^2 \alpha \\
R_{\tau \tau}^{\rm (VBF)} \  = \ \frac{\sigma(pp \to hjj) \times {\rm BR} (h \to \tau \tau)}{\sigma(pp \to h_{SM}jj) \times {\rm BR} (h_{SM} \to \tau \tau)} \ \sim \ c_V^2 \times \frac{c_\tau^2}{c_b^2} & \to & \sin^2 \alpha \\
R_{\tau \tau}^{\rm (inc.)} \ \approx \ \frac{\sigma(gg \to h) \times {\rm BR} (h \to bb)}{\sigma(gg \to h_{SM}) \times {\rm BR} (h_{SM} \to bb)} \ \sim \ c_t^2 \times \frac{c_\tau^2}{c_b^2} & \to & \beta^{-2} \cos^2 \alpha 
\eeq
Thus we can easily understand why the diphoton contour narrows at small $\beta$ while the $VV$ mode, whose VBF-initiated contribution vanishes in this limit, occupies a finite region of the $\beta = 0$ line.  The asymmetry in the $VV$ contour at small $\beta$ reflects the fact that $\cos \alpha>0$ by definition. 
\end{itemize}

In the Type 2 2HDM case we see a more dynamic interplay between the separate channels, with $VV$ providing a powerful constraint due to its sensitivity to the vector coupling.  We note the following limiting behaviors:
\begin{itemize}
\item At large $\beta$, we have
\beq
c_V, c_t &\to& \cos \alpha \\
c_b, c_\tau &\to& \frac{1}{\beta - \pi/2} \times \sin \alpha
\eeq
Rates for each channel thus scale as follows:
\beq
R_{\gamma \gamma}, R_{VV} &\to& \left( \beta -\frac{\pi}{2}\right)^2 \frac{\cos^4 \alpha}{\sin^2 \alpha}  \\
R_{bb}, R_{\tau \tau} & \to & \cos^2 \alpha
\eeq
Thus the $\gamma \gamma$ and $VV$ likelihoods vanish, while $bb$ and $\tau \tau$ remain finite at large $\beta$.  Notice that cancellation between large couplings ($c_b, c_\tau$) in branching fractions is accurate only asymptotically, or in regions of parameter space where the vector coupling and top coupling are significantly diminished.  Incidentally, this is the reason for the slight depression in the $bb/\tau \tau$ fit contour at large $\beta$ and small $\cos(\beta -\alpha)$, where values of vector and top couplings remain near their SM values.
\item At small $\beta$ the couplings scale as
\beq
c_V, c_b, c_\tau &\to& -\sin \alpha \\
c_t &\to& \beta^{-1} \cos \alpha
\eeq
Likelihoods therefore tend to vanish at small $\beta$ since rates become arbitrarily large, i.e.
\beq
R_{\gamma \gamma}, R_{VV}, R_{bb}^{(ttH)}, R_{\tau \tau}^{(inc.)} &\to& \beta^{-2}  \cos^2 \alpha \\
R_{bb}^{(VH)}, R_{\tau \tau}^{\rm (VBF)} & \to & \sin^2 \alpha
\eeq
\end{itemize}
Notice also that the diphoton contour shows some mild preference for positive values of $\cos (\beta - \alpha)$.  In this region of the fit  the bottom coupling is suppressed relative to the SM, thus allowing an enhanced $\gamma \gamma$ from the reduced $b \bar b$ branching fraction; this is the region that would be favored by current diphoton results from ATLAS \cite{ATLASgaga}.  This contour is cut off at small $\beta$ by an excessive top coupling (and thus excessive production via gluon fusion); at somewhat larger $\beta$ it is cut off by the reduction in the bottom coupling's magnitude overshooting the preferred enhancement in diphoton.  Additional channels sensitive to the vector coupling therefore become important in shaping the combined fit contour for these intermediate values of $\beta$. 

With these coupling fits in hand, we may investigate the prospects for evidence of extended electroweak symmetry breaking by studying the range of variations in SM-like Higgs couplings that remain poorly constrained by present measurements as well as the range of signals available to the heavier scalars $H, A$, and $H^\pm$.

\section{Signals of the Second Higgs Doublet \label{sec4}}

At present, the coupling fits of the SM-like Higgs are dominated by a few signal channels that have meaningful sensitivity in the discovery-level data set. Beyond these channels, there are a variety of signal channels whose measurements are poorly constrained at present but will improve with additional integrated luminosity. This makes it useful to study the possible variation in future measurements consistent with current fits. Deviations in these channels may provide the first indication of the presence of additional Higgs scalars. Of course, these additional scalars may also be observed directly, either in standard Higgs search channels or through cascade decays involving multiple scalars. Coupling fits of the SM-like Higgs serve to pick out the most promising of these discovery channels.

In what follows, we will explore the range of 2HDM signals in various modes consistent with current best fits to the SM-like Higgs couplings. To do so, we adopt the following procedure to determine the appropriate product of production cross section times branching ratio:

For the case of the SM Higgs boson, we take the NLO production cross sections for gluon fusion, vector boson fusion, and production in association with a vector boson or top quarks from the LHC Higgs Cross Section Working Group \cite{Dittmaier:2011ti}. For 2HDM, we calculate the ratio of LO production partial widths in each 
production channel for $h$ and $H$ relative to a SM Higgs boson of the same mass analytically from the couplings presented in Section 2 as functions of the mixing parameters $\alpha$ and $\beta$. The NLO SM Higgs production cross sections in each production channel are then rescaled by these factors to obtain an estimate for the NLO cross sections; for instance we take the $\alpha, \beta$ dependent cross section for gluon fusion production of $H$ to be 
\begin{equation}
\sigma_{\rm NLO}(gg \to H)|_{\alpha,\beta} = \sigma_{\rm NLO}(gg \to h_{\rm SM}) ~
  \frac{\Gamma_{\rm LO}(H \to gg){\big |}_{\alpha,\beta}}{\Gamma_{\rm LO}(h_{\rm SM} \to gg)}
\end{equation}
where $H$ and $h_{SM}$ are of the same mass.
The same procedure of normalizing to SM Higgs boson NLO cross sections through the $\alpha$ and $\beta$ dependent ratios of LO production partial widths is used for production of $A$ by gluon fusion or in association 
with top quarks. This is expected to be a good approximation since the fractional size of NLO corrections in these cases is not 
strongly dependent on the parity of the Higgs scalar away from the two-top threshold. Near the two-top threshold the NLO corrections to the pseudoscalar production cross section are larger than the NLO corrections to the scalar cross section, so our procedure is conservative.

Although the cross section for production of a SM Higgs boson in association with bottom quarks is quite small, it may play an important role in 2HDM with enhanced bottom couplings. In order to account for these contributions, we obtain the inclusive NLO production cross section $\sigma(p p \to (b \bar b) h_{SM} + X)$ for a SM Higgs boson produced via $b \bar b$ annihilation using \texttt{bbh@nlo} \cite{Harlander:2003ai} with the choice of equal factorization and renormalization scales: $\mu_F = \mu_R = m_h/4$. Note that \texttt{bbh@nlo} computes the inclusive cross section $\sigma(p p \to (b \bar b) h_{SM} + X)$ in a variable flavor number scheme in which the LO partonic process is $b \bar b \to h_{SM}$; the corresponding cross section does not include any tagging requirements on bottom quarks in the final state. To obtain the 2HDM NLO cross sections for this process we rescale the NLO SM Higgs production cross section by the ratio of LO partial widths as above.

We will often be interested in the ``inclusive'' production cross section for a given Higgs state. In practice, the experimental measurement of an inclusive Higgs process assigns different weights to various production modes. In the case of current measurements at the LHC, these weights are known and are used appropriately in determining the coupling fits in Section 3. However, the weights for many future measurements are unknown, and so in this section we approximate inclusive production rates by summing the appropriate production cross sections without any relative re-weighting due to differences in experimental acceptance.

Our procedure for decay modes is analogous. For the case of the SM Higgs boson, the NLO partial decay widths and branching ratios are taken from the LHC Higgs Cross Section Working Group \cite{Dittmaier:2011ti}. 
For the 2HDM spectra the ratio of LO partial decay widths for $h$ relative to a SM Higgs boson of the same mass are calculated analytically as functions of the mixing parameters $\alpha$ and $\beta$ using 
the couplings presented above. The NLO SM Higgs boson partial decay widths are then rescaled by these factors to obtain  estimates for the NLO partial widths; for instance we take the $\alpha, \beta$ dependent partial width for the SM-like Higgs $h$ to $b\bar{b}$  to be 
\beq
\Gamma_{\rm NLO}(h \to b \bar b)|_{\alpha,\beta} = \Gamma_{\rm NLO} (h_{\rm SM} \to b \bar b) ~ 
\frac{\Gamma_{\rm LO}(h \to b \bar b){\big |}_{\alpha,\beta}}{\Gamma_{\rm LO}(h_{\rm SM} \to b \bar b)}
\eeq
The same procedure of normalizing to SM Higgs boson NLO partial decay widths through the ratio of LO decay widths 
is used for the $H$ decay modes that are in common with the $h$ decay modes. For the pseudoscalar $A$, we instead use analytic results for the NLO partial widths into fermions as a function of $\alpha$ and $\beta$ \cite{Djouadi:1995gt} since the parity of the pseudoscalar is particularly important near the fermion pair thresholds.  For the charged Higgs $H^\pm$ we simply use the analytic results for the LO partial widths into fermions as a function of $\alpha$ and $\beta$ \cite{Djouadi:2005gj}. For the decay modes $H \to hh$ and $A \to Zh$ we use the analytic results for LO partial widths \cite{Djouadi:1995gv} as a function of  $\alpha$ and $\beta$. None of these decay modes involves strongly interacting particles, so LO widths are a reasonable approximation. The partial widths for all the open decay modes of each Higgs scalar are then used to calculate the $\alpha$ and $\beta$ dependent total widths and branching ratios.  

Note that certain processes such as $H \to hh$ and $A \to Zh$ constitute additional sources of production for the SM-like Higgs. In this work we do not simultaneously incorporate these potential new sources of $h$ production into the coupling fit. This is a reasonable approximation insofar as the acceptance for these production modes is currently unknown and, in general, the $\lesssim \mathcal{O}({\rm pb})$ cross-sections involved should not significantly distort the coupling fit given the signal channels that currently dominate the couplings. However, as sensitivity to $Vh$ exclusive production improves with additional integrated luminosity, a simultaneous fit to SM-like and heavy scalar production would be useful.

Throughout, we will neglect loop-level contributions from  $H^\pm$ to the $h,H \to \gamma \gamma$ rates. These contributions are suppressed relative to $W$ and top loops by a factor of $\mathcal{O}(m_W^2 / m_{H^\pm}^2)$ and decouple particularly rapidly near the alignment limit \cite{Gunion:2002zf}.

\section{Signs of the Second Higgs Doublet in the SM-like Higgs Couplings \label{sec5}}

Current signals are dominated by inclusive and VBF production of $h$ with $h \to \gamma \gamma$ as well as inclusive production of $h$ with $h \to VV^*$.  Sensitivity in inclusive and $Vh$/VBF associated production of $h$ with $h \to \tau^+ \tau^-$ and $Vh$ production of $h$ with $h \to b \bar b$ is improving, but the errors on these measurements remain large. There are also a variety of inclusive and associated measurements that may be made using the future LHC data set, but for which there is currently no sensitivity. Thus it is instrumental to consider the range of signals that might be anticipated in the following future measurements given the state of current fits:
\begin{itemize}
\item {Inclusive production of $h$ with $h \to \tau^+ \tau^-$ or $h \to \mu^+ \mu^-$}
\item { $Vh$/VBF production of $h$ with $h \to \tau^+ \tau^-$} 
\item {$Vh$ production of $h$ with $h \to b \bar b$} 
\item { $Vh$/VBF production of $h$ with $h \to \gamma \gamma$}
\item { $Vh$/VBF production of $h$ with $h \to VV^*$} 
\item { Inclusive production of $h$ with $h \to Z \gamma$}
\item { $t\bar{t}$ associated production of $h$ with $h \to \gamma \gamma$ }
\end{itemize}

The parametric dependence of the production cross section times branching ratio on $\alpha$ and $\beta$ for each of these processes may be understood by considering the functional dependence of the production mode, the decay mode, and the total width. While the scaling of decay modes is fairly straightforward, the production mode and total width are slightly more subtle. In Type 1 and 3 2HDM, the production modes are fairly simple; $Vh$ and VBF associated production scale as $c_V^2$, while inclusive production is dominated by gluon fusion and scales as $c_t^2$. In Type 2 and 4 2HDM, however, at large $\tan \beta$ the inclusive production is dominated by $b\bar bh$ associated production, and the parametrics switch over from $\propto c_t^2$ to $\propto c_b^2$.  As for branching ratios, the decay mode is unambiguous but the dominant contribution to the total width may vary. In general, the dominant contributions to the total width come from the partial widths $\Gamma(h \to b \bar b)$ and $\Gamma(h \to VV^*)$. Unsurprisingly, $\Gamma(h \to b \bar b)$ dominates over much of the parameter space, but becomes sub-dominant in specific regions where $c_b \to 0$, which vary depending on the 2HDM type. For Type 1 and Type 3 2HDM,  $c_b \to 0$ as $\alpha \to \pm \frac{\pi}{2}$. For Type 2 and Type 4 2HDM, $c_b \to 0$ as $\alpha \to 0$. In these regions, the contribution of $\Gamma(h \to b \bar b)$ to the total width vanishes, and the parametric dependence of the total width is instead dominated by $\Gamma(h \to VV^*)$. Thus the parametric scaling of the total width is generally dominated by the scaling of $c_b^2$, except in special regions where $c_b \to 0$ and the width is instead dominated by $c_V^2$. In general these regions are disfavored by current fits.

The parametrics of many of these channels as a function of $\alpha$ and $\beta$ in the four types of 2HDM were discussed in detail in \cite{Craig:2012vn} and above in Section 3; here we focus on the specific channels in which future deviations may arise.

\subsection{ Inclusive production of $h$ with $h \to \tau^+ \tau^-$ or $h \to \mu^+ \mu^-$} 

The measurement of the inclusive $h \to \tau^+ \tau^-$ rate is improving in sensitivity at both ATLAS and CMS, but there remains substantial room for deviations. The inclusive $h \to \mu^+ \mu^-$ rate has identical parametric scaling in the 2HDM we consider, and at present is very poorly constrained at the LHC. 

\begin{figure}[htbp] 
   \centering
   \includegraphics[width=3in]{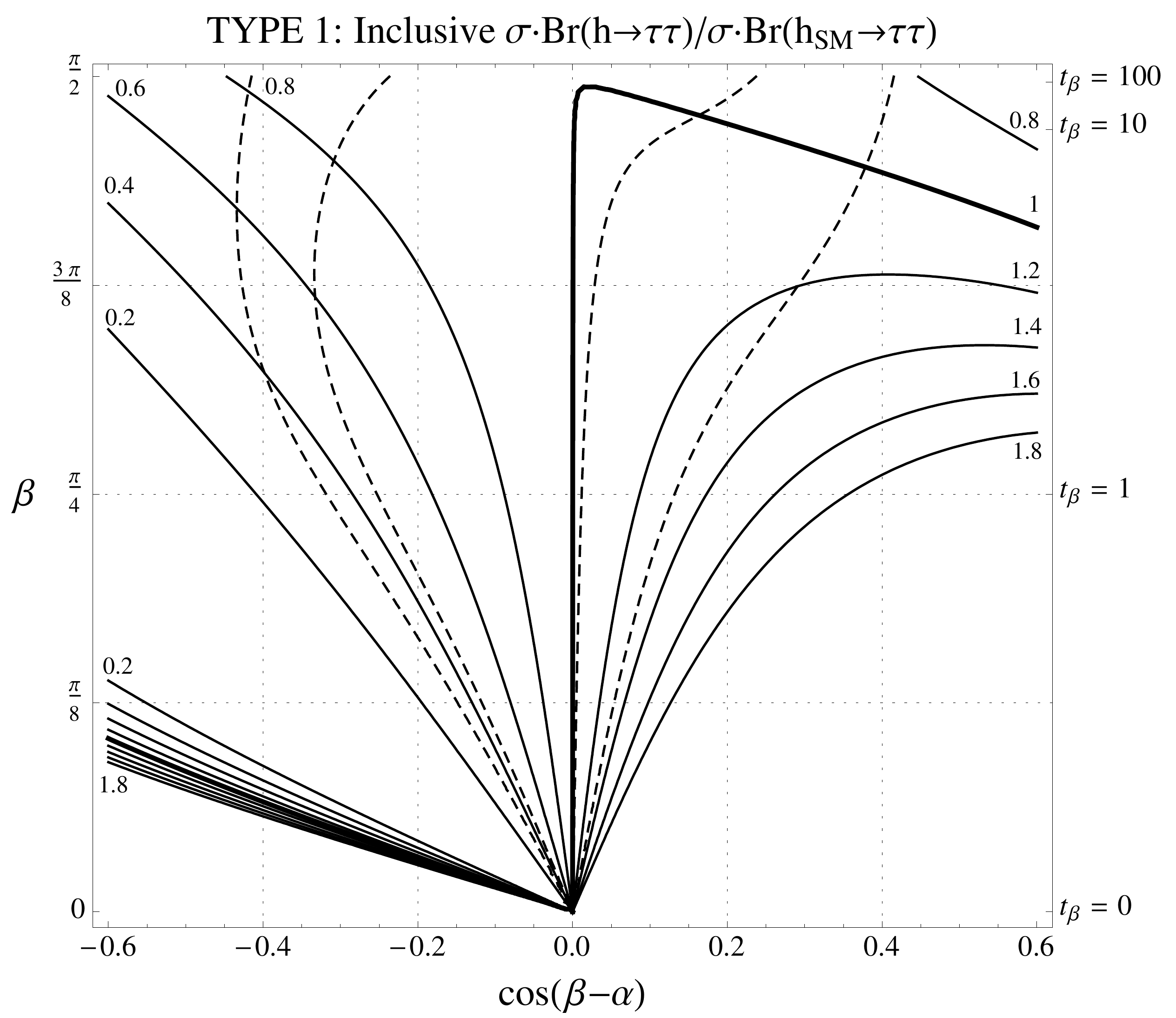} 
      \includegraphics[width=3in]{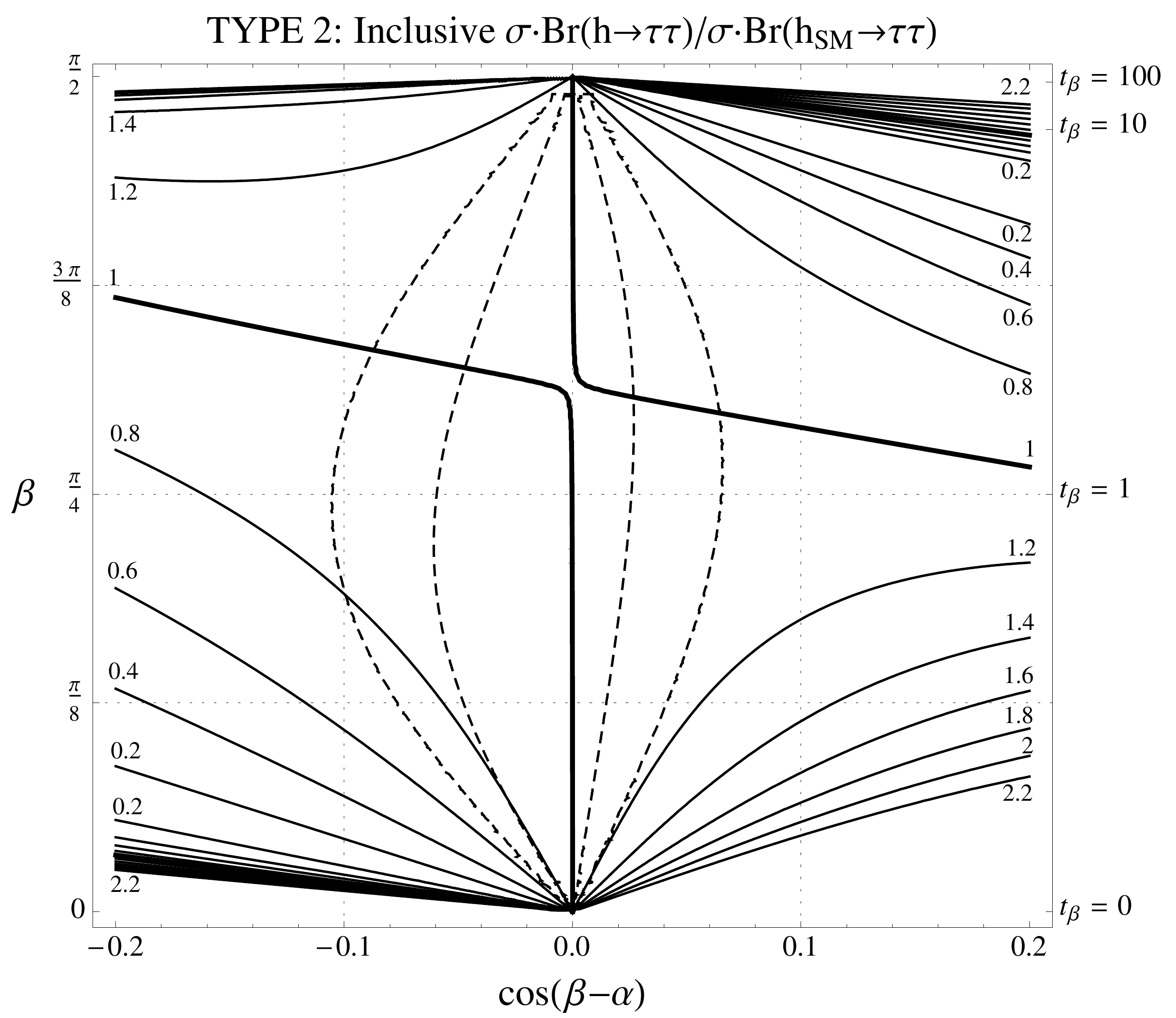} 
        \includegraphics[width=3in]{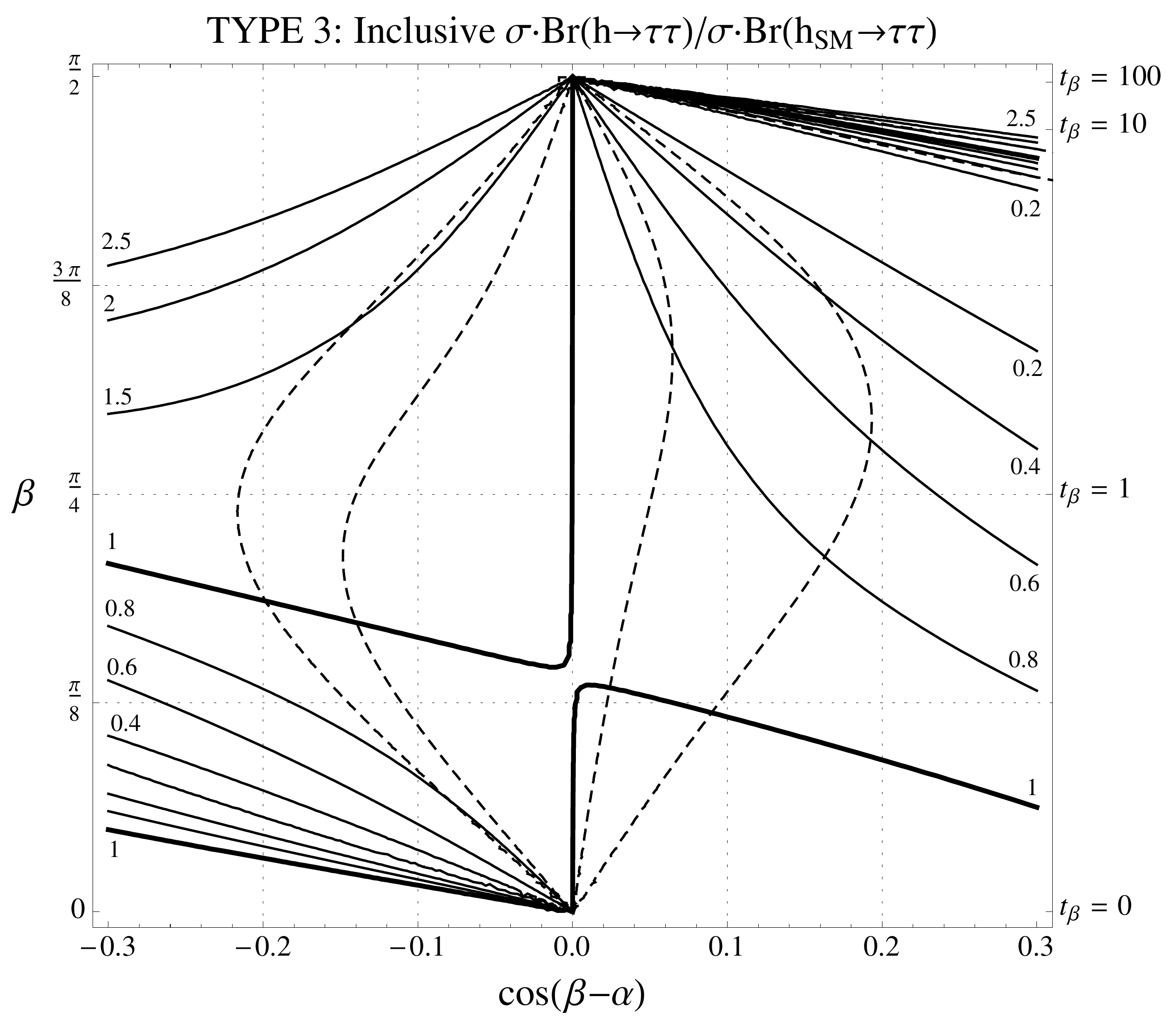} 
      \includegraphics[width=3in]{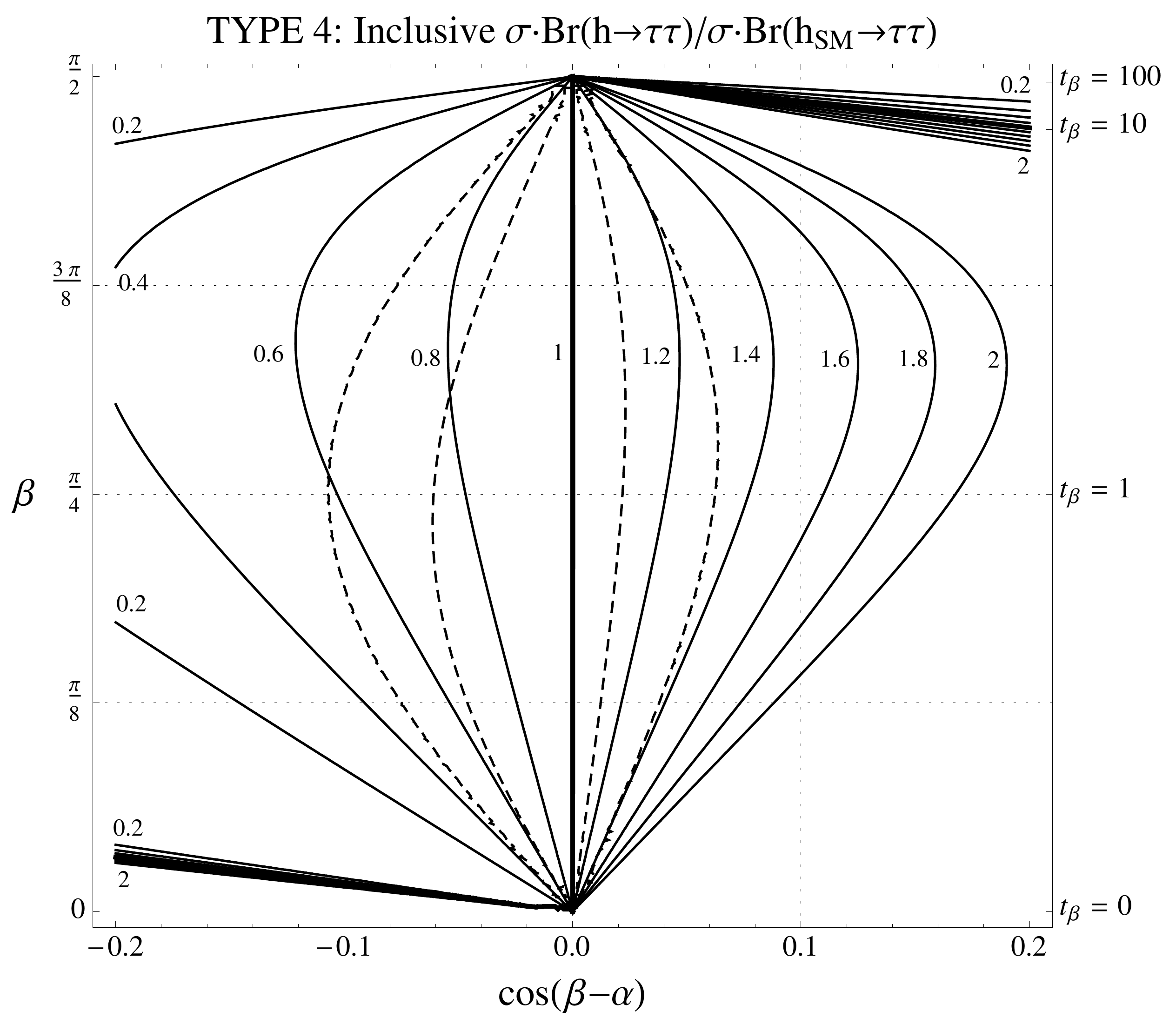} 
   \caption{Contours of the inclusive  $\sigma \cdot {\rm Br}(h \to \tau^+ \tau^-) / \sigma \cdot {\rm Br}(h_{SM} \to \tau^+ \tau^-) $ for 8 TeV $pp$ collisions for the SM-like Higgs boson, shown as a function of $\cos(\beta - \alpha)$ and $\beta$ for Type 1 (upper left), Type 2 (upper right), Type 3 (lower left), and Type 4 (lower right) 2HDM.  The inner (outer) dashed contour denotes the 68\% (95\%) CL best fit to the signals of the SM-like Higgs.}
   \label{fig:htautau}
\end{figure}

Contours of the inclusive ratio  $\sigma \cdot {\rm Br}(h \to \tau^+ \tau^-) / \sigma \cdot {\rm Br}(h_{SM} \to \tau^+ \tau^-) $ are shown in Fig.~\ref{fig:htautau}. The inclusive rates scale identically for both lepton flavors over the bulk of the 2HDM parameter space, $\sim c_t^2 \times c_\tau^2 / c_b^2$ except where $c_b \to 0$ and $\Gamma(h \to VV^*)$ takes over the total width. Thus in Type 1 models, around the alignment limit the inclusive rate scales as $\sim 1+ 2 \frac{\cos(\beta - \alpha)}{\tan \beta}$ and may be raised or lowered relative to the SM rate depending on the sign of $\cos(\beta - \alpha)$. The region of suppressed rate for $\cos(\beta - \alpha) < 0$ corresponds to $\alpha = - \pi/2$ where $c_b, c_\tau \to 0$ and is disfavored by current fits. 

In Type 2 models, for moderate $\tan \beta$ the inclusive rate scales similarly around the alignment limit, since any increase in coupling to leptons is offset by an increase in the total width due to enhanced bottom couplings. At large $\tan \beta$,  however, enhanced bottom couplings begin to dominate production modes (primarily through $b \bar b h$ production) and the scaling at large $\tan \beta$ scales as $\sim 1 - 2 \cos(\beta - \alpha) \tan \beta$. The region of suppressed rate for $\cos(\beta - \alpha) > 0$ corresponds to $\alpha = 0$ where $c_b, c_\tau \to 0$ and unsurprisingly is disfavored by current fits. 

In Type 3 models, the parametric scaling of production rates and total width cancel, leaving the enhancement of the lepton couplings to dominate the signal $\sim 1 - 2 \cos(\beta - \alpha) \tan \beta$. In Type 4 models, the scaling is $\sim 1 + 4 \frac{\cos(\beta - \alpha)}{\tan \beta} +2 \cos(\beta - \alpha) \tan \beta$. 

In general, there is not much room for significant deviations in the inclusive $h \to \tau^+ \tau^-$ and $h \to \mu^+ \mu^-$ rates given relatively good limits on the fermionic couplings of the Higgs. In Type 1 (3) models there may be $\mathcal{O}(50\%)$ suppression (enhancement) of the inclusive rate consistent with the 68 \% CL best fit, but in Type 2 and Type 4 models the remaining room for deviations is not more than $\mathcal{O}(20\%)$ above or below the SM rate, which is smaller than the current sensitivity in this channel. 

\subsection{$Vh$/VBF production of $h$ with $h \to \tau^+ \tau^-$} 

\begin{figure}[htbp] 
   \centering
   \includegraphics[width=3in]{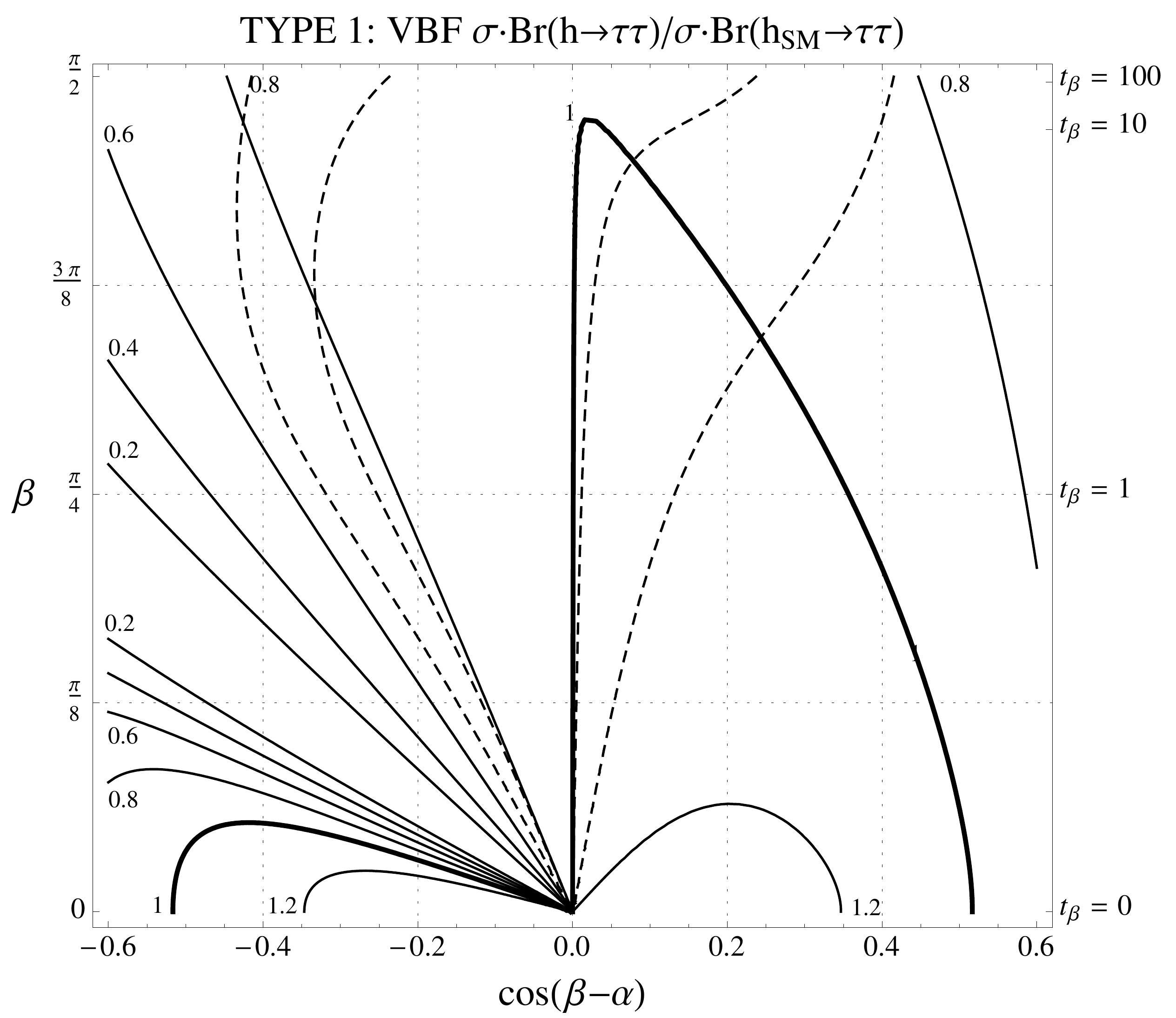} 
      \includegraphics[width=3in]{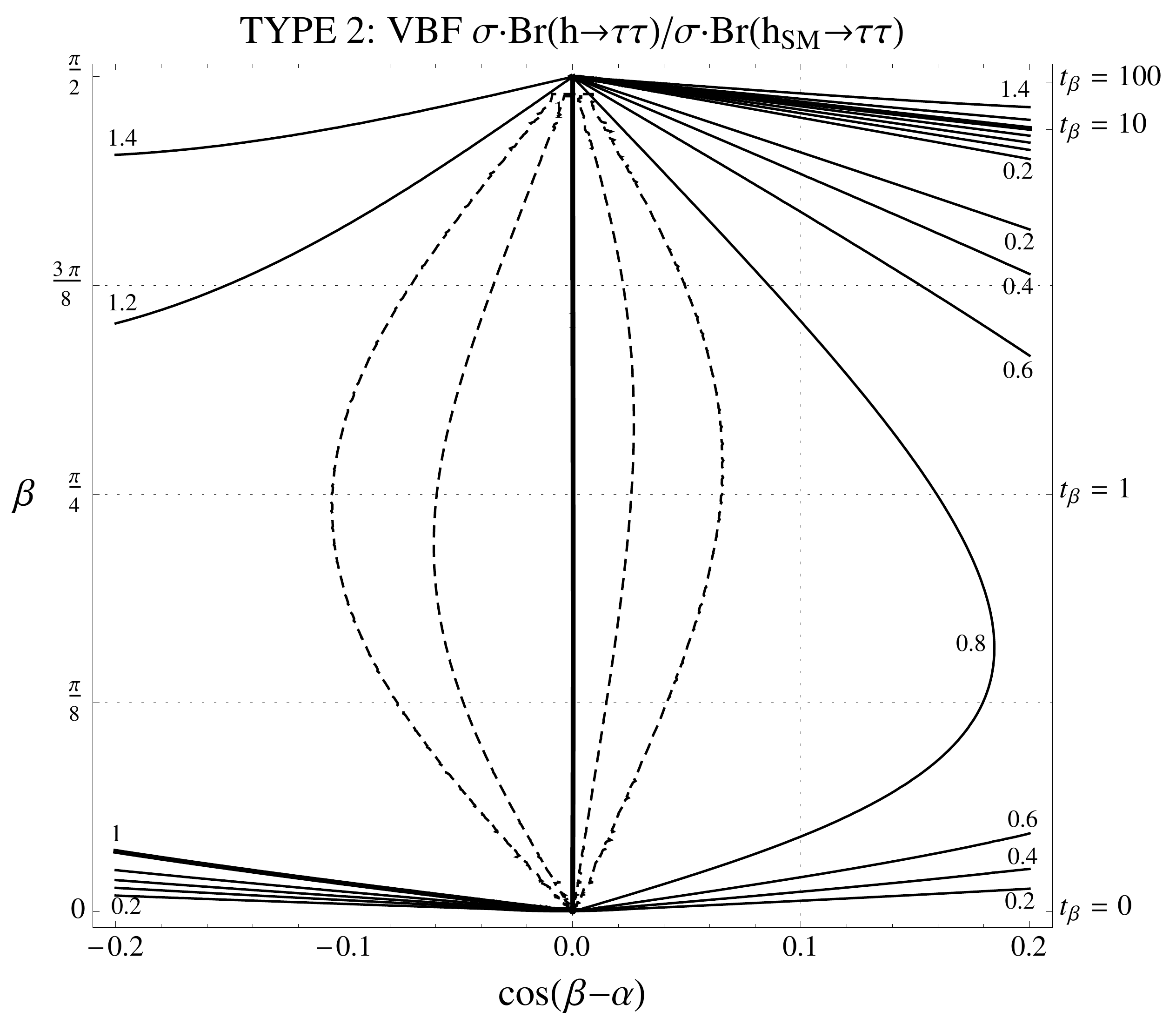} 
         \includegraphics[width=3in]{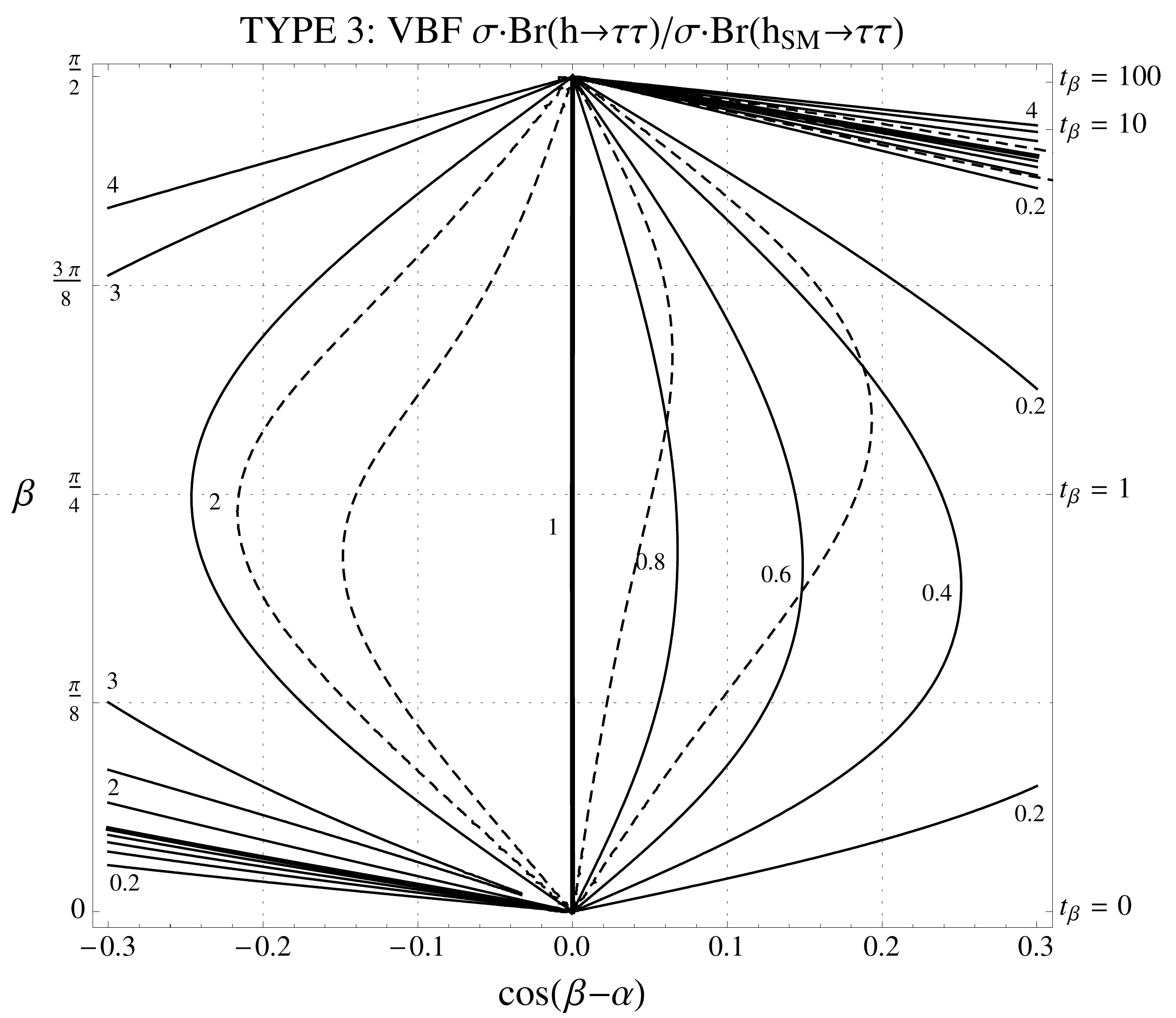} 
      \includegraphics[width=3in]{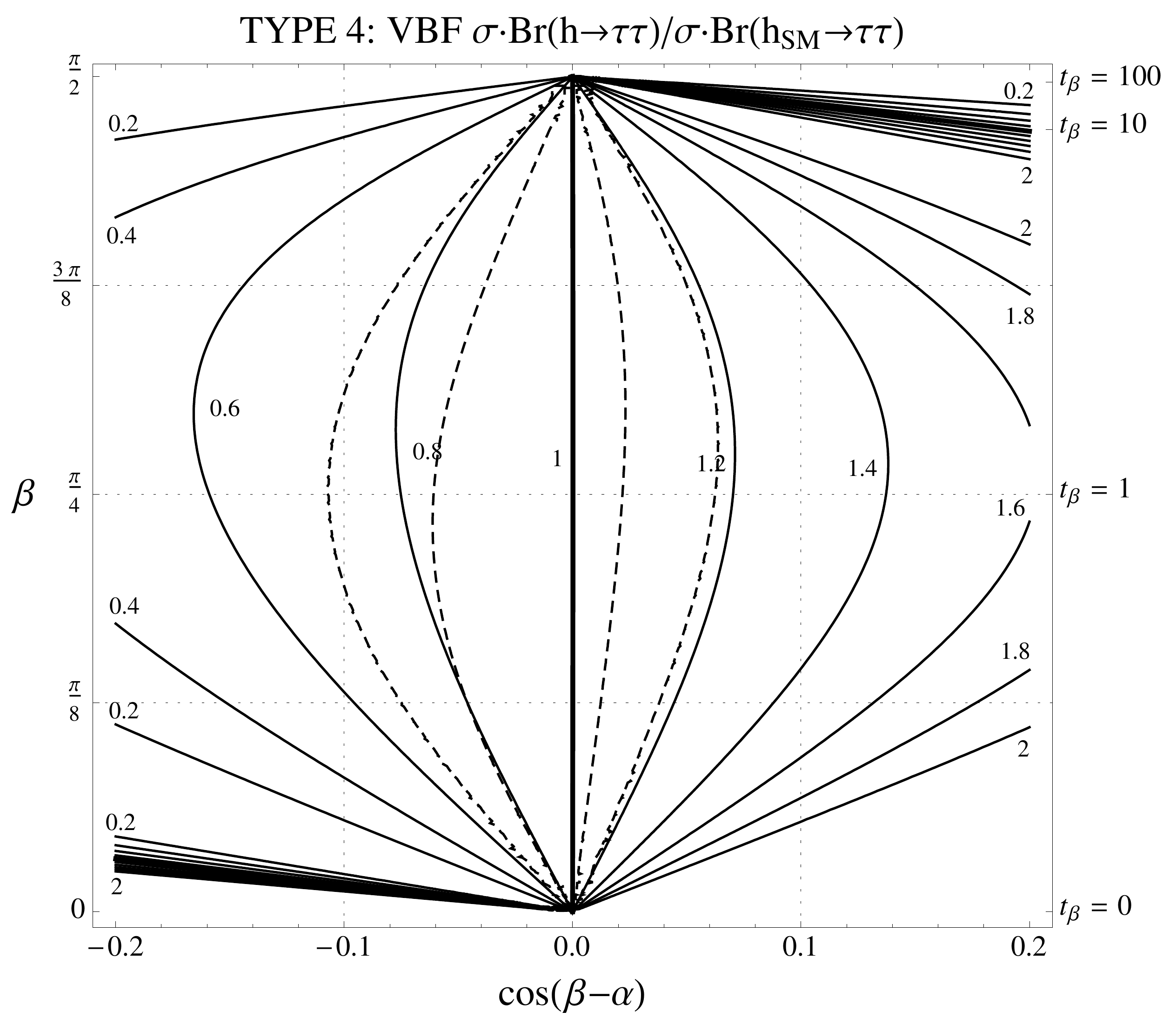} 
   \caption{Contours of $\sigma \cdot {\rm Br}({\rm VBF \; or \;} Vh  \to \tau^+ \tau^-) / \sigma \cdot {\rm Br}({\rm VBF \; or \;} Vh_{SM} \to \tau^+ \tau^-) $ for 8 TeV $pp$ collisions for the SM-like Higgs boson, shown as a function of $\cos(\beta - \alpha)$ and $\beta$ for Type 1 (upper left), Type 2 (upper right), Type 3 (lower left), and Type 4 (lower right) 2HDM.  The inner (outer) dashed contour denotes the 68\% (95\%) CL best fit to the signals of the SM-like Higgs.}
   \label{fig:VBFhtautau}
\end{figure}

Measurement of $h \to \tau^+ \tau^-$ in vector boson fusion currently has comparable sensitivity to the inclusive measurement, while $h \to \tau^+ \tau^-$ in $Vh$ associated production remains fairly poorly constrained due to  the low rate.  Contours of the exclusive ratio $\sigma \cdot {\rm Br}({\rm VBF \; or \;} Vh  \to \tau^+ \tau^-) / \sigma \cdot {\rm Br}({\rm VBF \; or \;} Vh_{SM} \to \tau^+ \tau^-) $ are shown in Fig.~\ref{fig:VBFhtautau}. Both processes have the same parametric scaling in 2HDM, and the exclusive rates scale as $\sim c_V^2 \times c_\tau^2 / c_b^2$ over most of the parameter space, with the above-mentioned features associated with $c_b, c_\tau \to 0$ lying outside the current best fit. In both Type 1 and Type 2 models this implies that the rates are constant at $\CO(\cos(\beta - \alpha))$, and only vary significantly far from the alignment limit. Indeed, in both types the variation consistent with current fits is no more than $\sim \pm 20 \%$ of the SM value.

However, in Type 3 and Type 4 models the parametric freedom is greater due to the difference between lepton and bottom quark couplings.  In Type 3 models the inclusive rate scales as $1 - 2 \frac{\cos(\beta - \alpha)}{\tan \beta}$ at small $\tan \beta$ and as $1 - 2 \cos(\beta - \alpha) \tan \beta$ at large $\tan \beta$, allowing up to 50\% enhancement of this exclusive rate consistent with the 68\% fit. In Type 4 models the sign of the variation changes: $1 + 2 \frac{\cos(\beta - \alpha)}{\tan \beta}$ at small $\tan \beta$ and $1 + 2 \cos(\beta - \alpha) \tan \beta$ at large $\tan \beta$.

\subsection{$Vh$ production of $h$ with $h \to b \bar b$}

Measurement of $Vh$ associated production of $h$ with $h \to b \bar b$ remains challenging at the LHC and current direct limits are fairly loose.  However, the exclusive rate for $b \bar b$ production scales identically to the rate for $Vh$ production of $h$ with $h \to \tau^+ \tau^-$ in Type 1 and Type 2 models and are essentially constant near the alignment limit; contours are identical to those shown in Fig.~\ref{fig:VBFhtautau}. In fact, all four types of 2HDM scale similarly as long as $h \to b \bar b$ still dominates the total width of the Higgs.

\subsection{$Vh$/VBF production of $h$ with $h \to \gamma \gamma$}

VBF production of $h$ with $h \to \gamma \gamma$ is measured by both collaborations through dijet-tagged categories, but error bars remain fairly large. The 8 TeV data set allows preliminary measurements of $Vh$ associated production with $h \to \gamma \gamma$, but current error bars are too large to place a meaningful constraint. 

\begin{figure}[htbp] 
   \centering
   \includegraphics[width=3in]{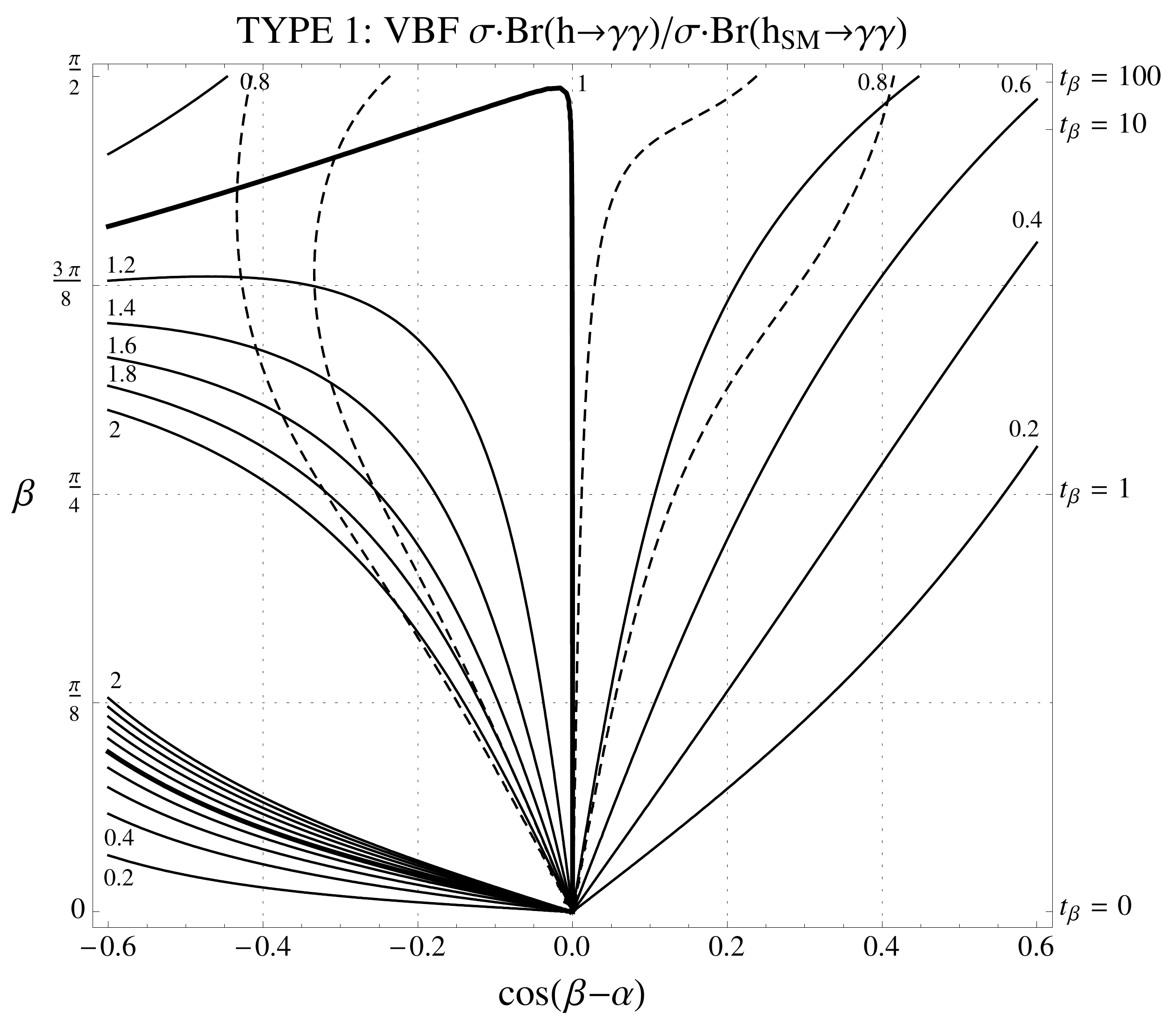} 
      \includegraphics[width=3in]{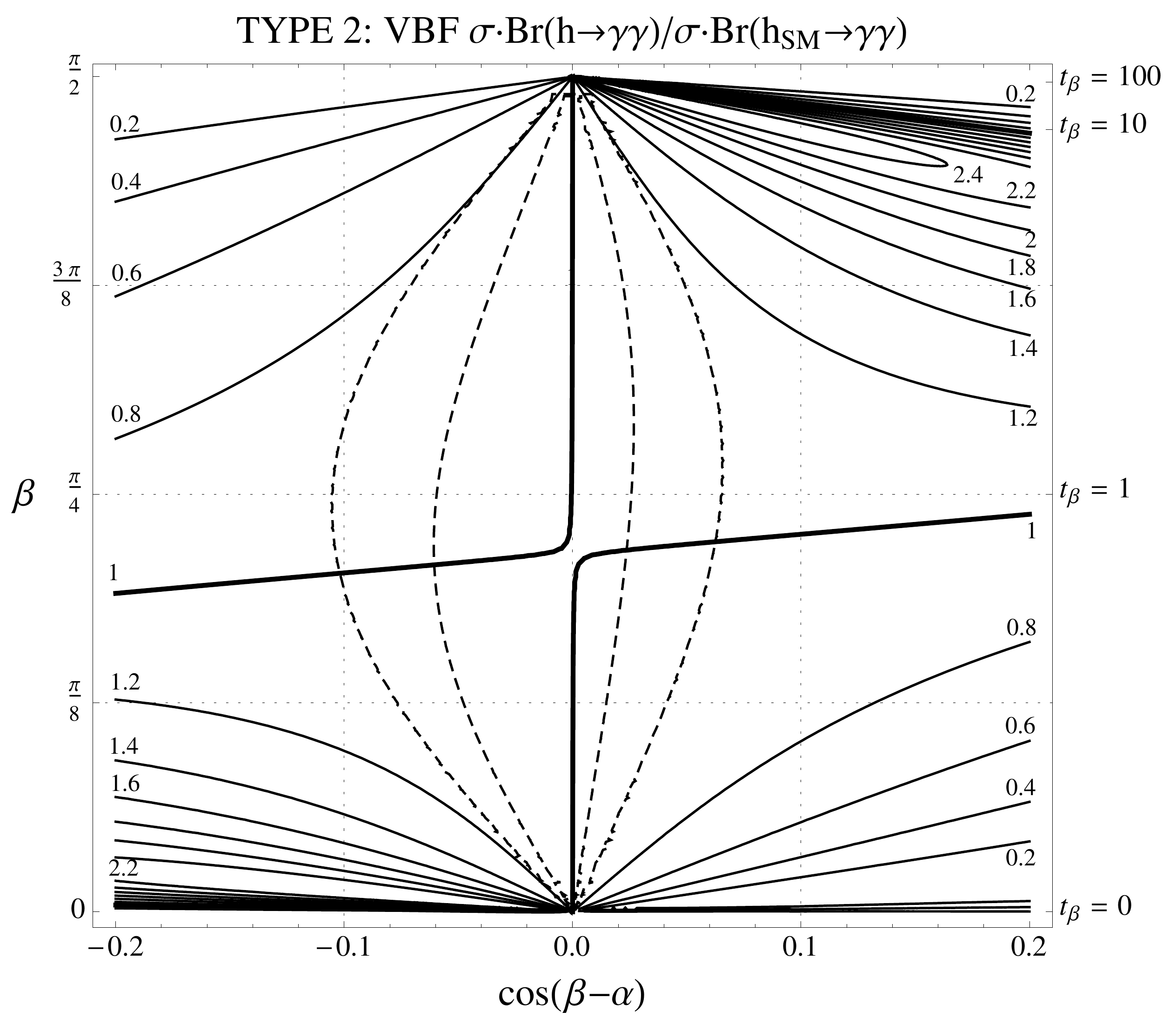} 
   \caption{Contours of $\sigma \cdot {\rm Br}({\rm VBF \; or \;} Vh \to \gamma \gamma) / \sigma \cdot {\rm Br}({\rm VBF \; or \;} Vh_{SM} \to \gamma \gamma) $ for 8 TeV $pp$ collisions for the SM-like Higgs boson, shown as a function of $\cos(\beta - \alpha)$ and $\beta$ for Type 1 (left) and Type 2 (right) 2HDM. The inner (outer) dashed contour denotes the 68\% (95\%) CL best fit to the signals of the SM-like Higgs.}
   \label{fig:VBFhgammagamma}
\end{figure}

 Contours of the exclusive ratio $\sigma \cdot {\rm Br}({\rm VBF \; or \;} Vh \to \gamma \gamma) / \sigma \cdot {\rm Br}({\rm VBF \; or \;} Vh_{SM} \to \gamma \gamma) $ are shown in Fig.~\ref{fig:VBFhgammagamma}. Both processes share the same parametric scaling in 2HDM.  Unless the  coupling to vectors is highly suppressed, the exclusive rates scale as $\sim c_V^2 \times c_V^2 / c_b^2$ since the $W$ loop dominates the effective $h \gamma \gamma$ coupling. In Type 1 models the parametric scaling is dominated by the change in the total width since $c_V \sim 1$, and so over the bulk of parameter space scales as $1 - 2 \frac{\cos(\beta - \alpha)}{\tan \beta}$. The region of enhancement for $\cos(\beta - \alpha) < 0$ again corresponds to $c_b, c_\tau \to 0$, but for this process it leads to an enhancement (rather than the suppression apparent in inclusive or VBF/$Vh$ production with $h \to \tau^+ \tau^-$) because only the total width decreases while the production and decay modes remain constant. In Type 2 models the scaling is again dominated by the total width, $\sim 1 + 2 \cos(\beta - \alpha) \tan \beta$, and the region of enhancement for $\cos(\beta - \alpha) > 0$ corresponds to $c_b, c_\tau \to 0$ in analogy with the Type 1 case. 
 
 In Type 1 2HDM there is still considerable room for an enhanced rate of VBF production with $h \to \gamma \gamma$, with an enhancement by as much as $\sim 80$\% consistent with the 68\% CL fit. Although the recent CMS 8 TeV measurement of dijet-tagged diphoton categories now shows a deficit with respect to the SM, the CMS 7 TeV measurement and ATLAS 7 and 8 TeV measurements remain high, and this remains an interesting channel to probe in future measurements. In Type 2 2HDM the situation is much more tightly constrained, with  no more than $\pm 20\%$ variation consistent with the 68\% CL fit.

\subsection{$Vh$/VBF production of $h$ with $h \to VV^*$}

$Vh$ and VBF associated production of $h$ with $h \to VV^*$ is currently poorly constrained by ATLAS and CMS, though both collaborations currently measure tagged categories for $WW^*$ final states, and CMS now includes a dijet-tagged category for $ZZ^*$.  These exclusive rates scale as $c_V^2 \times c_V^2 / c_b^2$, much as the $Vh$/VBF production of $h$ with $h \to \gamma \gamma$, and unsurprisingly the parametrics are essentially identical; we do not show them explicitly. 

As with $Vh$/VBF production of $h$ with $h \to \gamma \gamma$, in Type 1 2HDM there is room for up to $\sim 80$\% enhancement in these channels consistent with the 68\% CL fit. Should the VBF diphoton rate remain high, ancillary measurements in VBF and $Vh$ with $h \to VV^*$ could provide a useful validation.

\subsection{$t\bar{t}$ associated production of $h$ with $h \to \gamma \gamma$}

\begin{figure}[htbp] 
   \centering
   \includegraphics[width=3in]{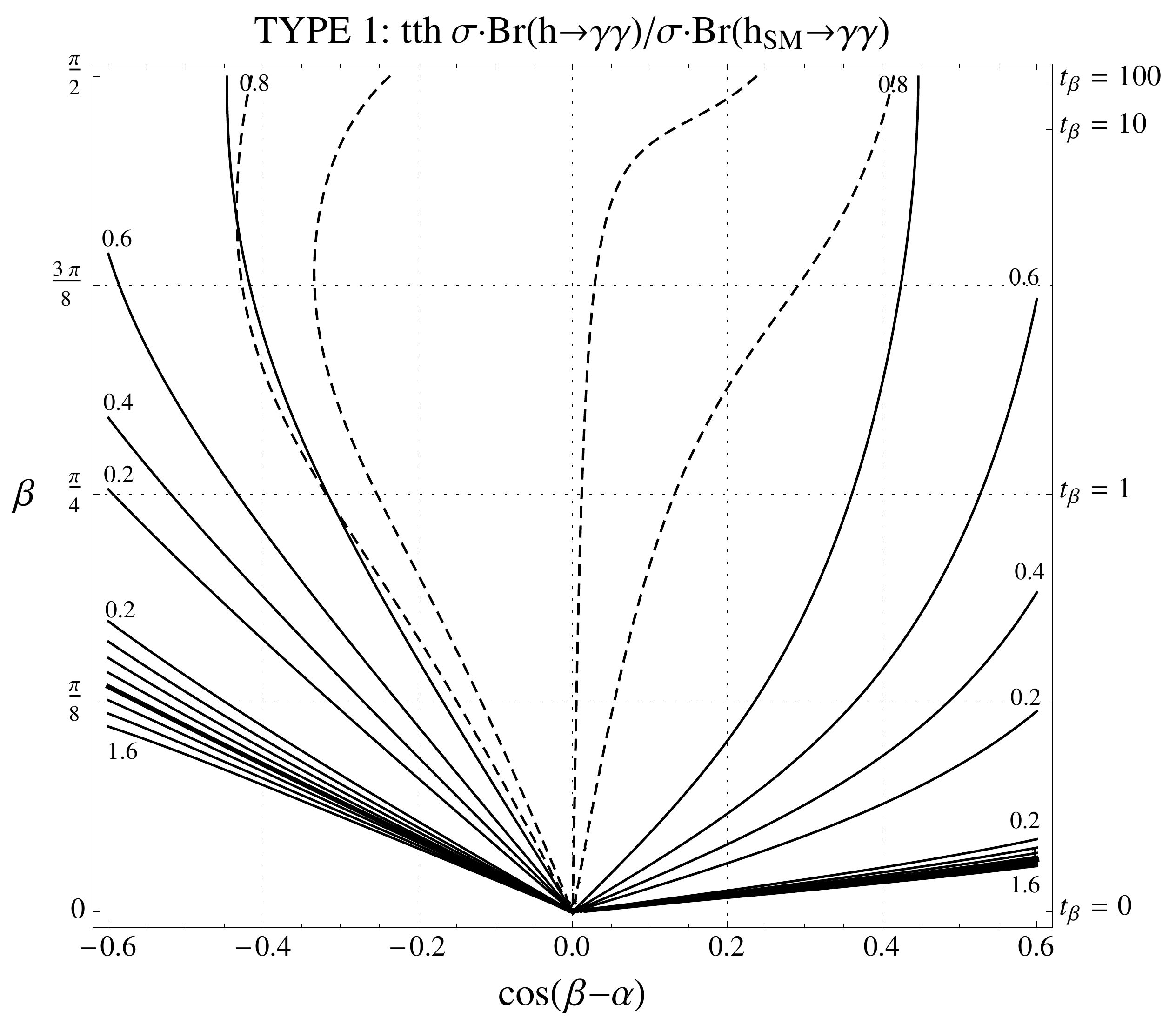} 
      \includegraphics[width=3in]{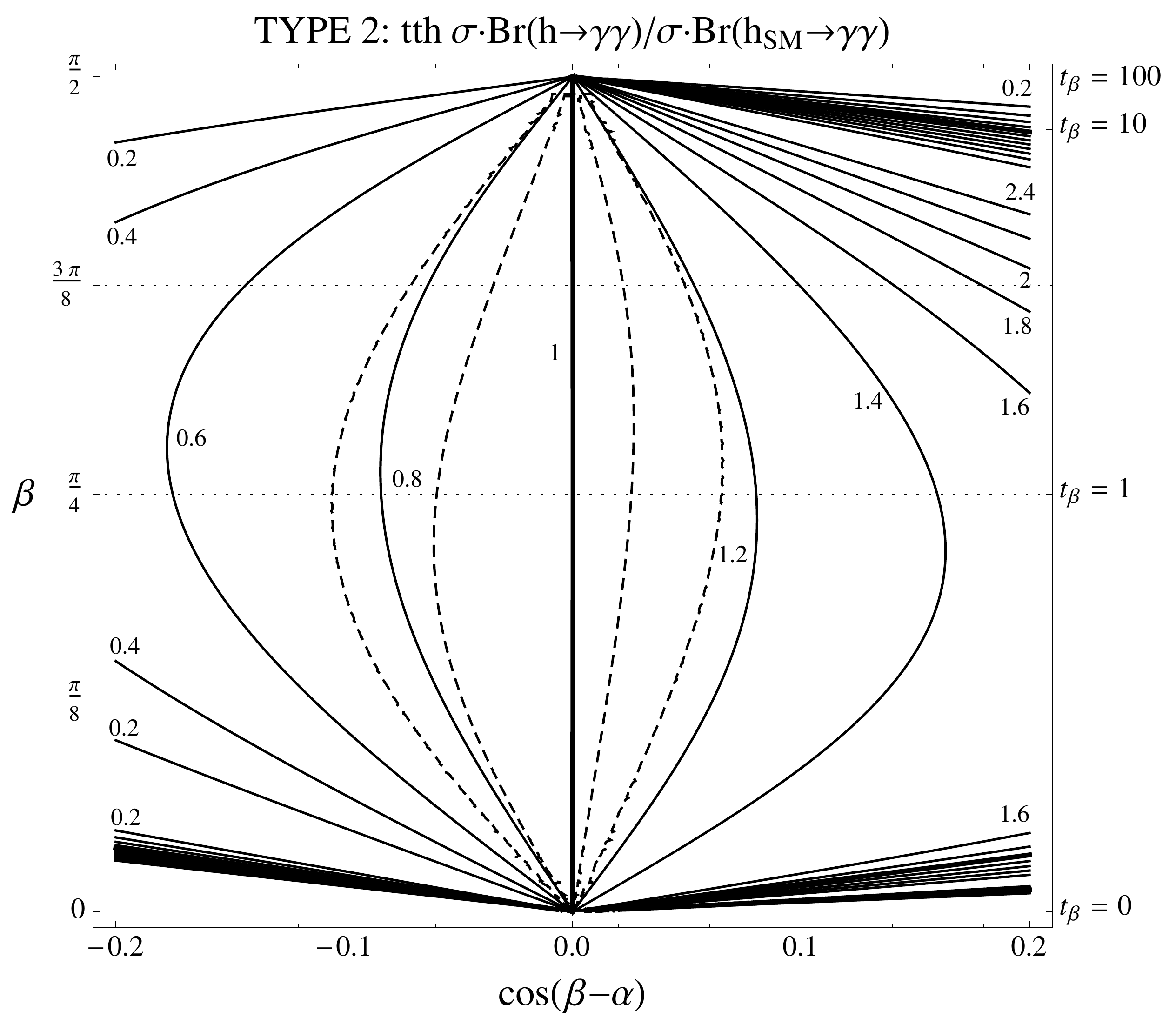} 
   \caption{Contours of $\sigma \cdot {\rm Br}(t \bar t h  \to \gamma \gamma) / \sigma \cdot {\rm Br}(t \bar t h_{SM} \to \gamma \gamma) $ for 8 TeV $pp$ collisions for the SM-like Higgs boson, shown as a function of $\cos(\beta - \alpha)$ and $\beta$ for Type 1 (left) and Type 2 (right) 2HDM. The inner (outer) dashed contour denotes the 68\% (95\%) CL best fit to the signals of the SM-like Higgs.}
   \label{fig:tthgammagamma}
\end{figure}

There is currently no meaningful measurement of $t \bar t$ associated production of $h$, though both ATLAS and CMS quote preliminary measurements of $t \bar t h$ with $h \to b \bar b$ with low sensitivity. The $t \bar t$ associated production of $h$ with $h \to \gamma \gamma$ should be measurable with considerable integrated luminosity at 14 TeV. Contours of the exclusive $\sigma \cdot {\rm Br}(t \bar t h  \to \gamma \gamma) / \sigma \cdot {\rm Br}(t \bar t h_{SM} \to \gamma \gamma) $  ratio are shown in Fig.~\ref{fig:tthgammagamma}. The rate scales as $c_t^2 \times c_V^2 / c_b^2$ over much of the parameter space. These parametrics are similar to the parametrics for inclusive production with $h \to VV^*, \gamma \gamma$, and so unsurprisingly the signal contours trace the fit contours closely. In Type 1 2HDM this renders the rate flat around the alignment limit to $\mathcal{O}(\cos(\beta - \alpha))$, with mild modulation from sub-leading production modes. In Type 2 2HDM the parametric scaling around the alignment limit is $\sim 1 + 2 \frac{\cos(\beta - \alpha)}{\tan \beta} + 2 \cos(\beta - \alpha) \tan \beta$, so that the overall rate varies rapidly at large and small $\tan \beta$. This leaves little room for significant deviations in $t\bar{t}$ associated production of $h$ with $h \to \gamma \gamma$ in these 2HDM.

\subsection{Inclusive production of $h$ with $h \to Z \gamma$}

Inclusive production of $h$ with $h \to Z \gamma$ is poorly constrained by current measurements, but may be measured at the 14 TeV LHC provided considerable integrated luminosity. The rate again scales as $c_t^2 \times c_V^2 / c_b^2$ over much of the parameter space since the $h Z \gamma$ coupling is dominated by vector loops, so that the contours of $h \to Z \gamma$ are similar to those of $t \bar t h$ with $h \to \gamma \gamma$ and trace the contours of the fit. Given the parametric similarity to $t \bar t h$ with $h \to \gamma \gamma$, we do not show the contours explicitly as a function of $\alpha$ and $\beta$.  Since $h \to Z \gamma$ tracks the overall fit contours so closely, there is little room for substantial deviations from the SM prediction in these 2HDM.

\subsection{Future prospects}

For the most part, many decays of the SM-like Higgs that are poorly constrained by current data but measurable at 14 TeV are not expected to deviate significantly from the SM expectation, with typically $\pm 20 \%$ deviation consistent with the current $68 \%$ CL coupling fits. This is due to the fact that these modes scale similarly to well-measured production and decay modes that dominate the coupling fits. There are three notable exceptions in a Type 1 2HDM, which should be correlated in the event that substantial deviations are observed. An enhancement of up to $\sim 80\%$ in VBF or $Vh$ associated production of $h$ with $h \to \gamma \gamma, VV^*$ is consistent with the current $68 \%$ CL coupling fits at low $\tan \beta$. If such enhancement is present, it would be accompanied by as much as a $50\%$ diminution in the inclusive $h \to \tau^+ \tau^-$ rate, with the suppression in $h \to \tau^+ \tau^-$ proportional to the enhancement in $h \to \gamma \gamma, VV^*$. The prospects are similar in a Type 3 2HDM,  except that enhancement of VBF or $Vh$ associated production with $h \to \gamma \gamma, VV^*$ would be accompanied by enhancement in both inclusive and exclusive processes with $h \to \tau^+ \tau^-$. 

Given that most remaining channels are fairly well constrained by current coupling fits, the possible direct signals of heavier Higgs scalars take on additional importance.

\section{Direct Signs of the Second Higgs Doublet \label{sec6}}

In addition to looking for evidence of additional states through deviations from SM predictions for the couplings of the SM-like Higgs, it is instrumental to search directly for additional scalars, both in standard Higgs channels and in cascade decays involving multiple scalars. The most fruitful channels typically involve gluon fusion production of $H$ and $A$, both because this offers a large production cross section and because the proximity to the alignment limit favored by current fits suppresses VBF and $VH$ associated production of $H$. The production rate for $H^\pm$ is small unless $m_{H^\pm} < m_t$, in which case $H^\pm$ appears in $t \bar t$ pair production followed by the decay $t \to H^+ b$. In this work we will not consider the case $m_{H^\pm} < m_t$, where there is a small rate for $t \bar b H^\pm$ associated production.

\begin{figure}[htbp] 
   \centering
   \includegraphics[width=3in]{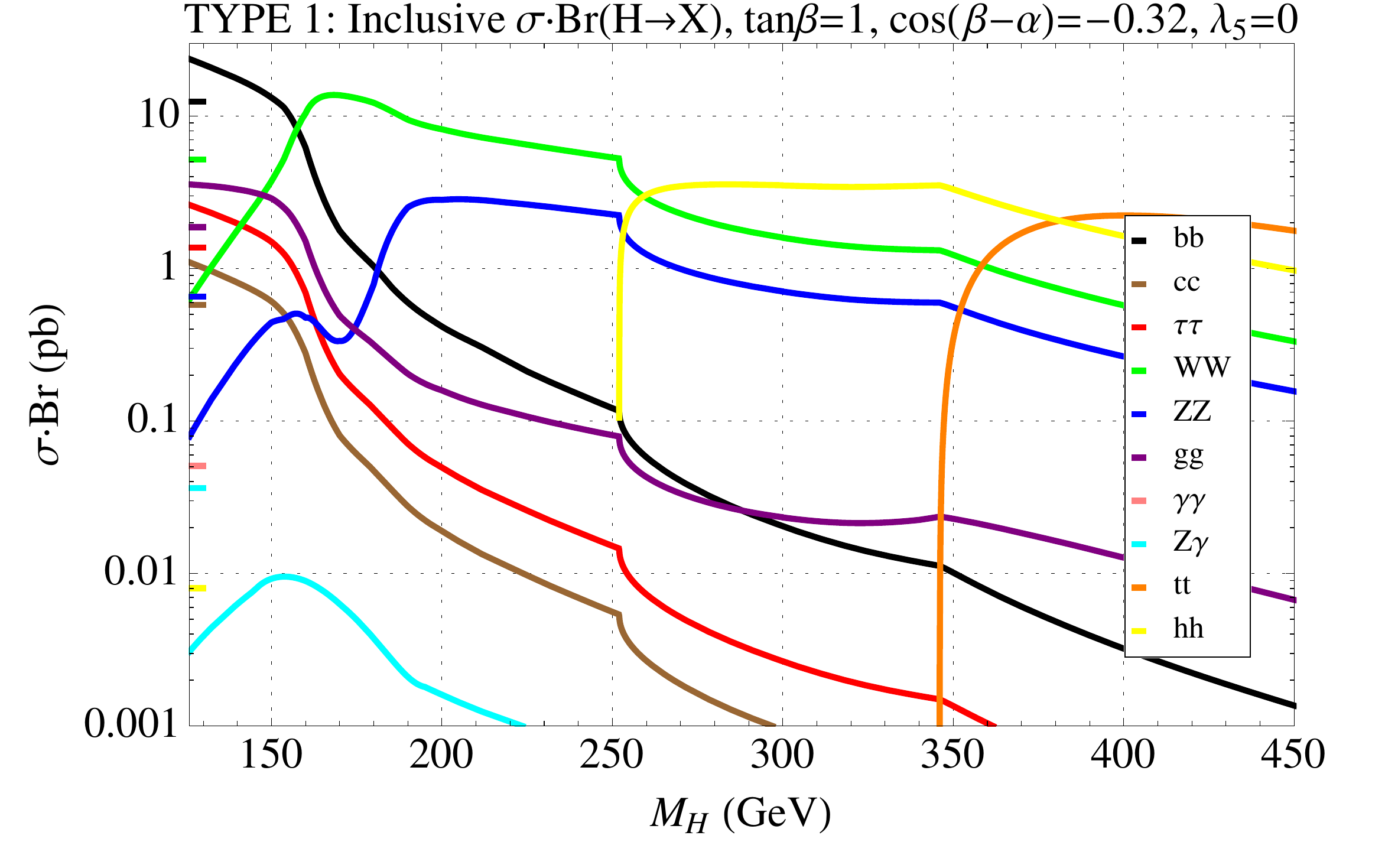} 
      \includegraphics[width=3in]{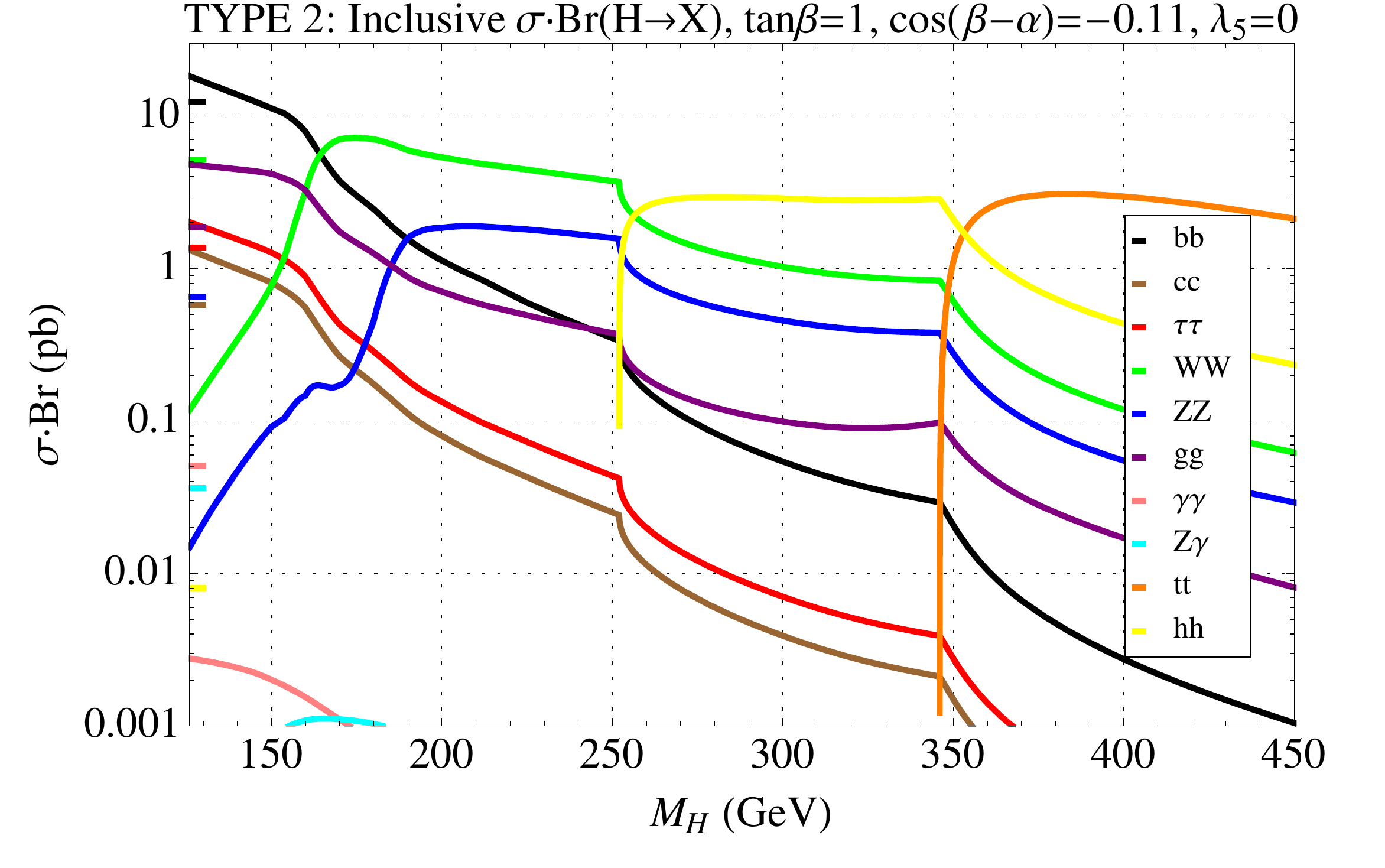} 
         \includegraphics[width=3in]{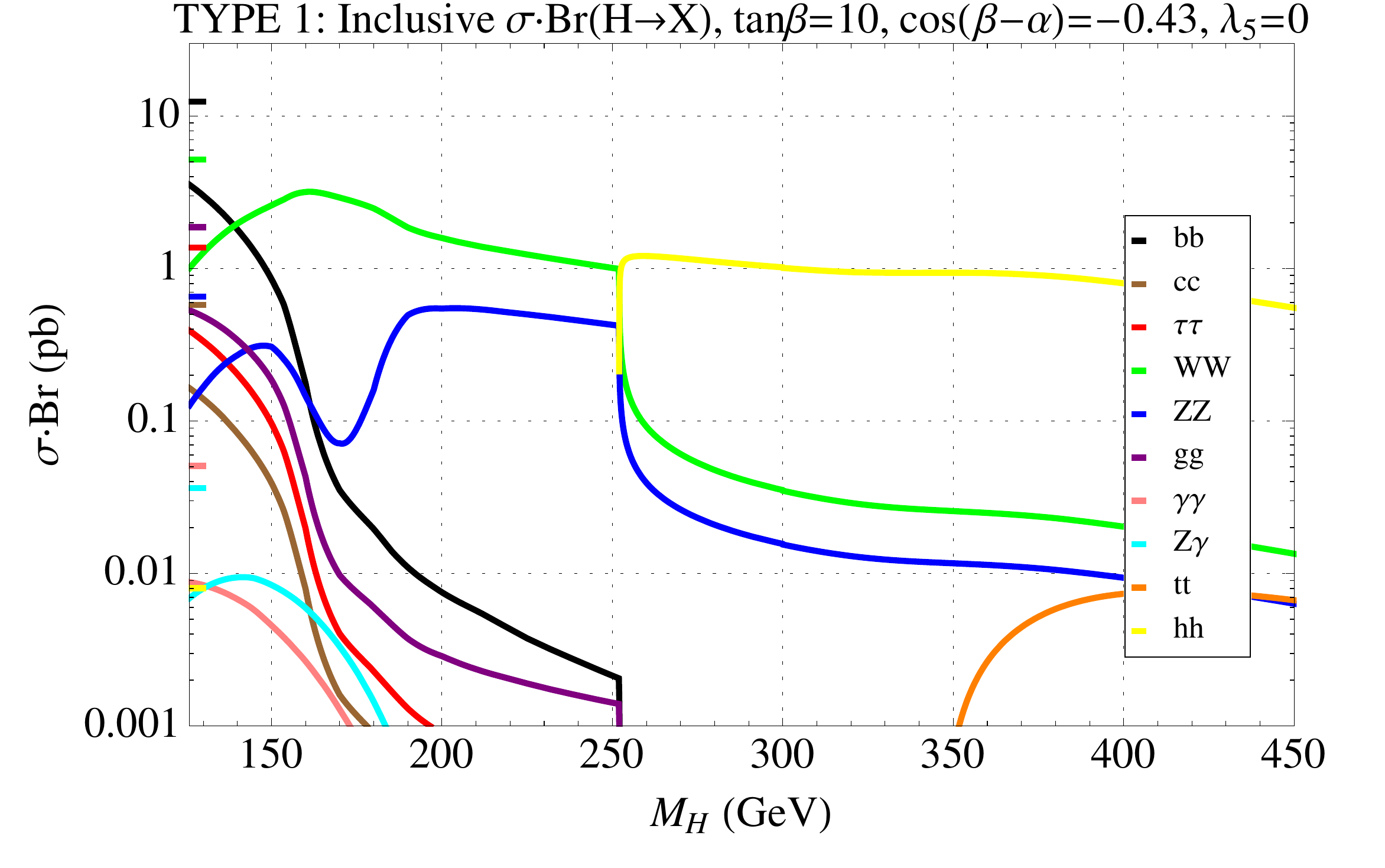} 
      \includegraphics[width=3in]{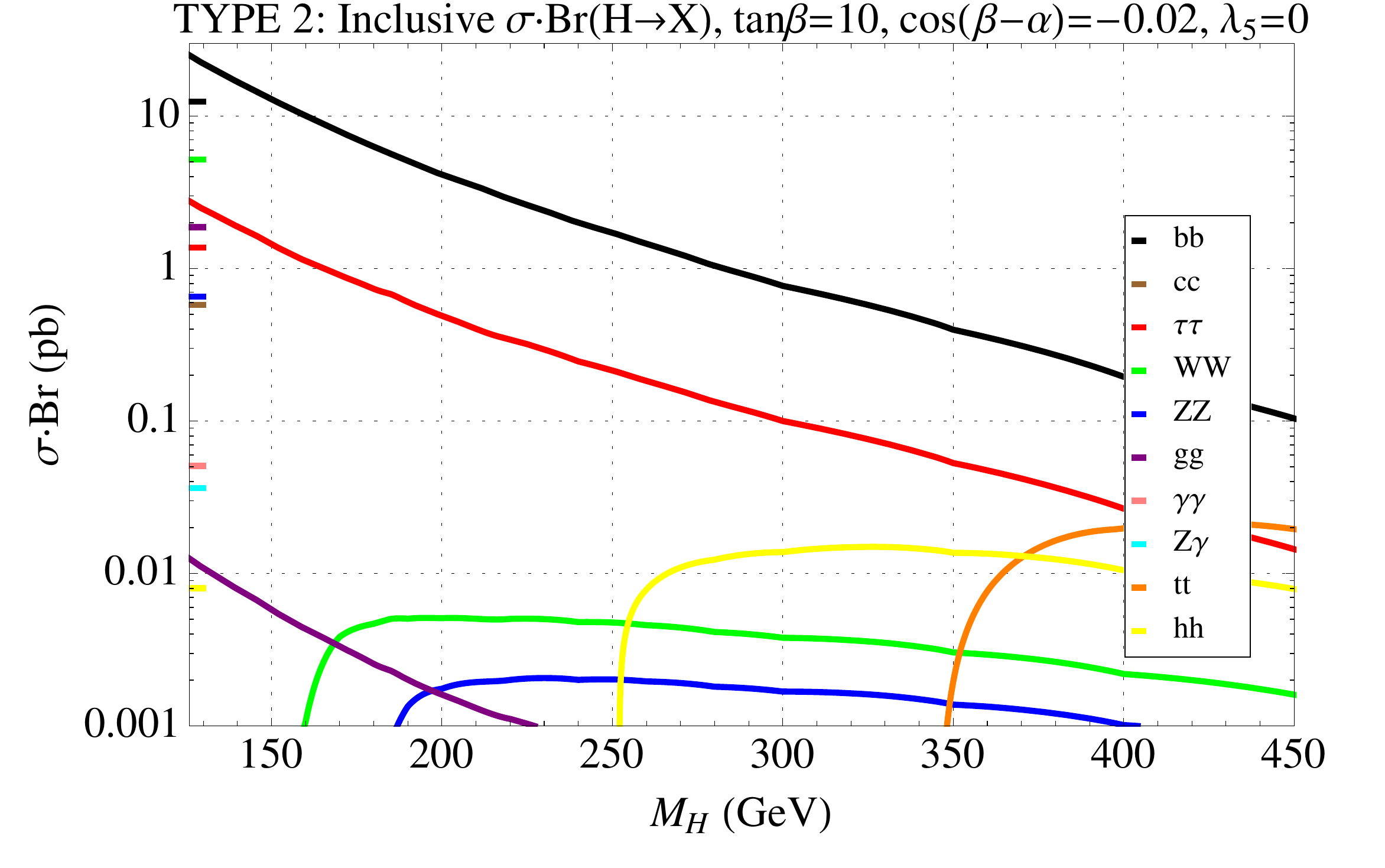} 
   \caption{Cross section times branching ratio $\sigma \cdot {\rm Br}(H \to X)$ to available final states in units of pb for 8 TeV $pp$ collisions for the non-SM-like scalar Higgs boson, shown as a function of $m_H$. Upper left: $\tan \beta = 1$, $\cos(\beta - \alpha) = -0.32 $ for Type 1 2HDM. Upper right: $\tan \beta = 1$, $\cos(\beta - \alpha) = -0.11$ for Type 2 2HDM. Lower left:  $\tan \beta = 10$, $\cos(\beta - \alpha) = -0.43$ for Type 1 2HDM. Lower right: $\tan \beta = 10$, $\cos(\beta - \alpha) = -0.02$ for Type 2 2HDM. In each case we have chosen $\lambda_{5,6,7} = 0$ and $m_A = m_H$. }
   \label{fig:HBRs}
\end{figure}

\begin{figure}[htbp] 
   \centering
   \includegraphics[width=3in]{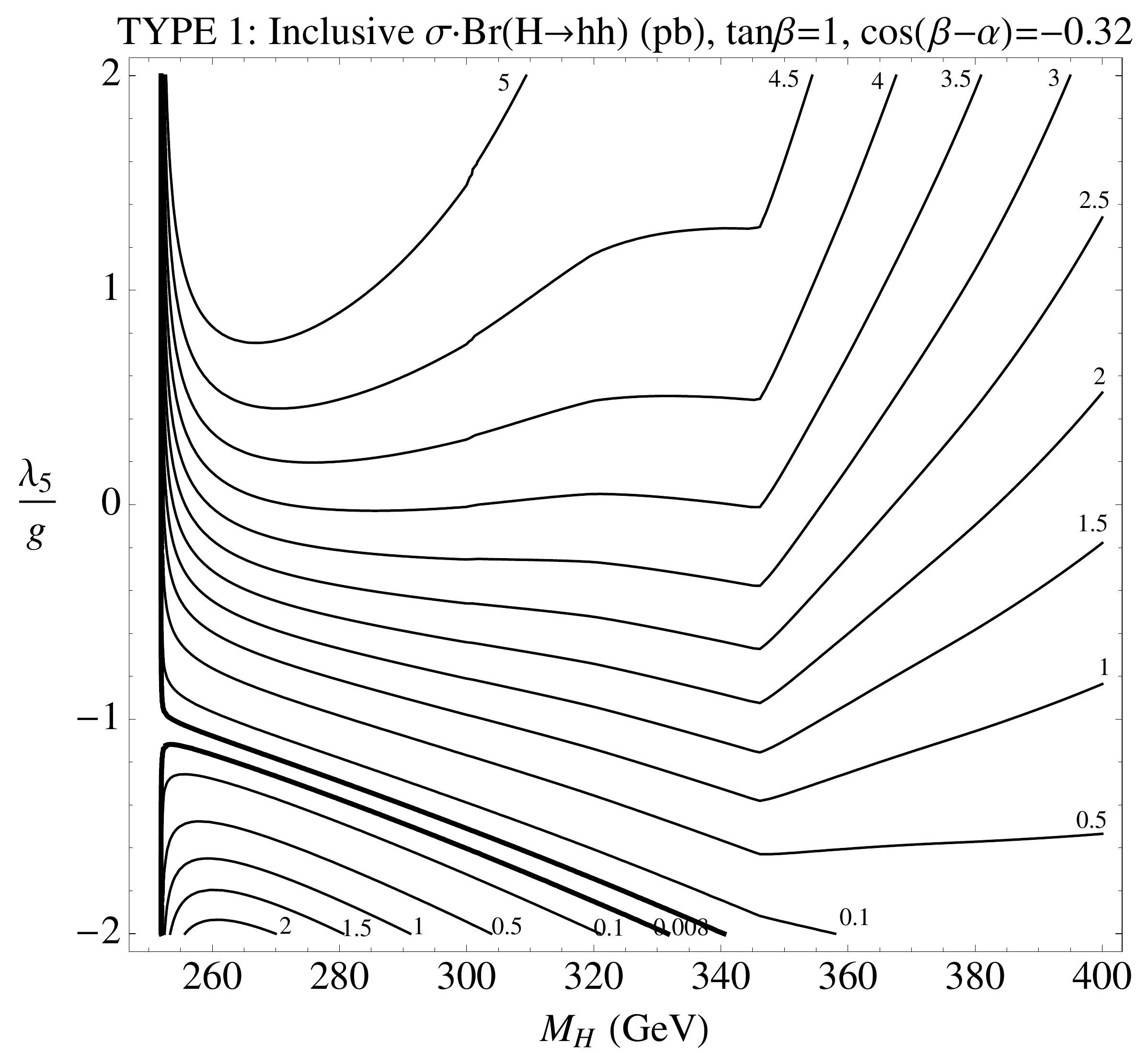} 
      \includegraphics[width=3in]{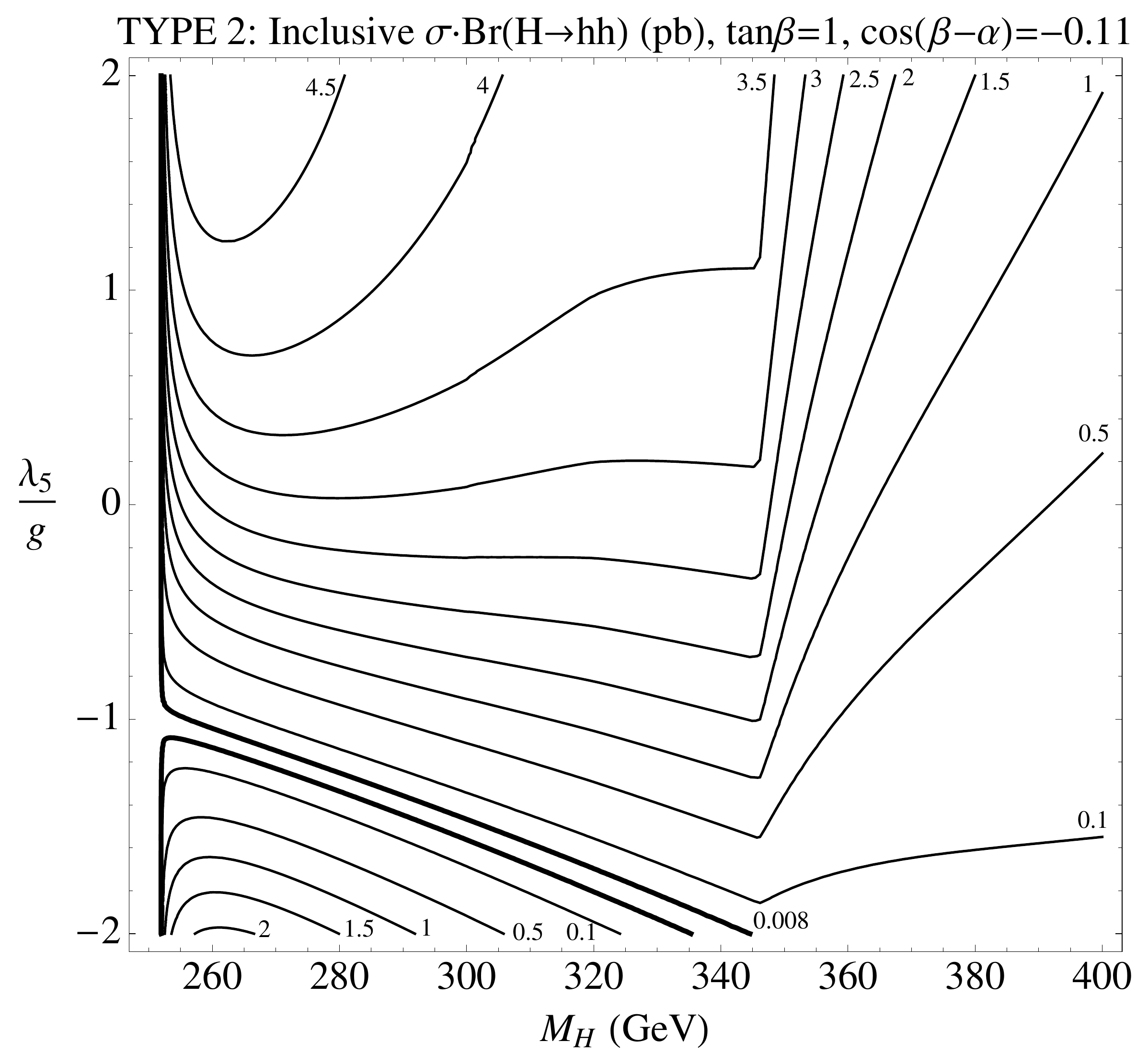} 
         \includegraphics[width=3in]{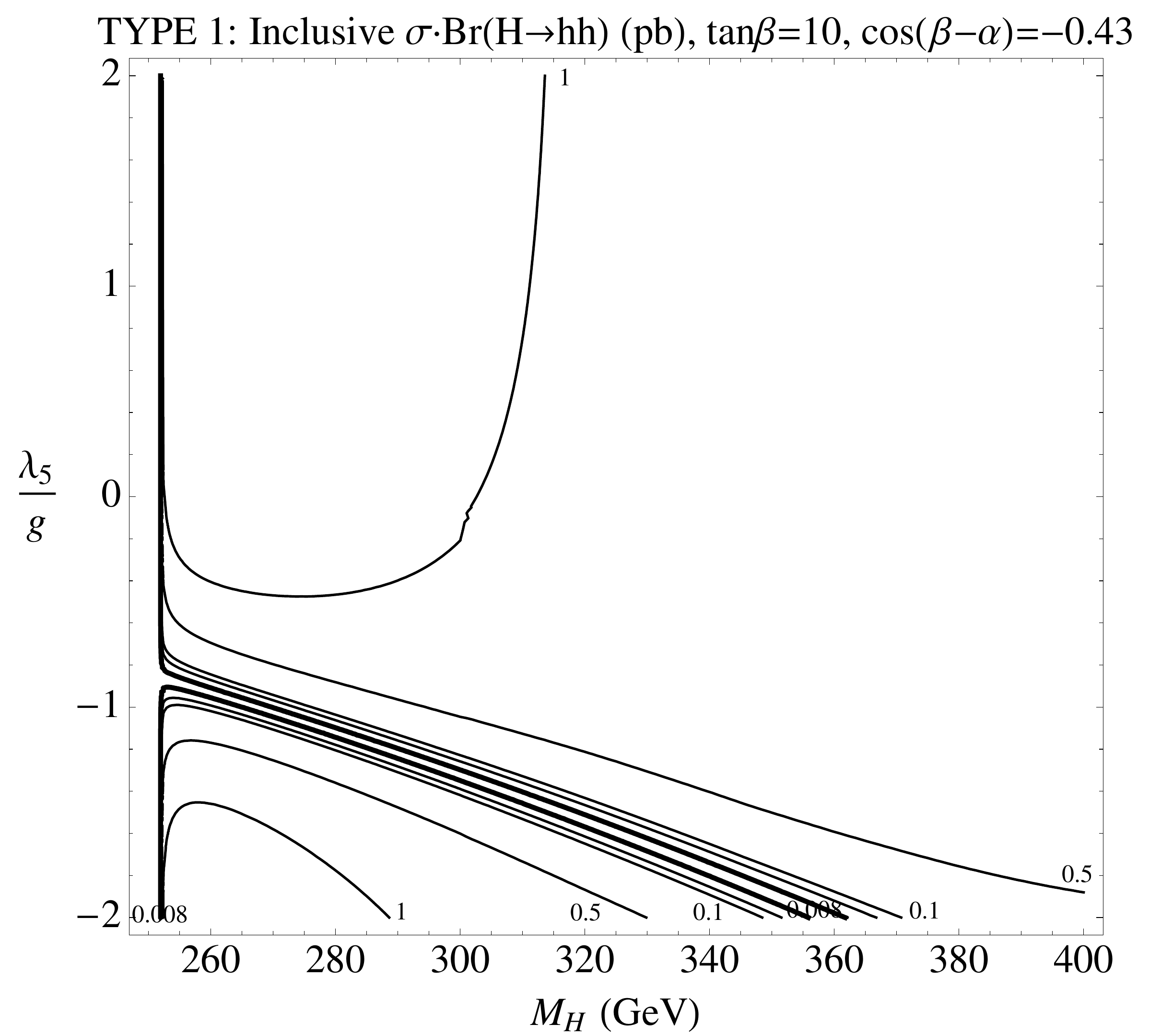} 
      \includegraphics[width=3in]{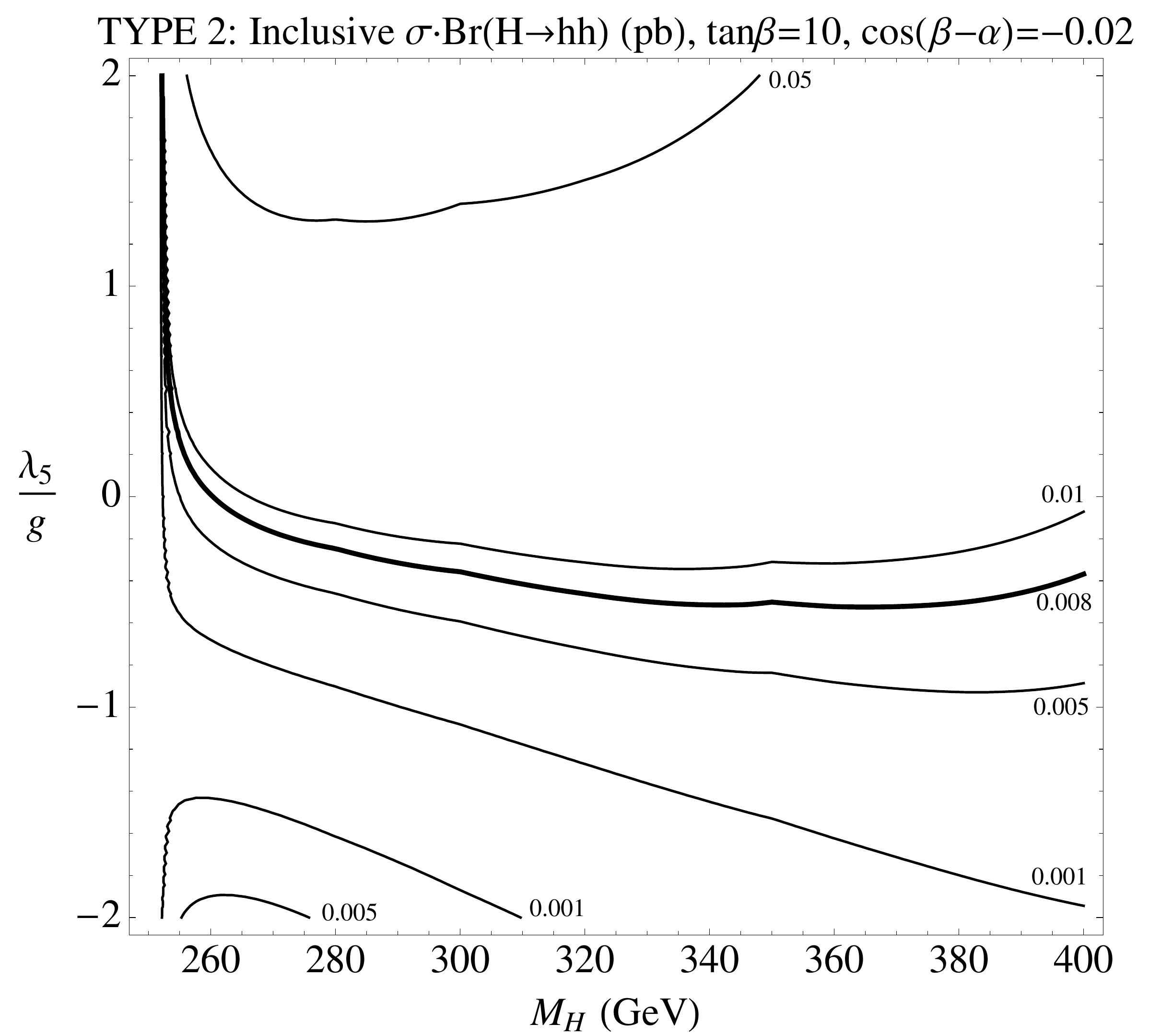} 
   \caption{Contours of the inclusive $\sigma \cdot {\rm Br}(H \to hh)$ in units of pb for 8 TeV $pp$ collisions for the non-SM-like scalar Higgs boson, shown as a function of $m_H$ and $\lambda_5$. Upper left: $\tan \beta = 1$, $\cos(\beta - \alpha) = -0.32 $ for Type 1 2HDM. Upper right: $\tan \beta = 1$, $\cos(\beta - \alpha) = -0.11$ for Type 2 2HDM. Lower left:  $\tan \beta = 10$, $\cos(\beta - \alpha) = -0.43$ for Type 1 2HDM. Lower right: $\tan \beta = 10$, $\cos(\beta - \alpha) = -0.02$ for Type 2 2HDM.  Here we have chosen $\lambda_{6,7} = 0$ and $m_A = m_H$.}
   \label{fig:HhhBRs}
\end{figure}

\begin{figure}[htbp] 
   \centering
   \includegraphics[width=3in]{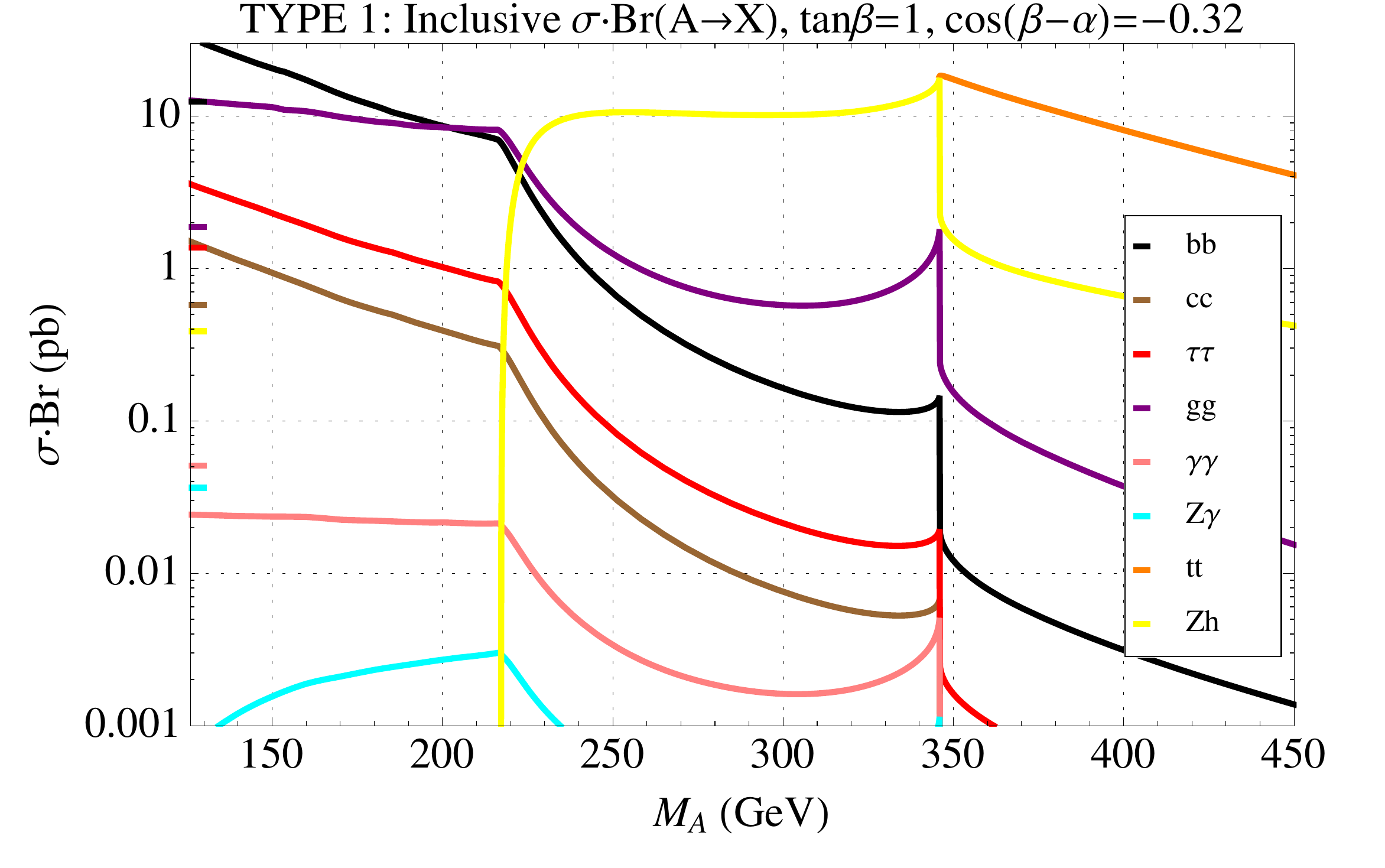} 
      \includegraphics[width=3in]{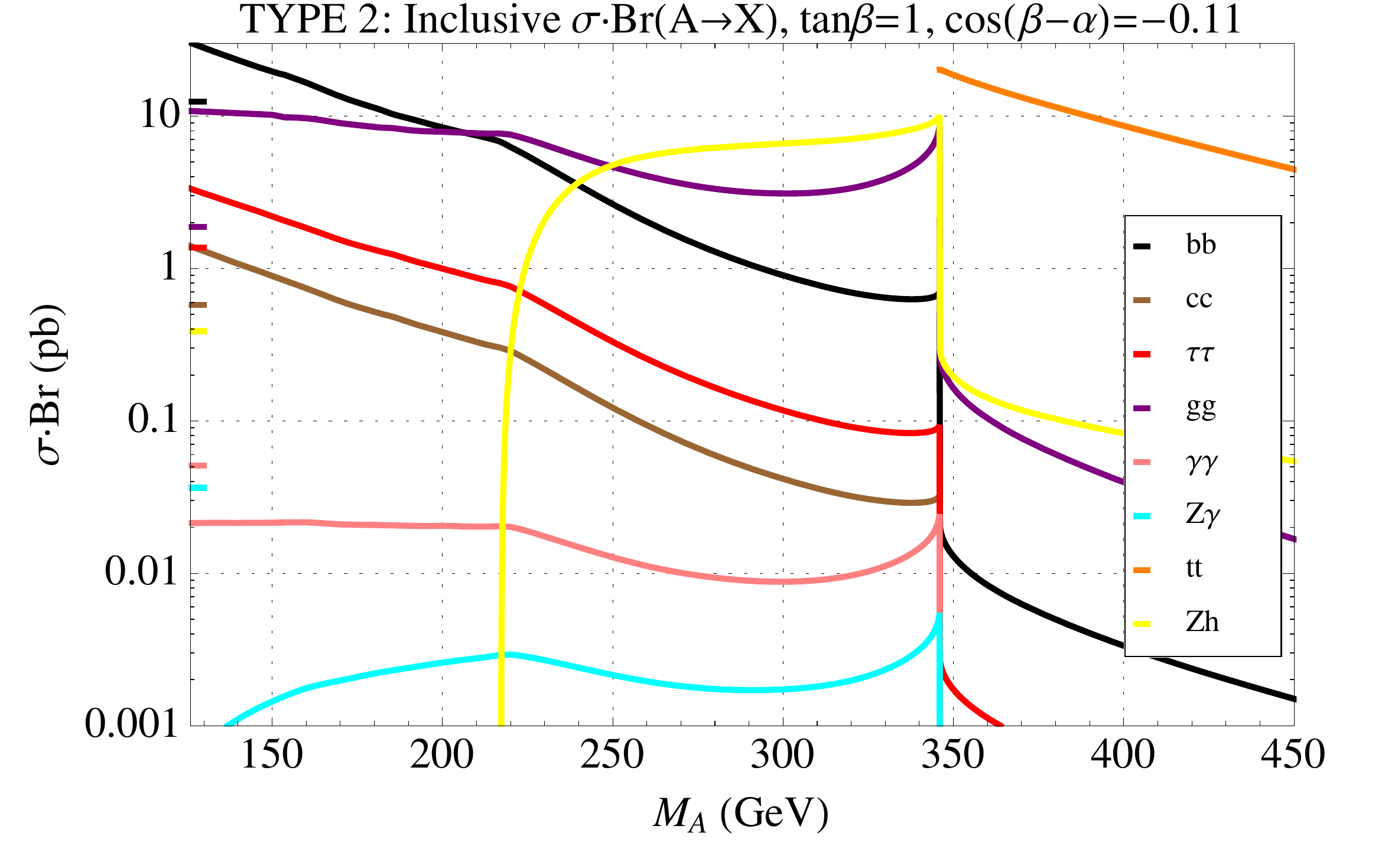} 
         \includegraphics[width=3in]{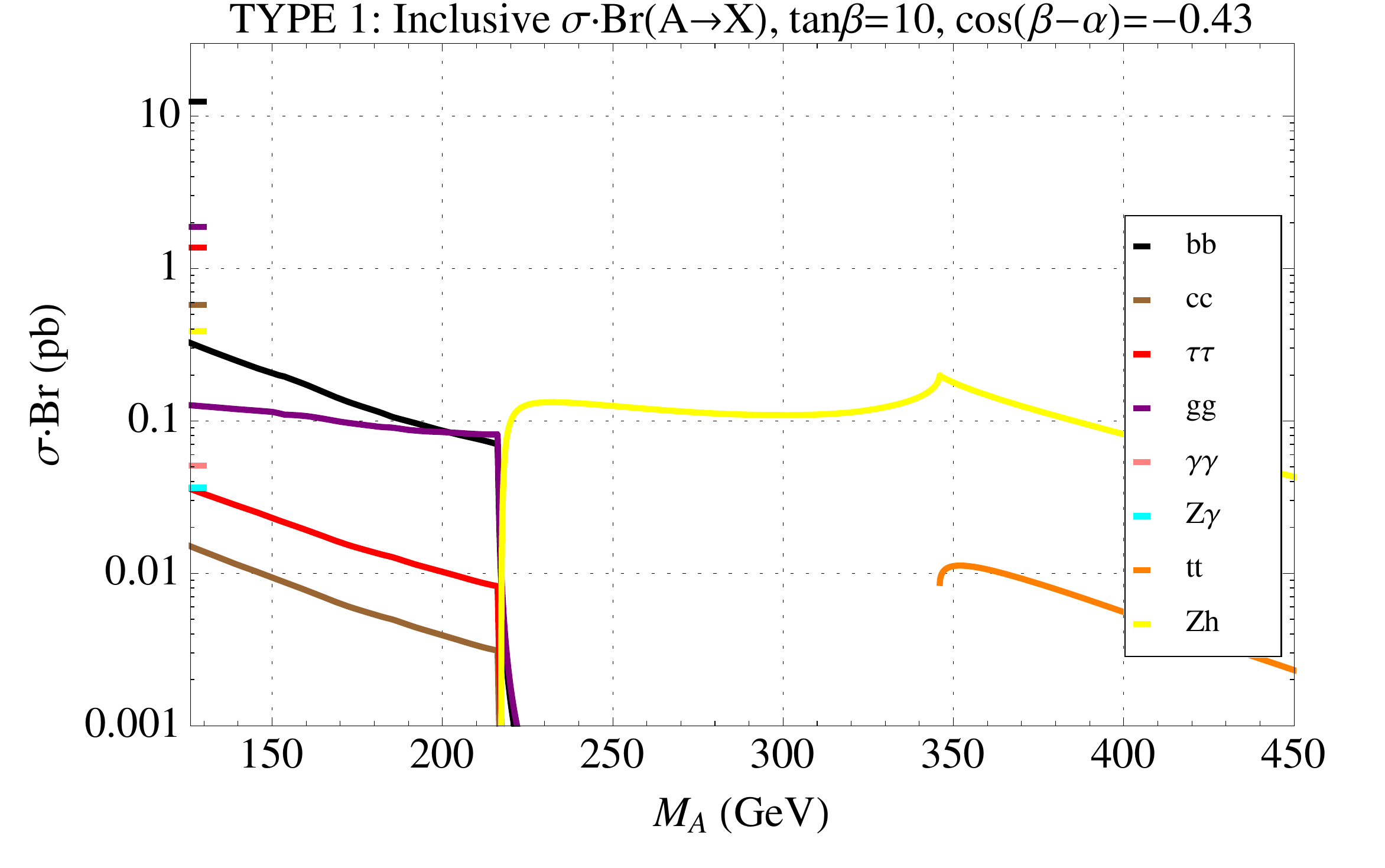} 
      \includegraphics[width=3in]{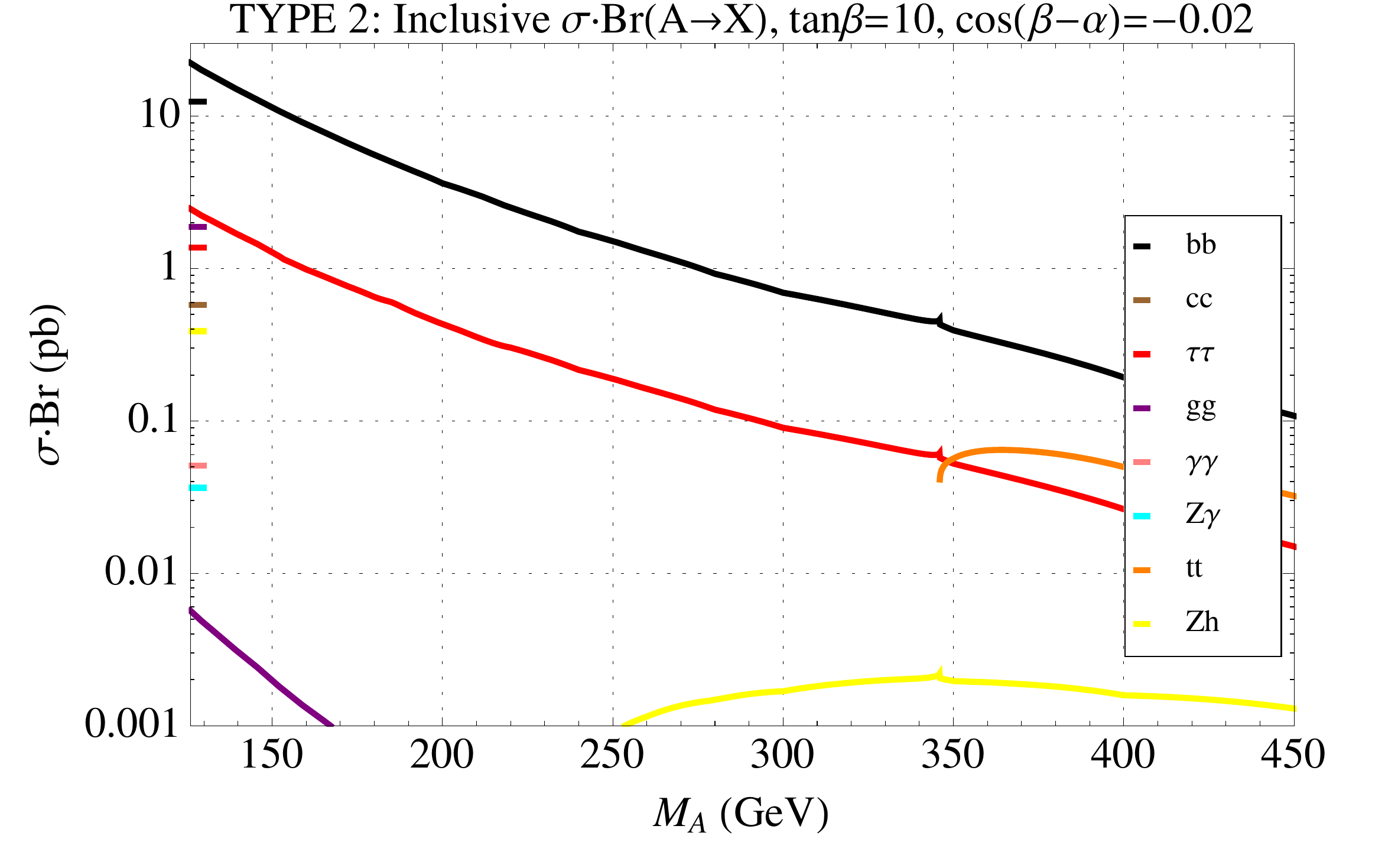} 
   \caption{Cross section times branching ratio $\sigma \cdot {\rm Br}(A \to X)$ to available final states in units of pb for 8 TeV $pp$ collisions for the pseudoscalar Higgs boson, shown as a function of $m_A$. Upper left: $\tan \beta = 1$, $\cos(\beta - \alpha) = -0.32 $ for Type 1 2HDM. Upper right: $\tan \beta = 1$, $\cos(\beta - \alpha) = -0.11$ for Type 2 2HDM. Lower left:  $\tan \beta = 10$, $\cos(\beta - \alpha) = -0.43$ for Type 1 2HDM. Lower right: $\tan \beta = 10$, $\cos(\beta - \alpha) = -0.02$ for Type 2 2HDM. In each case we have chosen $\lambda_{5,6,7} = 0$ and $m_A = m_H$.}
   \label{fig:ABRs}
\end{figure}

To illustrate the possible range of branching ratios as a function of mass consistent with current coupling fits, we first choose benchmark values of $\alpha$ and $\beta$ corresponding to the 95\% CL boundary of the Higgs fits as shown in Table \ref{tab:95bounds} and plot the inclusive production cross section times branching ratios as a function of $m_H$ or $m_A$. Although many  cascade decays are possible among scalars, in the limit $m_h < m_A \sim m_H \sim m_{H^\pm}$ there is a natural ordering of available decay modes when kinematically available; only $H \to hh, A \to Zh$, and $H^\pm \to W^\pm h$ are open. In what follows we assume this approximate mass ordering and neglect other possible decays among scalars. Fig.~\ref{fig:HBRs} shows the cross section times branching ratios of $H$ as a function of $m_H$ in Type 1 and Type 2 2HDM for two reference points: a ``low $\tan \beta$'' point with $\tan \beta = 1$ and $\cos(\beta - \alpha)$ on the boundary of the  95\% CL fit, and a ``high $\tan \beta$'' point with $\tan \beta = 10$ and $\cos(\beta - \alpha)$ again on the boundary of the  95\% CL fit. For these branching ratios we have set $\lambda_5 = \lambda_6 = \lambda_7 = 0$. Dependence of ${\rm Br}(h \to hh)$ on the value of $\lambda_5$ is illustrated in Fig.~\ref{fig:HhhBRs} for $\lambda_6 = \lambda_7 = 0$. Note that when $\lambda_5 \neq 0$, $\Gamma(H \to hh)$ may be parametrically enhanced and further dominate over $\Gamma(H \to VV)$. For the pseudoscalar, Fig.~\ref{fig:ABRs} shows the production cross section times branching ratios of $A$ as a function of $m_A$ in Type 1 and Type 2 2HDM for the same two reference points; here there is no additional parametric freedom beyond the dependence on $\alpha$ and $\beta$. 

Several features are immediately apparent, though these branching ratio plots are not necessarily generic. In Type 1 2HDM, away from the exact alignment limit the dominant modes for $H$ are $H \to VV, H \to hh,$ and $H \to t \bar t$. When kinematically available, the di-Higgs decay $H \to hh$ often dominates over $H \to VV$ due to $\mathcal{O}(1)$ numerical factors as well as $\tan \beta$-enhancement away from the alignment limit. Whether or not $H \to t \bar t$ dominates the branching ratio when kinematically available depends on $\tan \beta$. Since $y_{Htt}$ is suppressed at large $\tan \beta$ in a Type 1 2HDM, it may remain subdominant relative to $H \to hh$ provided sufficient distance from the alignment limit. Thus $H \to hh$ may remain the dominant decay mode of the non-SM-like heavy Higgs even when $m_H > 2 m_t$. For $A$ the dominant modes are $A \to b \bar b, A \to gg, A \to \tau^+ \tau^-, A \to Zh,$ and $A \to t \bar t$. When kinematically available, $A \to Z h$ dominates for $m_A < 2 m_t$; when it becomes available, $A \to t \bar t $ may then dominate at low $\tan \beta$, but remains sub-dominant at high $\tan \beta$ due to the suppression of $y_{Att}$. Much like $H \to hh$, $A \to Zh$ may then remain the dominant decay mode even when $m_A > 2 m_t$.

In Type 2 2HDM, the important modes are again $H \to VV, H \to hh, H \to t \bar t$ as well as $H \to \tau^+ \tau^-$ and $H \to b \bar b$, both of which become important at large $\tan \beta$. In this case $H \to hh$ is rarely important once $m_H > 2 m_t$, since either $H \to t \bar t$ takes over at low $\tan \beta$ or $H \to b \bar b, \tau^+ \tau^-$ take over at high $\tan \beta$. For $A$ the dominant modes are again $A \to b \bar b, A \to gg, A \to \tau^+ \tau^-, A \to Zh,$ and $A \to t \bar t$. When kinematically available, $A \to Z h$ dominates for $m_A < 2 m_t$ at low $\tan \beta$ but is highly suppressed at high $\tan \beta$. Once it becomes kinematically available, $A \to t \bar t $ dominates at low $\tan \beta$, but is suppressed relative to $A \to b \bar b$ at high $\tan \beta$ due to the diminution of $y_{Att}$ and enhancement of $y_{Abb}$.

In Type 3 2HDM, the parametric scaling is similar to Type 1 2HDM except at large $\tan \beta$, where eventually $H \to \tau^+ \tau^-$ and $A \to \tau^+ \tau^-$ take over the total widths. In Type 4 2HDM, the scaling is similar to Type 2 2HDM, except $H, A \to \tau^+ \tau^-$ no longer play a role at large $\tan \beta$.

Of course, the dominant modes in the branching ratios of $H$ and $A$ are not necessarily the best modes for discovery. In general the most fruitful standard channels (i.e., involving direct decays to SM final states) are those that may be distinguished above SM backgrounds, primarily:

\begin{itemize}

\item { Inclusive production of $H$ with $H \to VV^{(*)}$ or  $H \to \gamma \gamma$}
\item { Inclusive production of $A$ with $A \to \gamma \gamma$}
\item {Inclusive production of $H$ or $A$ with $H, A \to \tau^+ \tau^-$ or $H, A \to \mu^+ \mu^-$}
\item {$t \bar t$ production with $t \to H^\pm \bar b$ and $H^\pm \to \tau^\pm \nu$}

\end{itemize}

Beyond standard channels, it is useful to search for additional scalars through their decays to $h$ and other states. Again assuming the approximate mass ordering $m_h < m_A \sim m_H \sim m_{H^\pm}$, the kinematically available channels with promising search prospects are:

\begin{itemize}

\item { Inclusive production of $H$ with $H \to hh$}
\item {Inclusive production of $A$ with $A \to Zh$}
\item {$t \bar b$ associated production of $H^\pm$ with $H^\pm \to W^\pm h$}

\end{itemize}

Note the latter mode of $t \bar b$ associated production of $H^\pm$ with $H^\pm \to W^\pm h$ is not kinematically available when $m_{H^\pm} < m_t$, and so the associated production cross section is bound to be fairly small. We will not consider this case in detail here,\footnote{For a recent discussion of prospects for discovering $H^\pm \to W^\pm h$ at the LHC, see \cite{Sher:2012xb}.} but will instead focus on inclusive production of $H$ and $A$ with $H \to hh$ and $A \to Zh$, which may have sizable production cross sections and appreciable branching ratios.

While choosing benchmark values of $\alpha, \beta$ consistent with current signal fits and studying the branching ratios as a function of mass gives useful intuition, a more detailed study is required to understand the signals as $\alpha$ and $\beta$ are varied. For simplicity, we focus on the benchmark mass $m_H = m_A = 300$ GeV where $H \to hh$ and $A \to Zh$ are kinematically available. The relative parametrics remain similar for $m_H < 2 m_h$ and $m_A < m_Z + m_h$, save that the modes $H \to hh, A \to Zh$ are inaccessible and $H \to VV, A \to b \bar b$ become  dominant. For $m_H, m_A > 2 m_t$ the decays $H,A \to t \bar t$ become kinematically available, and may or may not dominate the total width depending on the 2HDM type and the value of $\tan \beta$.

As with the SM-like Higgs $h$, the parametric behavior of the production cross section times branching ratio for the processes of interest is governed by the scaling of the production mode, the decay mode, and the total width. For most of the cases we consider, production is entirely dominated by gluon fusion and scales as $\sim y_{Htt}^2$. The only exception is in Type 2 and Type 4 2HDM at large $\tan \beta$, where $b\bar bH$ and $b\bar bA$ associated production take over and production scales as $\sim y_{Hbb}^2$ and $\sim y_{Abb}^2,$ respectively. 

On the decay side, the total width may vary considerably across the parameter space as different processes contribute in different regions. For the heavy CP even scalar $H$ in a Type 1 2HDM, precisely in the alignment limit $H \to VV$ and $H \to hh$ vanish, leaving $h \to gg$ and $h \to b \bar b$ to constitute the bulk of the total width. For $m_H \gtrsim 2 m_h$ the two widths are comparable, with $\Gamma(h \to gg) \gtrsim \Gamma(h \to b \bar b)$ at $m_H = 300$ GeV. Thus in the strict alignment limit the total width scales as $\sim y_{Htt}^2 \sim y_{Hbb}^2$. 

\begin{figure}[t] 
   \centering
      \includegraphics[width=3in]{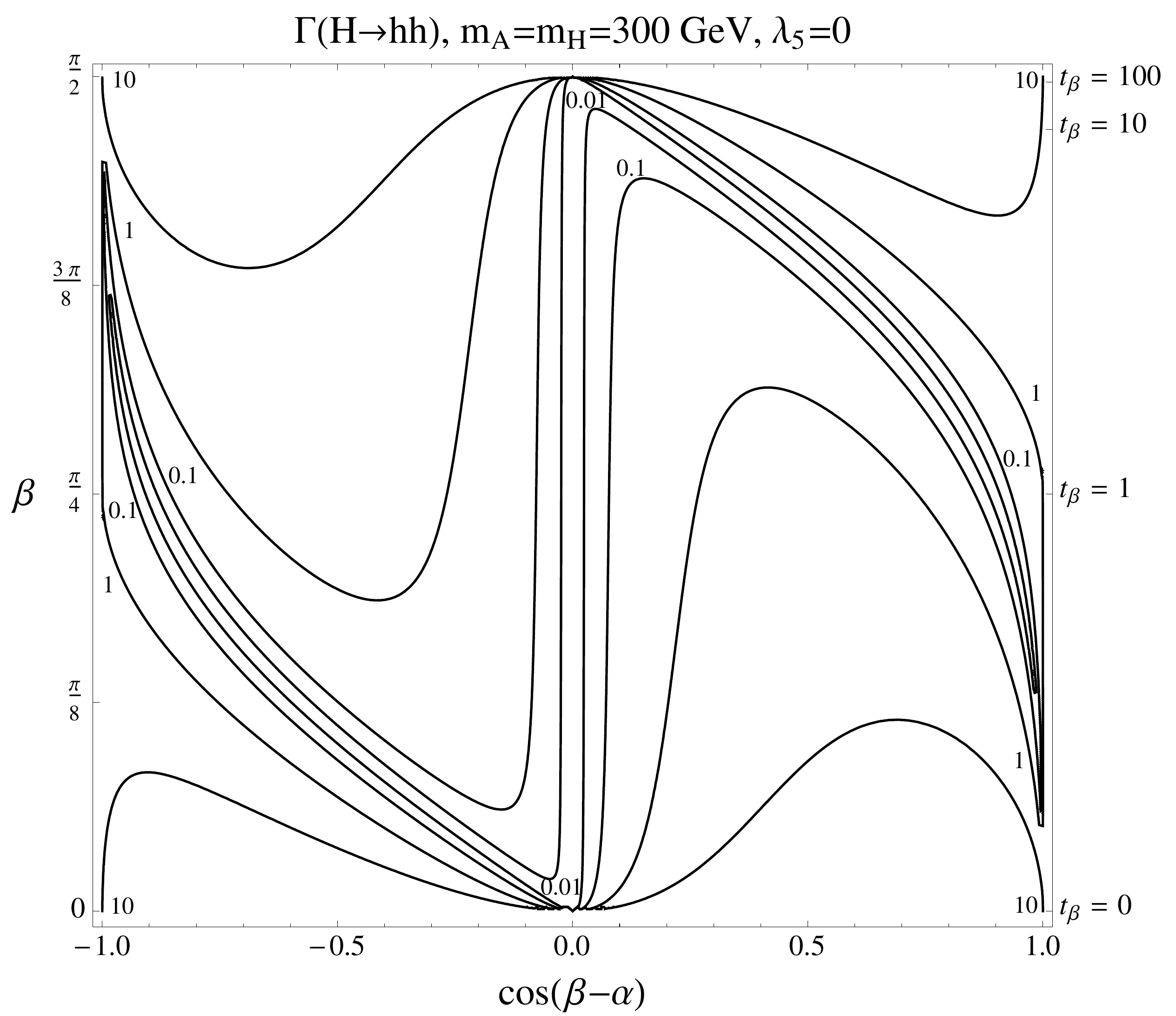} 
   \includegraphics[width=3in]{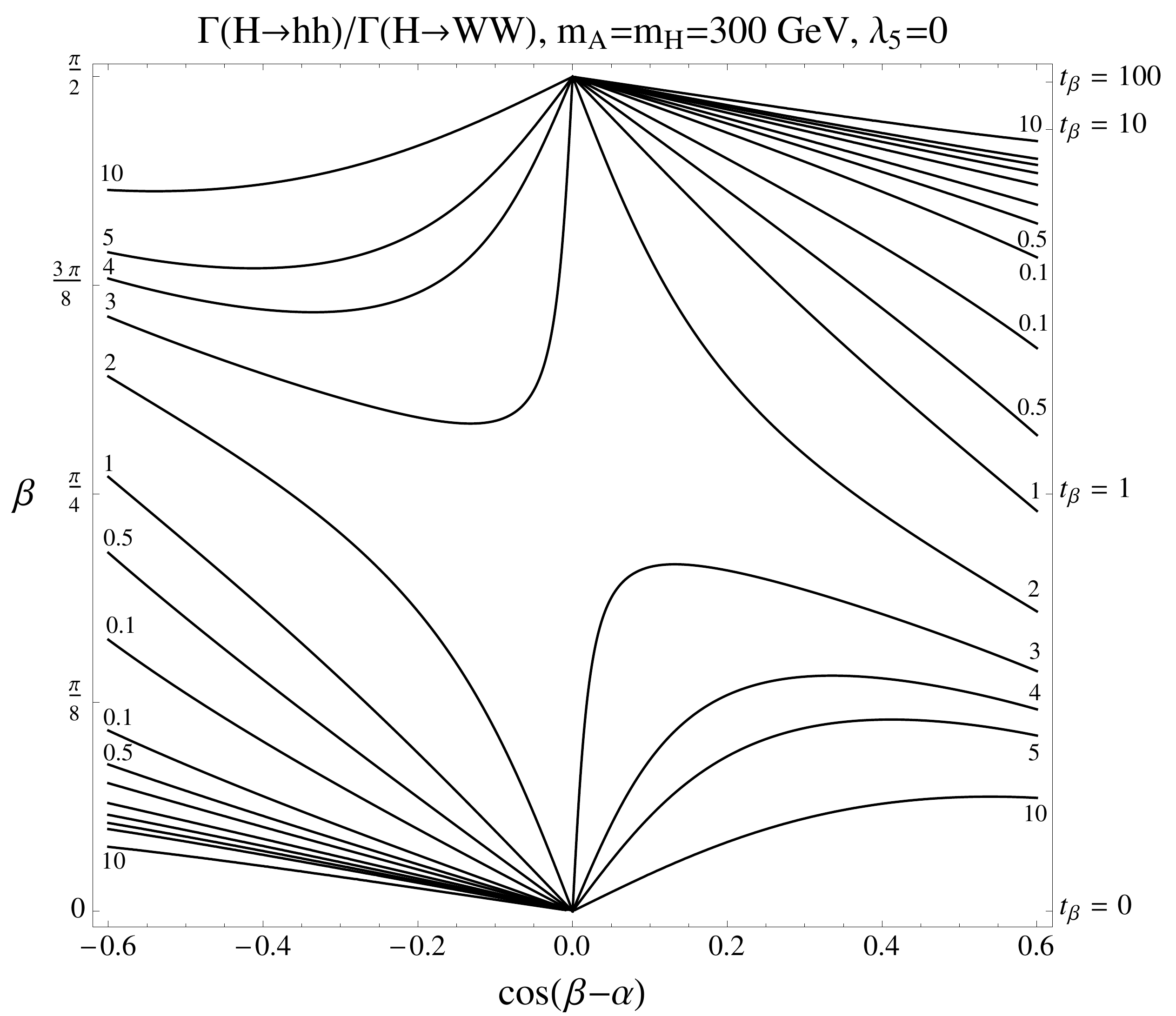} 
   \caption{Left: Contours of $\Gamma(H\to hh)$ in units of pb for the non-SM-like scalar Higgs boson $H$ in all four 2HDM types, shown as a function of $\cos(\beta - \alpha)$ and $\beta$. Right: Contours of $\Gamma(H\to hh)/\Gamma(H\to WW)$ for the non-SM-like scalar Higgs boson $H$ in all four 2HDM types, shown as a function of $\cos(\beta - \alpha)$ and $\beta$. In both plots we have chosen $\lambda_{5,6,7} = 0$ and $m_A = m_H = 300$ GeV. The partial width $\Gamma(H \to hh)$ typically exceeds $\Gamma(H \to WW)$ over a wide range of parameter space.}
   \label{fig:Hhhcoup}
\end{figure}

However, with even a small deviation from the alignment limit the partial widths for $H \to VV, hh$ rapidly come to dominate. Parametrically, the partial widths scale as
\begin{equation} \label{eq:widthratio}
\frac{\Gamma(H \to WW, hh)}{\Gamma(H \to b \bar b)} \propto \frac{g_{HVV}^2}{y_{Hbb}^2} \frac{m_H^2}{m_b^2}
\end{equation}
In a Type 1 2HDM, the ratio of coefficients scales as $\sim \cos^2(\beta - \alpha)$ around the alignment limit when $\tan \beta$ is not too large, which suggests that $H \to VV, hh$ take over the partial width when $|\cos(\beta - \alpha)| \gtrsim m_b / m_H \sim 0.02.$ This turns out to be a good approximation; for $m_H = 300$ GeV,  numerically we find $\Gamma(H \to WW) > \Gamma(H \to b \bar b)$ for $|\cos(\beta - \alpha) | \gtrsim 0.03$ and $\tan \beta = 1$. The $y_{Hbb}$ coupling is suppressed in Type 1 2HDM when  $\tan \beta$ is large, so that $H \to VV, hh$ dominate even closer to the alignment limit in this case. Thus even very small deviations lead to $\Gamma(H \to VV)$ and $\Gamma(H \to hh)$ dominating the total width. Whether $H \to VV$ or $H \to hh$ dominates depends in detail on numerical coefficients; we see from Fig.~\ref{fig:Hhhcoup} that it is typically $\Gamma(H \to hh)$ that is greater by an $\mathcal{O}(1)$ numerical factor, with additional $\tan \beta$ enhancement away from the alignment limit. However, since $g_{Hhh}$ also has additional zeroes away from the alignment limit, in these regions $\Gamma(H \to VV)$ takes over the total width.

For Type 2 2HDM the story is similar, except at large $\tan \beta$ where $y_{Hbb}$ is $\tan \beta$-enhanced and $H \to b \bar b$ takes over the partial width when $\lambda_{6,7} = 0$ (though as discussed above, when $\lambda_{6,7} \neq 0$ then $g_{Hhh}$ is $\tan \beta$-enhanced at leading order and continues to dominate the total width). This is again apparent from (\ref{eq:widthratio}), which suggests that $\Gamma(h \to b\bar b)$ dominates the partial width for $|\cos(\beta - \alpha)| \lesssim (m_b / m_H) \tan \beta$. Since current fits constrain $|\cos(\beta - \alpha)| \lesssim 0.1$ in Type 2 models, this suggests $H \to b \bar b$ dominates the total width as soon as $\tan \beta \gtrsim 5$. In practice, this is again a good approximation, and for $\cos(\beta - \alpha) = -0.1$ we find $\Gamma(H \to WW) \lesssim \Gamma(H \to b \bar b)$ for $\tan \beta \gtrsim 3.6$.

For the pseudoscalar $A$, the competitive modes are $A \to Zh$, $A \to b \bar b$, and $A \to gg$. Parametrically, the partial widths scale as 
\begin{equation} \label{eq:widthratioA}
\frac{\Gamma(A \to Z h)}{\Gamma(H \to b \bar b)} \propto \frac{g_{AZh}^2}{y_{Abb}^2} \frac{m_A^2}{m_b^2}
\end{equation}
In Type 1 2HDM, for moderate $\tan \beta$ we again have $\frac{g_{AZh}^2}{y_{Abb}^2}  \propto \cos^2(\beta - \alpha)$ and expect $A \to Zh$ to dominate when  $|\cos(\beta - \alpha)| \gtrsim m_b / m_A \sim 0.02$;  numerically  this is the case when $|\cos(\beta - \alpha)| \gtrsim m_b / m_A \sim 0.04$ for $\tan \beta = 1$, and $A \to Z h$ takes over even closer to the alignment limit when $\tan \beta$ is large. In Type 2 2HDM, the situation parallels the case for $H \to hh$; $y_{Abb}$ is $\tan \beta$-enhanced, and so $A \to b \bar b$ dominates when $\tan \beta \gtrsim 3$. 

With this parametric understanding of the production and decay modes, it is straightforward to study the variation in potential signals of the scalars $H$ and $A$ as a function of $\alpha$ and $\beta$.

\subsection{Inclusive production of $H$ with $H \to VV^{(*)}$ }

\begin{figure}[htbp] 
   \centering
   \includegraphics[width=3in]{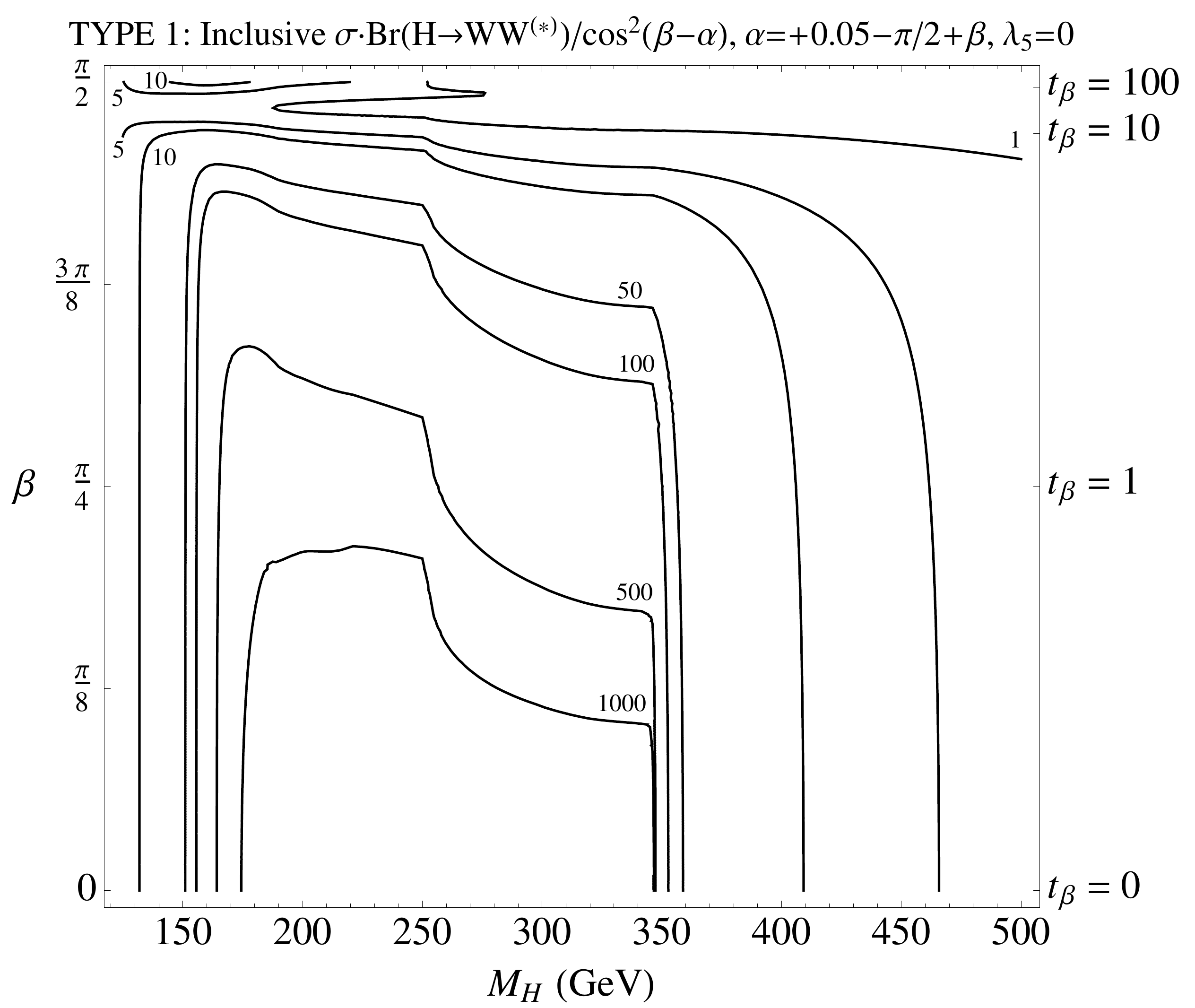} 
      \includegraphics[width=3in]{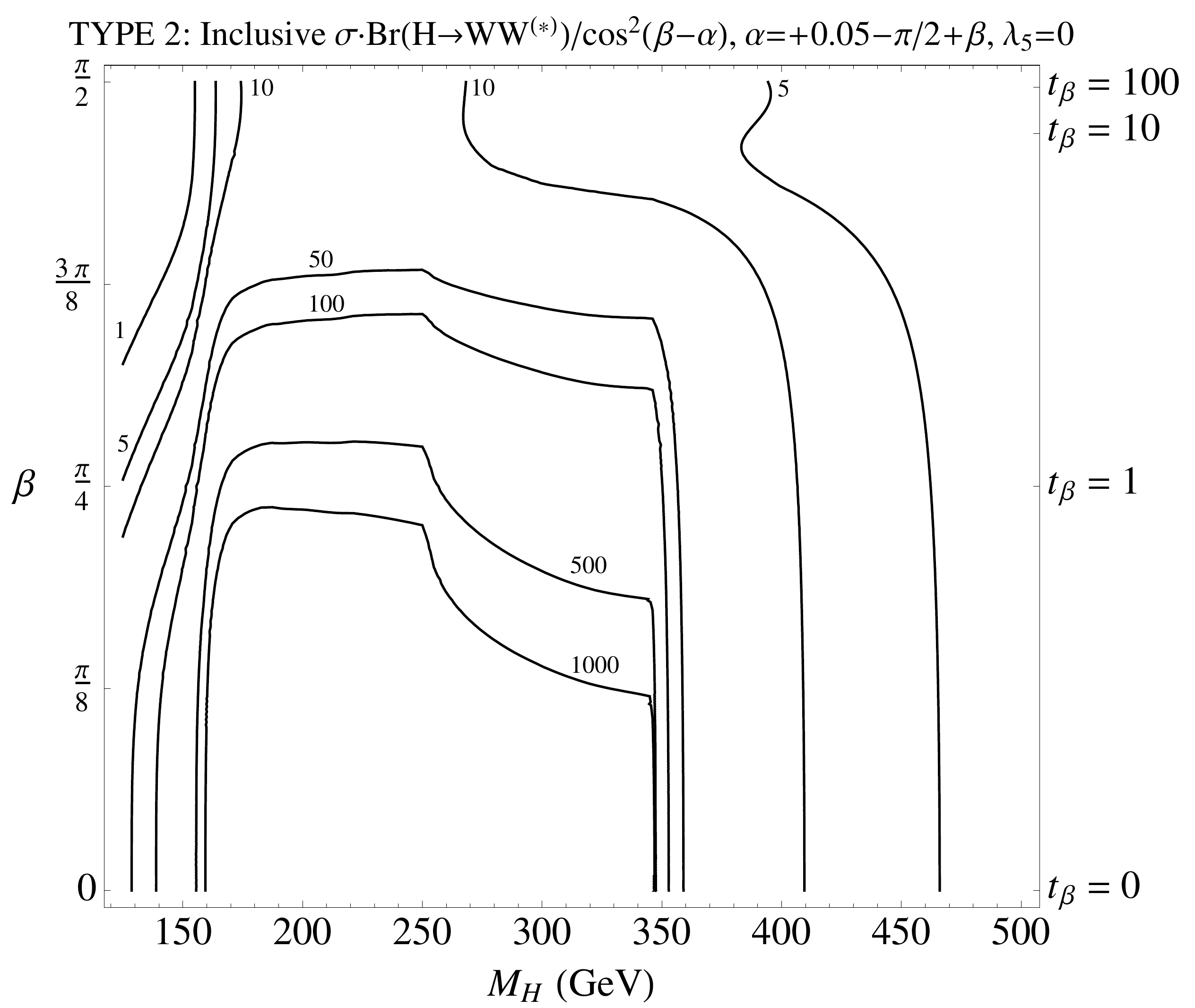}
   \caption{Contours of the inclusive $\sigma \cdot {\rm Br}(H \to WW) / \cos^2(\beta - \alpha)$ in units of pb for 8 TeV $pp$ collisions for the non-SM-like scalar Higgs boson, shown as a function of $m_H$ and $\beta$ for Type 1 (left) and Type 2 (right) 2HDM. Here we have chosen $\alpha = 0.05 - \pi/2 + \beta$, $\lambda_{5,6,7} = 0$ and $m_A = m_H$.}
   \label{fig:HWWcos2}
\end{figure}

Current searches for additional Higgs boson are primarily focused on $H \to VV^{(*)}$ at high mass. Although $g_{HVV}$ is $\cos(\beta - \alpha)$-suppressed around the alignment limit, since $\Gamma(H \to hh,VV^{(*)})$ comprise the majority of the total width this does not necessarily imply suppression of the branching ratio except very close to the exact alignment limit or, for Type 2 2HDM, at large $\tan \beta$ when $\Gamma(H \to b \bar b)$ takes over. For $H \to VV$, we find it useful to illustrate the size of the available signal as a function of mass and $\tan \beta$, accounting for most of the $\alpha$ dependence by plotting contours of the inclusive $\sigma \cdot {\rm Br}(H \to WW) / \cos^2(\beta - \alpha)$ as shown in Fig.~\ref{fig:HWWcos2}. This illustrates how $\sigma \cdot {\rm Br}(H \to WW)$ varies as a function of $m_H, \tan \beta$, and one may easily infer the value of $\sigma\cdot {\rm Br}$ for any specific choice of $\cos(\beta - \alpha)$. Note that in general there is still weak dependence on $\alpha$ due to the variation of the total width, but Fig.~\ref{fig:HWWcos2} captures the leading parametric dependence. In both types of 2HDM, $\sigma \cdot {\rm Br}(H \to WW) / \cos^2(\beta - \alpha)$ is suppressed at low $m_H$ because one $W$ boson is off-shell, and the rate rises rapidly when both $W$'s go on-shell. The rate falls at $m_H = 2 m_h$ when $H \to hh$ becomes kinematically available, and likewise at $m_H = 2 m_t$ when $H \to t \bar t$ opens. In Type 1 2HDM the rate falls with increasing $\tan \beta$ simply because the production coupling $y_{Htt}$ is $\tan \beta$-suppressed, while in Type 2 2HDM the rate falls more rapidly because the production mode is suppressed and because the total width grows with $\tan \beta$ due to the enhancement of $\Gamma(H \to b \bar b)$.

\begin{figure}[htbp] 
   \centering
   \includegraphics[width=3in]{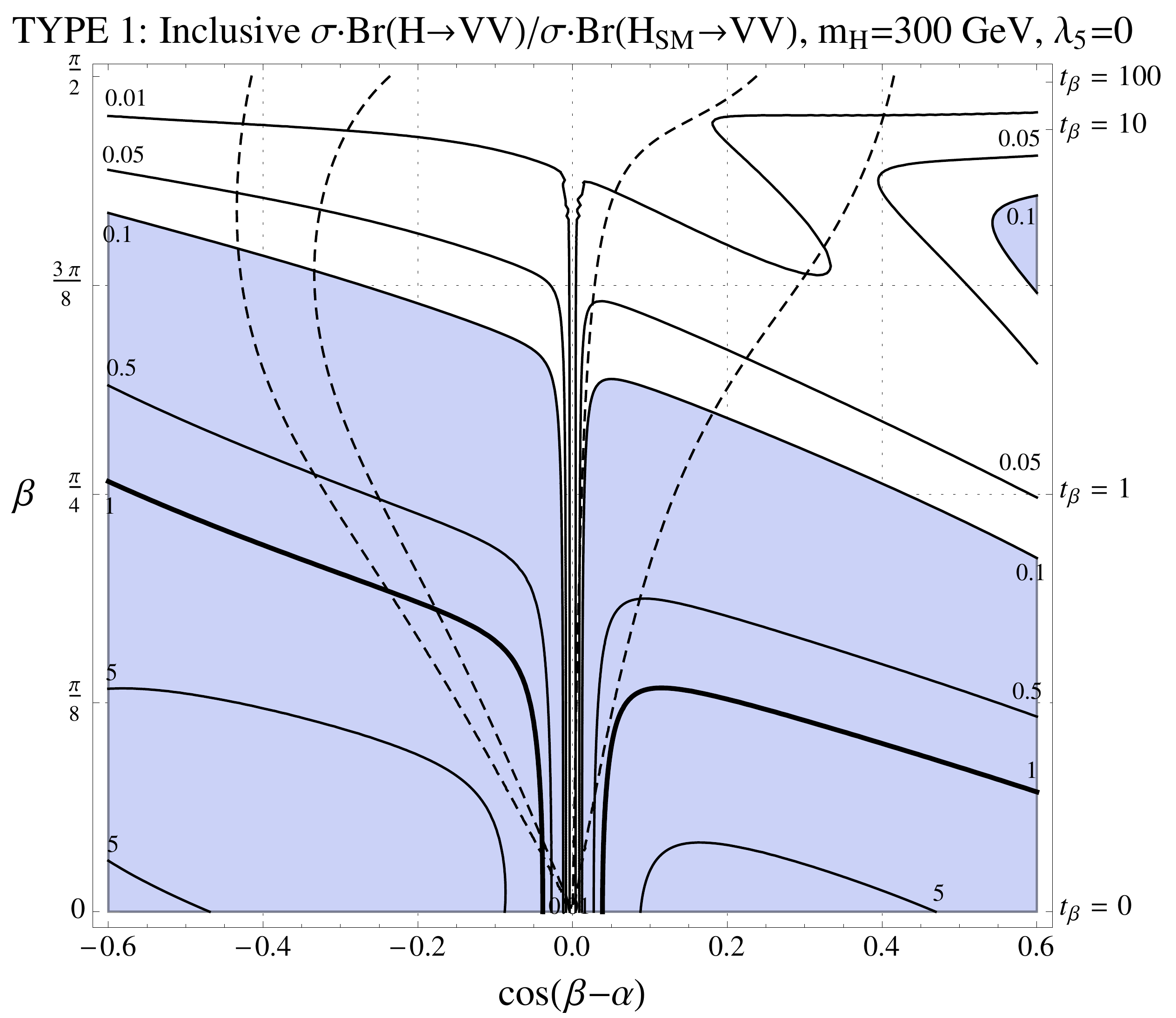} 
      \includegraphics[width=3in]{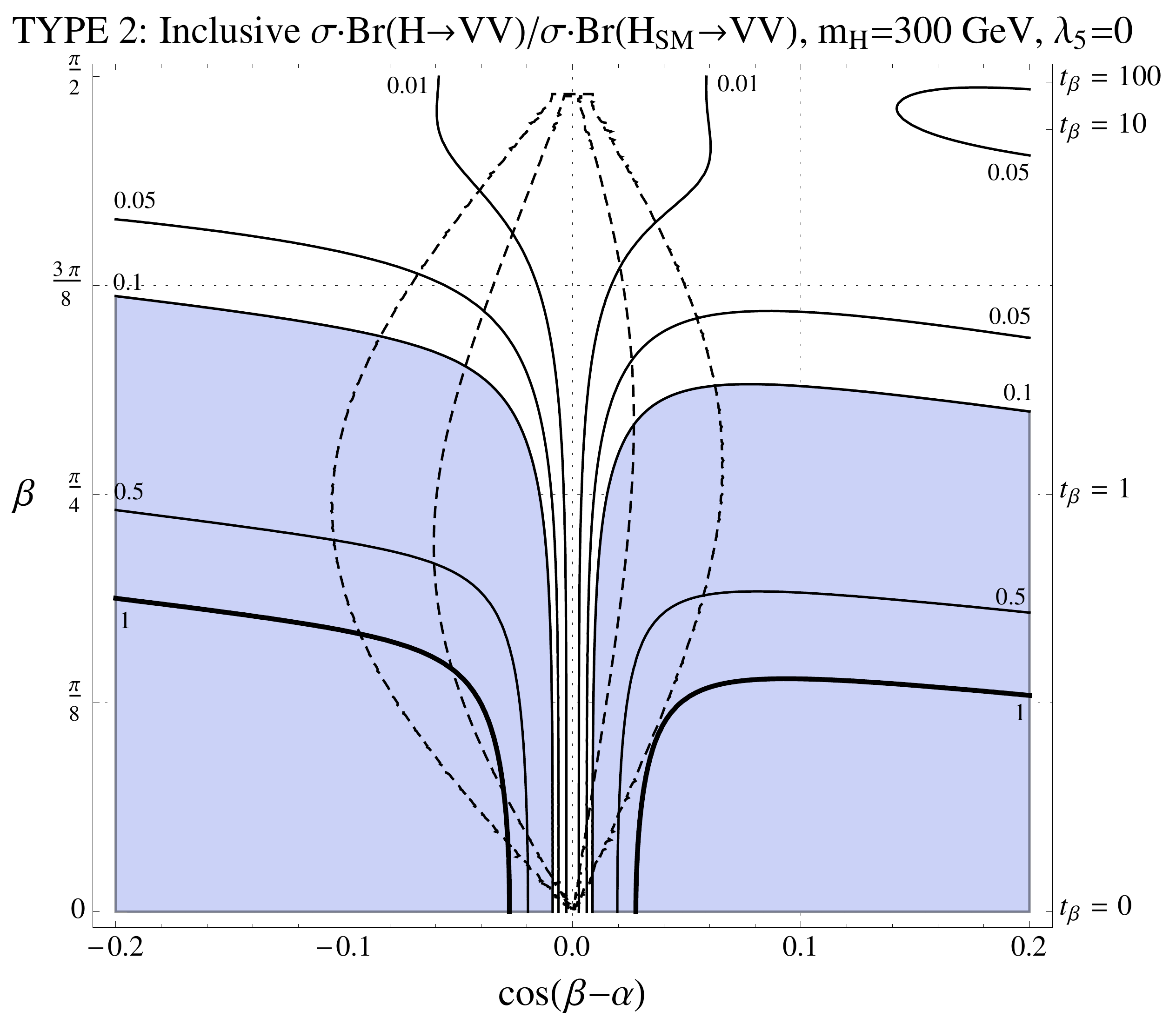}
   \caption{Contours of the inclusive $\sigma \cdot {\rm Br}(H \to VV^{(*)}) / \sigma \cdot {\rm Br}(H_{SM} \to VV^{(*)})$ for 8 TeV $pp$ collisions for the non-SM-like scalar Higgs boson, shown as a function of $\cos(\beta - \alpha)$ and $\beta$ for Type 1 (left) and Type 2 (right) 2HDM. Here we have chosen $\lambda_{5,6,7} = 0$ and $m_A = m_H$. The inner (outer) dashed contour denotes the 68\% (95\%) CL best fit to the signals of the SM-like Higgs. The blue shaded region denotes the parameter space excluded by the most recent LHC searches for a heavy SM-like Higgs \cite{Chatrchyan:2013yoa}.}
   \label{fig:HVVlim}
\end{figure}

Given that current LHC searches for additional Higgs scalars focus on the $H \to VV^{(*)}$ mode, it is also useful to consider the impact of present limits on the 2HDM parameter space. Fig.~\ref{fig:HVVlim} illustrates the parametric variation of $\sigma \cdot {\rm Br}(H \to VV^{(*)})$ as a function of $\alpha, \beta$ for the benchmark mass $m_H = 300$ GeV, normalized to the cross section times branching ratio of a SM-like Higgs of the same mass. Limits from current direct searches \cite{Chatrchyan:2013yoa} for heavy SM-like Higgs bosons are also shown. Unsurprisingly, the rate depends primarily on $\tan \beta$ away from the alignment limit, since here $H \to hh, VV^{(*)}$ dominate the total width and the parametric dependence of the cross section times branching ratio comes from the production coupling $y_{Htt}$. In Type 2 2HDM the more rapid falloff at large $\tan \beta$ again arises because the total width grows with $\tan \beta$ due to the enhancement of $\Gamma(H \to b \bar b)$. Very close to the alignment limit, $H \to b \bar b$ dominates the total width and ${\rm Br }(H \to VV^{(*)})$ is diminished. For this particular benchmark with $\lambda_{5,6,7} = 0$, current searches exclude $\tan \beta \lesssim 2.5$ in both 2HDM types away from the alignment limit for $m_H = 300$ GeV.  But note that this does not amount to an invariant bound, since if $\lambda_{5,6,7} \neq 0$ then the resulting increase in $\Gamma(H \to hh)$ further reduces ${\rm Br }(H \to VV^{(*)})$ and softens the limit. 

\subsection{Inclusive production of $H$ with $H \to \gamma \gamma$}

\begin{figure}[htbp] 
   \centering
   \includegraphics[width=3in]{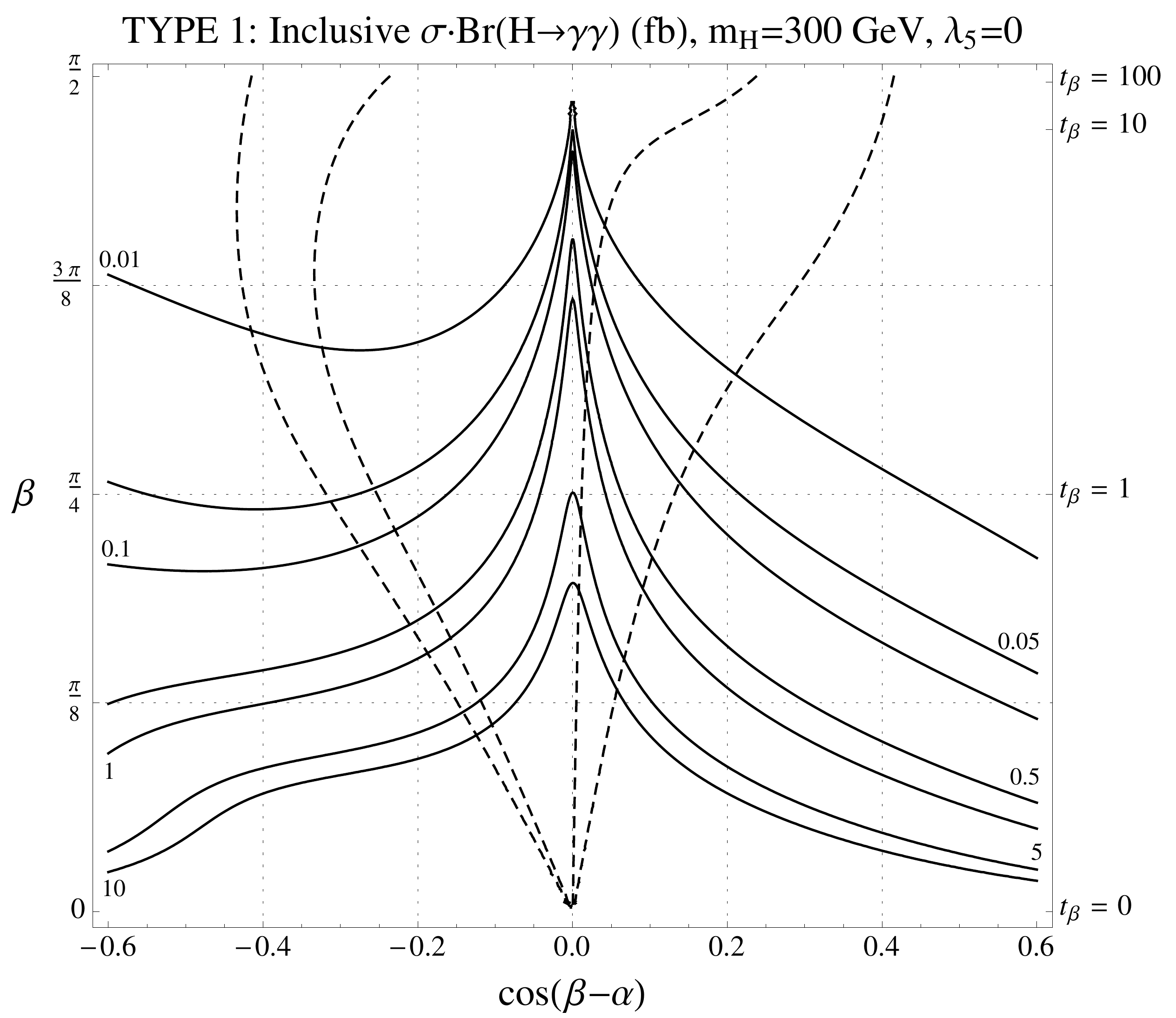} 
      \includegraphics[width=3in]{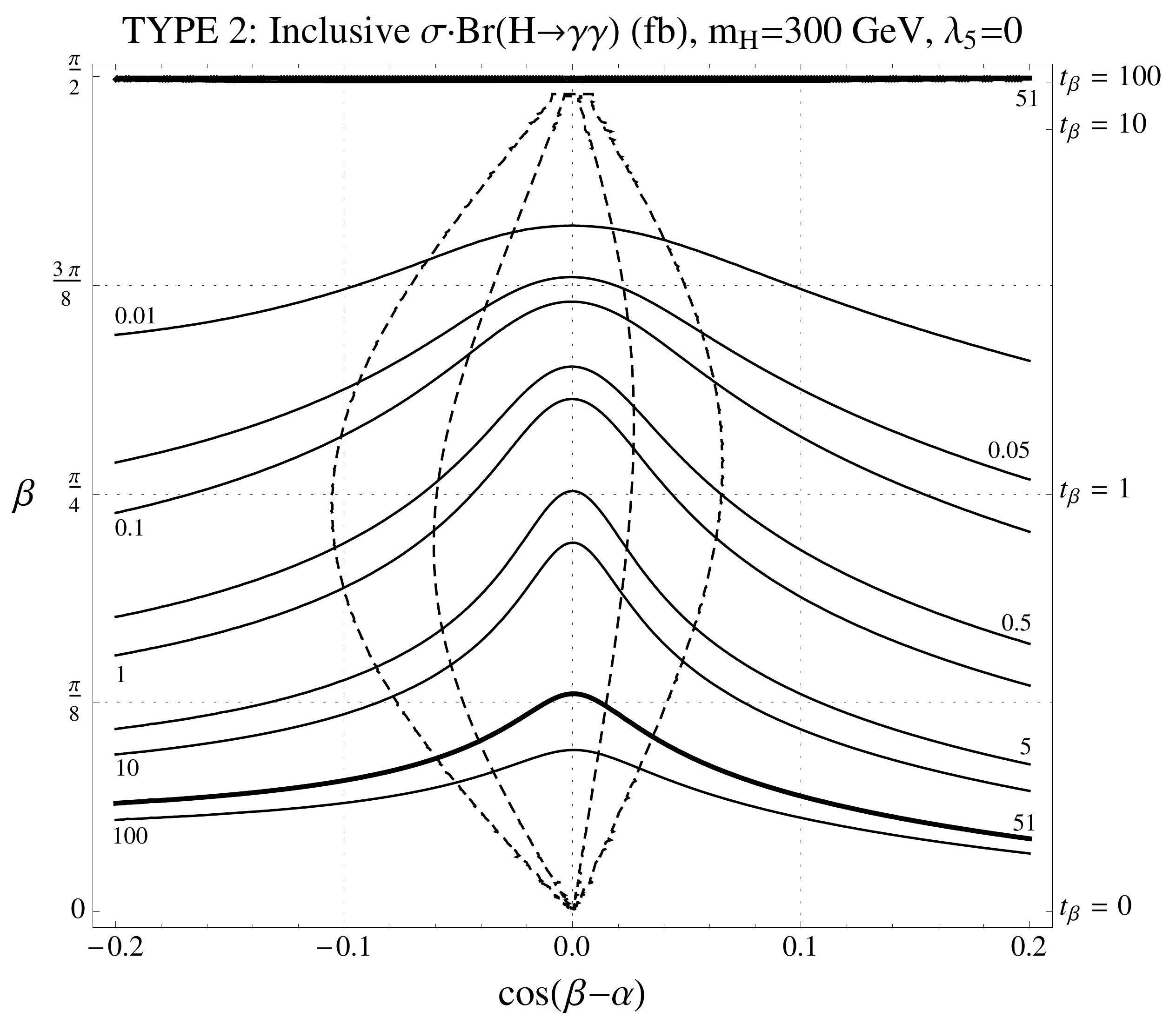} 
   \caption{Contours of the inclusive $\sigma \cdot {\rm Br}(H \to \gamma \gamma)$ in units of fb for 8 TeV $pp$ collisions for the non-SM-like scalar Higgs boson with $m_H = 300$ GeV, shown as a function of $\cos(\beta - \alpha)$ and $\beta$ for Type 1 (left) and Type 2 (right) 2HDM. Here we have chosen $\lambda_{5,6,7} = 0$ and $m_A = m_H$. The inner (outer) dashed contour denotes the 68\% (95\%) CL best fit to the signals of the SM-like Higgs.}
   \label{fig:Hgammagamma}
\end{figure}

Current searches for $H \to \gamma \gamma$ are truncated at relatively low mass, $m_H \sim 160$ GeV, since the already-small branching ratio  for a SM-like Higgs falls rapidly with increasing mass. Nonetheless, $H \to \gamma \gamma$ may be important at high mass if the 2HDM is very close to the alignment limit, where the contributions from $H \to VV$ and $H \to hh$ vanish from the total width. Contours of the inclusive $\sigma \cdot {\rm Br}(H \to \gamma \gamma)$ are shown in Fig.~\ref{fig:Hgammagamma}. For Type 1 2HDM, the rate falls with increasing $\tan \beta$ due to the $\tan \beta$ suppression of the production mode, while it peaks in the alignment limit because the total width drops precipitously as $H \to VV$ and $H \to hh$ decouple. The peaking is particularly sharp at large $\tan \beta$ because fermion partial widths are all suppressed by $\tan \beta$.

In Type 2 2HDM, the rate falls more rapidly with increasing $\tan \beta$ due to the suppressed production and the growing total width, though at large $\tan \beta$ the contributions from $b\bar bH$ associated production cause the rate to increase once again. The rate peaks in the alignment limit due to the decoupling of $H \to VV$ and $H \to hh$, but the peaking is less pronounced than in Type 1 2HDM because the non-decoupling contribution from $\Gamma(H \to b \bar b)$ is always a fairly important contribution to the total width.

\subsection{Inclusive production of $A$ with $A \to \gamma \gamma$}

\begin{figure}[htbp] 
   \centering
   \includegraphics[width=3in]{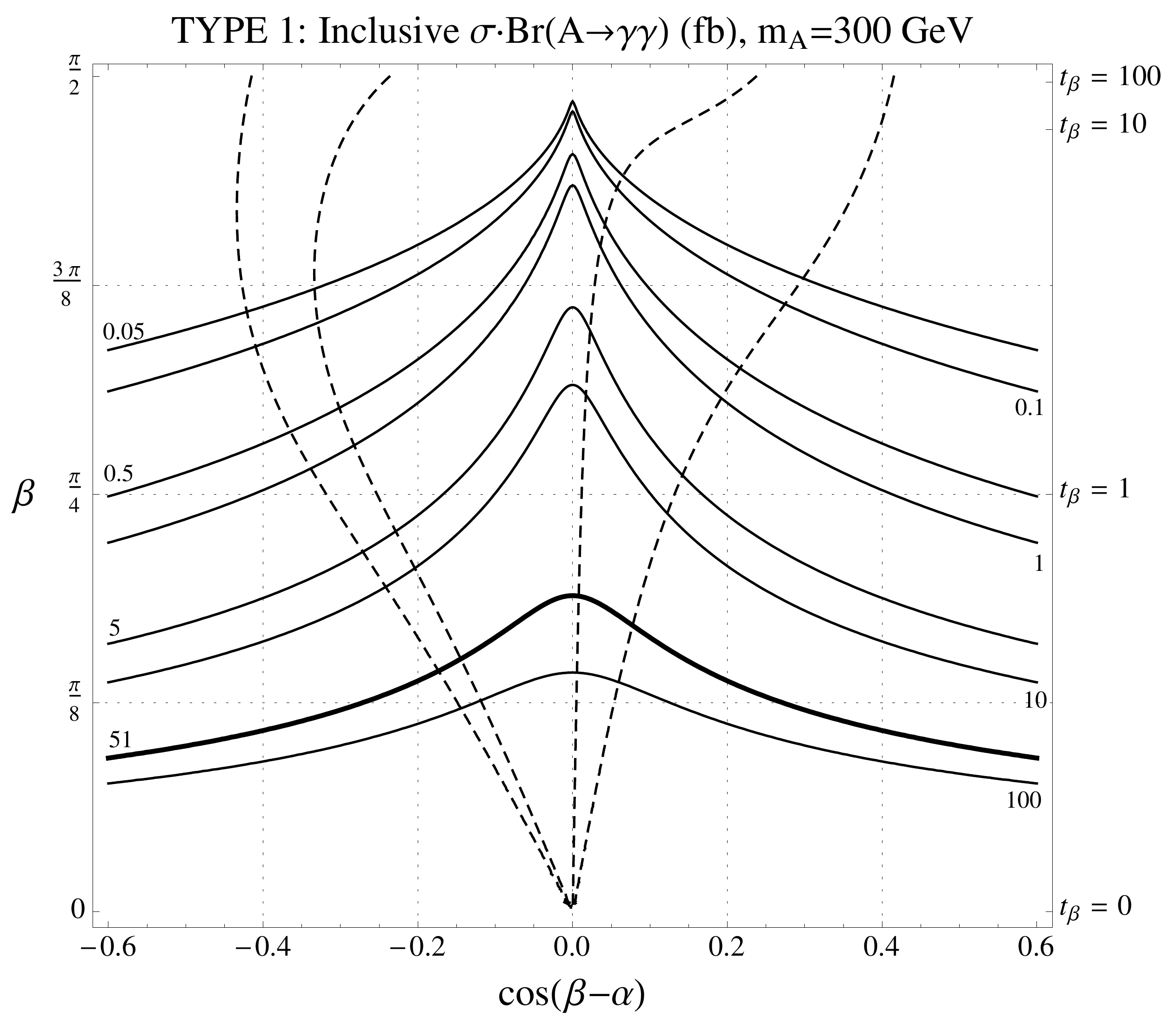} 
      \includegraphics[width=3in]{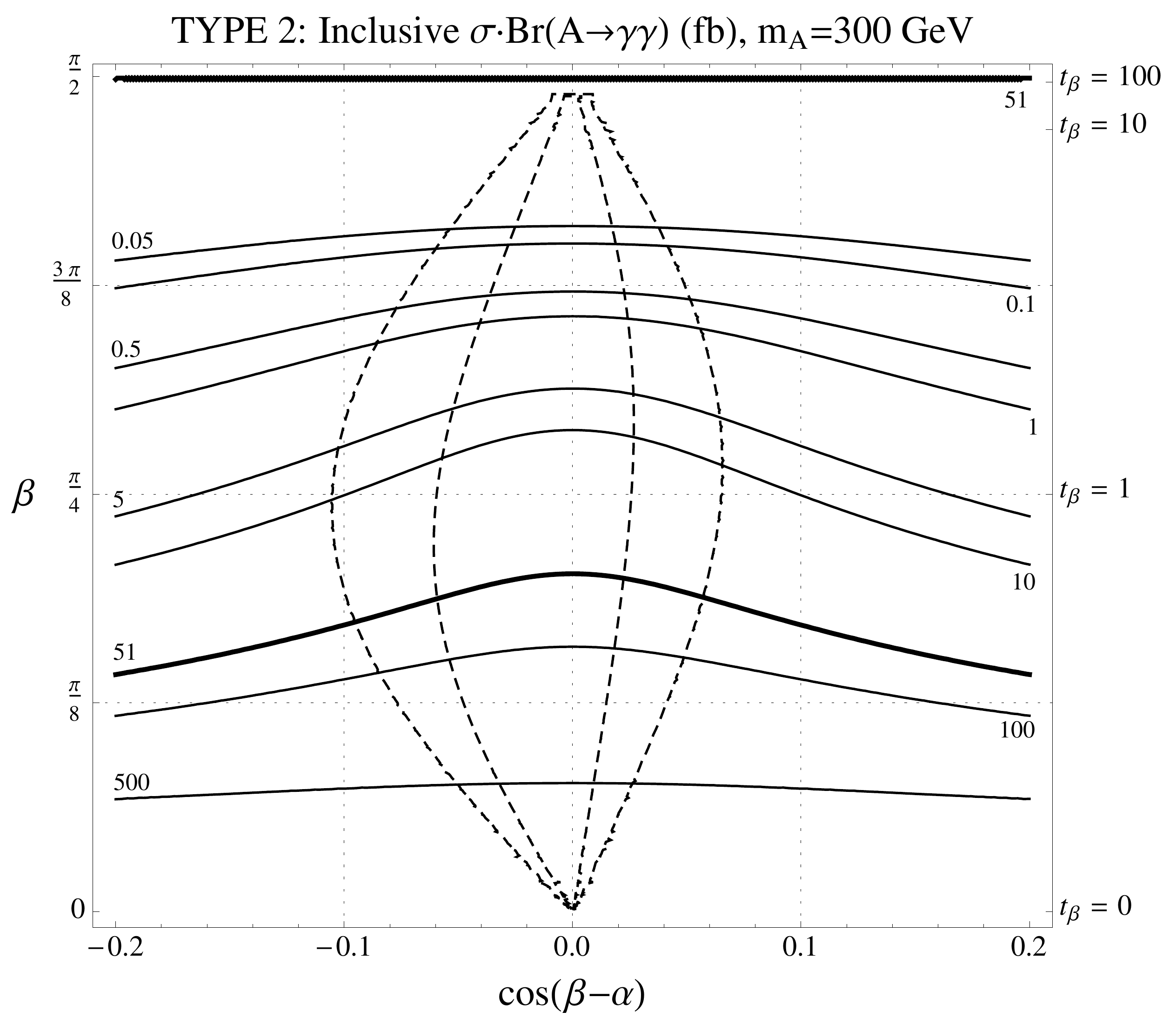} 
   \caption{Contours of the inclusive $\sigma \cdot {\rm Br}(A \to \gamma \gamma)$ in units of fb for 8 TeV $pp$ collisions for the pseudoscalar Higgs boson with $m_A = 300$ GeV, shown as a function of $\cos(\beta - \alpha)$ and $\beta$ for Type 1 (left) and Type 2 (right) 2HDM. The inner (outer) dashed contour denotes the 68\% (95\%) CL best fit to the signals of the SM-like Higgs.}
   \label{fig:Agammagamma}
\end{figure}

Inclusive production of $A$ with $A \to \gamma \gamma$  is entirely analogous to $H \to \gamma \gamma$ in its parametric scaling. Contours of the inclusive $\sigma \cdot {\rm Br}(A \to \gamma \gamma)$ are shown in Fig.~\ref{fig:Agammagamma} and scale as discussed above, with slightly different modulation as the alignment limit is approached due to the fact that there is no contribution from $W$ loops in the $A\gamma \gamma$ effective coupling.

\begin{figure}[htbp] 
   \centering
   \includegraphics[width=3in]{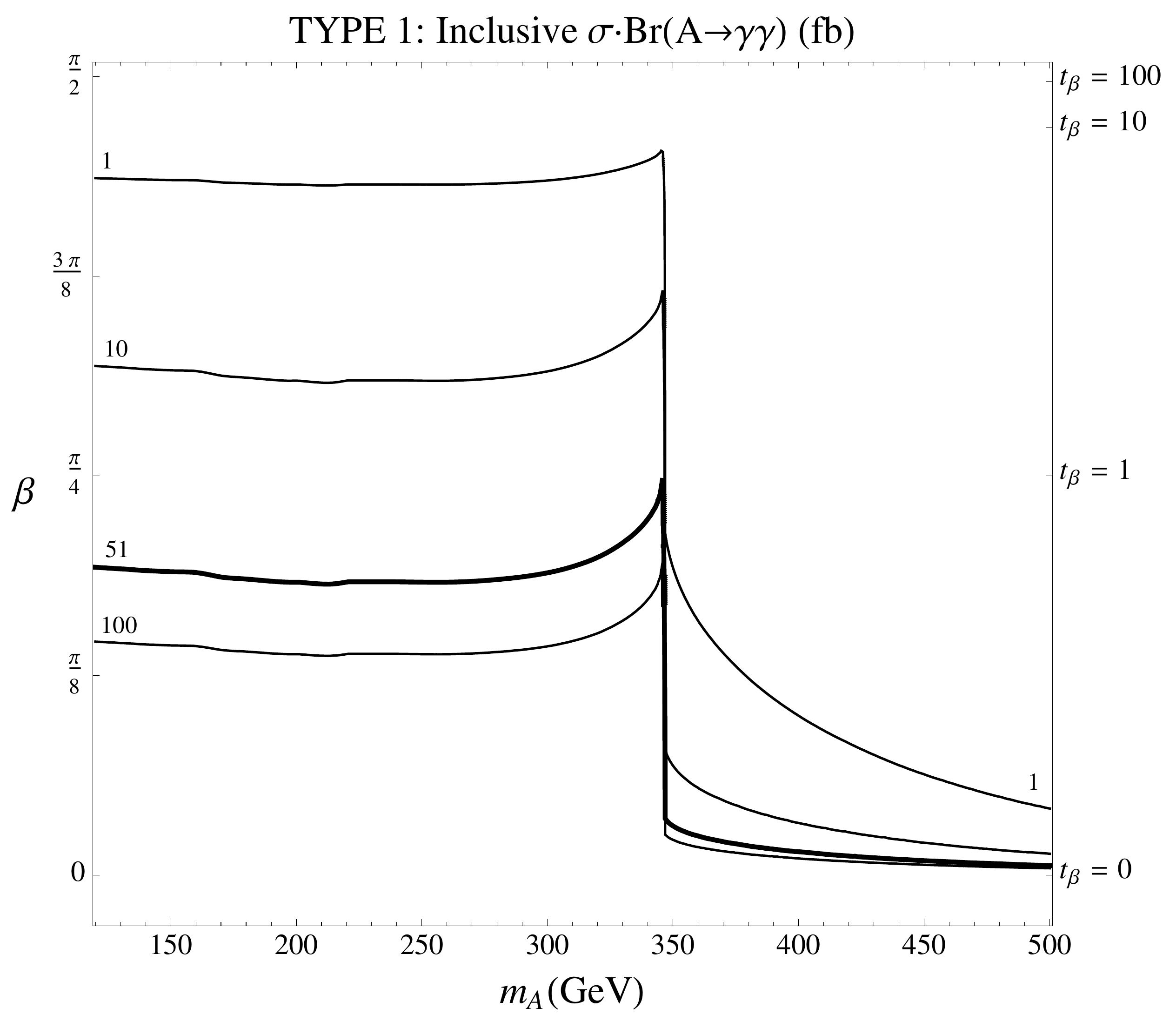} 
      \includegraphics[width=3in]{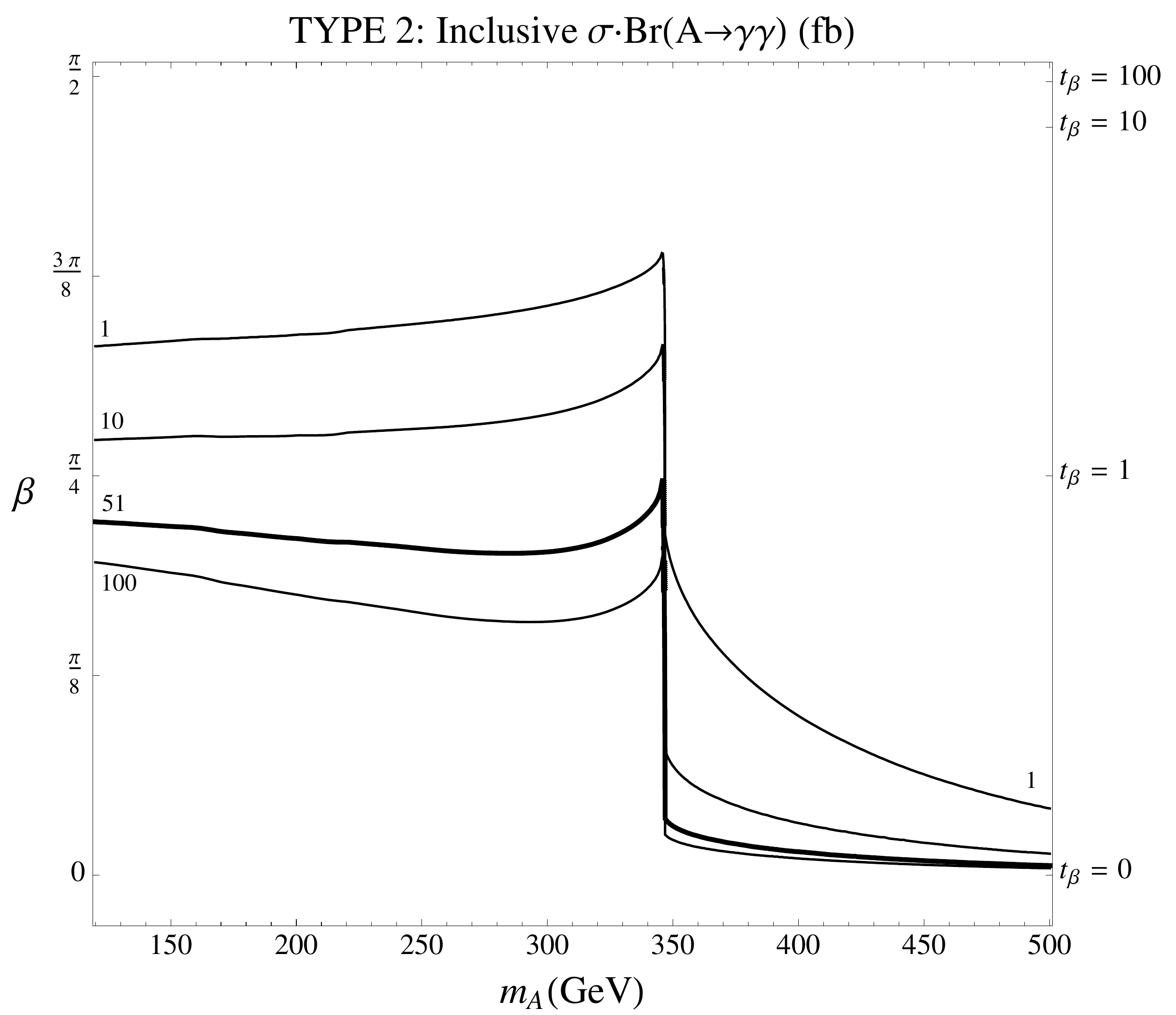} 
   \caption{Contours of the inclusive $\sigma \cdot {\rm Br}(A \to \gamma \gamma)$ in units of fb for 8 TeV $pp$ collisions for the pseudoscalar Higgs boson in the exact alignment limit, shown as a function of $m_A$ and $\tan \beta$ for Type 1 (left) and Type 2 (right) 2HDM. The thick line denotes the rate for a SM Higgs at 126 GeV of $\sim 51$ fb at 8 TeV $pp$ collisions.}
   \label{fig:Agammagammaalign}
\end{figure}

In Fig.~\ref{fig:Agammagammaalign} we also illustrate the variation of the inclusive production of $A$ with $A \to \gamma \gamma$ in the exact alignment limit as a function of mass and $\tan \beta$, focusing on low $\tan \beta$ where the rate is similar to the SM Higgs rate of $\sim 51$ fb at 8 TeV. In all 2HDM types, the rate falls rapidly with $\tan \beta$ due to the diminution of the $y_{Att}$ coupling. In Type 2 and Type 4 2HDM the inclusive rate rises again at large $\tan \beta$ due to $b\bar bA$ associated production, but not to the extent that the rate again reaches SM Higgs-like values. In all cases, once $m_A > 2 m_t$ the increase in total width due to $A \to t \bar t$ rapidly diminishes ${\rm Br}(A \to \gamma \gamma)$.

\subsection{Inclusive production of $H$ or $A$ with $H, A \to \tau^+ \tau^-$ or $H, A \to \mu^+ \mu^-$}

\begin{figure}[htbp] 
   \centering
   \includegraphics[width=3in]{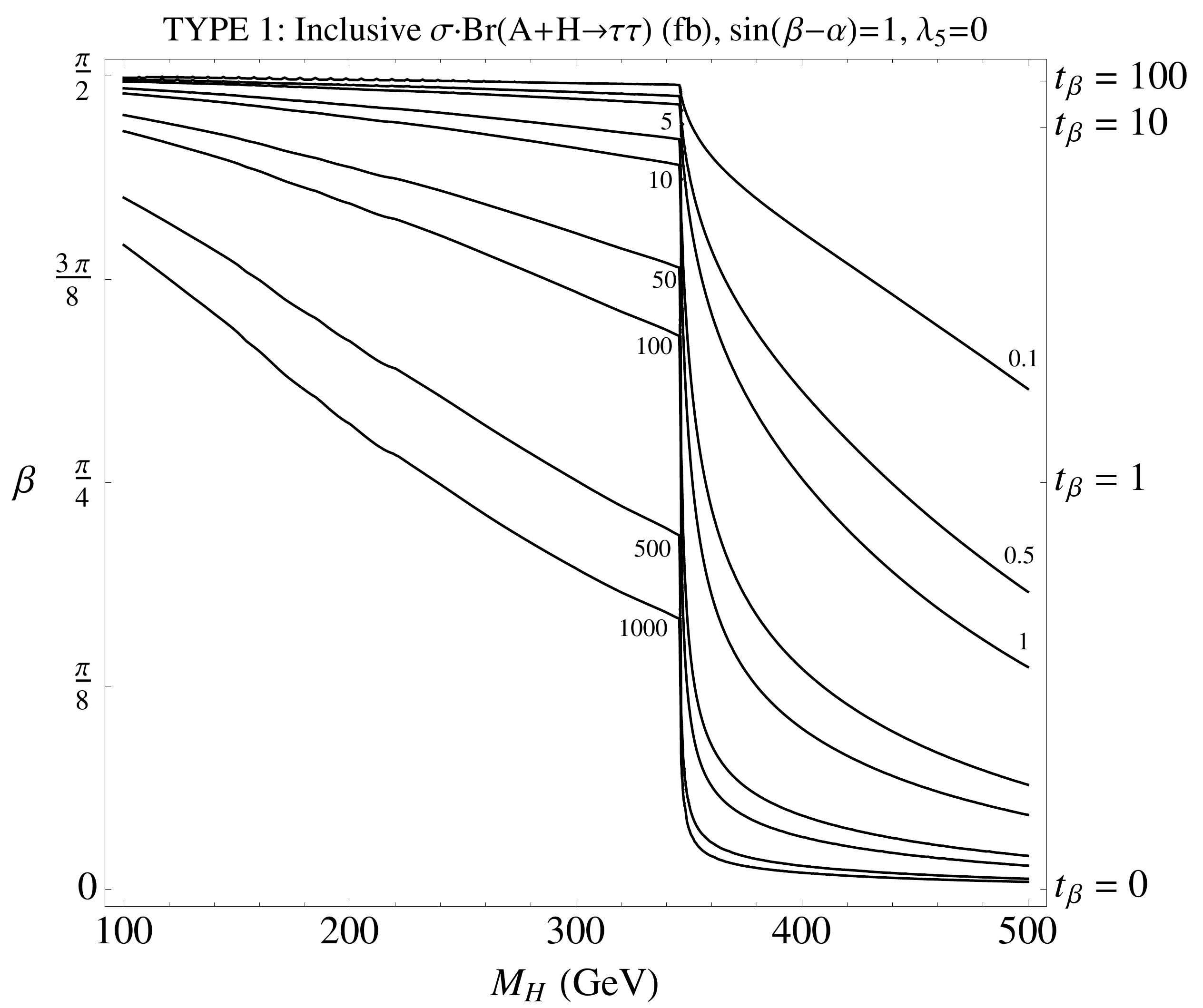} 
      \includegraphics[width=3in]{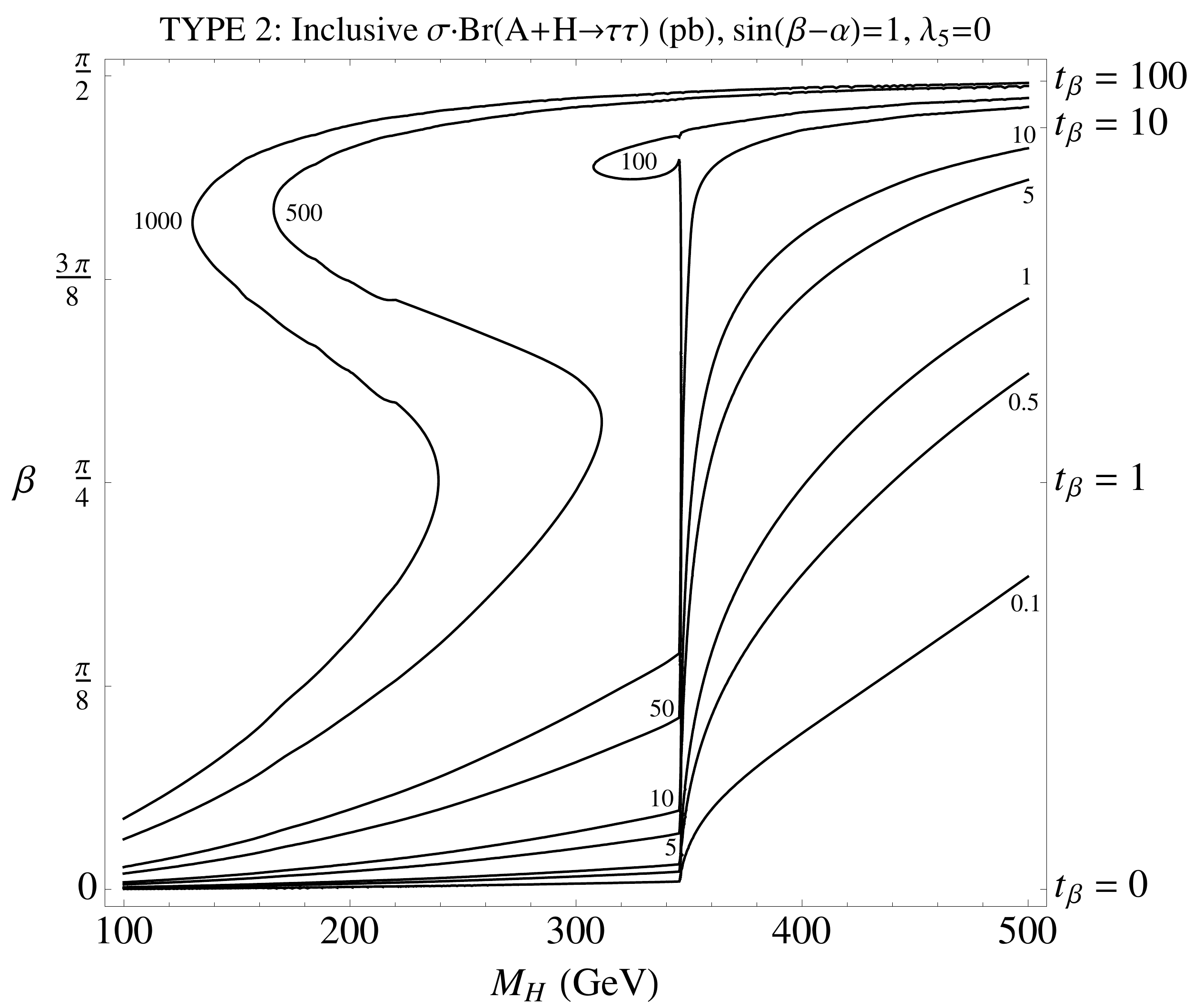} 
    \caption{Contours of the inclusive $\sigma \cdot {\rm Br}(A \to \tau^+ \tau^-)+\sigma \cdot {\rm Br}(A \to \tau^+ \tau^-)$ for 8 TeV $pp$ collisions for the sum of pseudocalar and non-SM-like scalar Higgs bosons in the alignment limit $\sin(\beta - \alpha) = 1$, shown as a function of $m_H$ and $\tan \beta$  in units of fb for Type 1 (left)  and in units of pb  for Type 2 (right) 2HDM. Here we have chosen $\lambda_{5,6,7} = 0$. }
   \label{fig:AHtautau}
\end{figure}

Inclusive production of $H$ or $A$ with $H,A \to \tau^+ \tau^-$ plays a particularly important role in searches for MSSM-like 2HDM, since $\Gamma(H \to \tau^+ \tau^-)$ is $\tan \beta$-enhanced. As such, high-mass searches for new scalars decaying to taus are already conducted at both ATLAS and CMS. Here we illustrate the parametric scaling as a function of $m_H, m_A$ and $\tan \beta$ in the exact alignment limit, $\sin(\beta - \alpha) = 1$, where this mode is expected to play a particularly strong role in discovery of new scalars. If both $H$ and $A$ are similar in mass, the poor mass resolution in the di-tau final state makes it useful to exhibit the collective signal from both states. Contours of the inclusive $\sigma \cdot {\rm Br}(A \to \tau^+ \tau^-)+\sigma \cdot {\rm Br}(A \to \tau^+ \tau^-)$ are shown in Fig.~\ref{fig:AHtautau}.

In Type 1 2HDM, the rate is particularly small, falling with $\tan \beta$ due to the suppression of the gluon fusion production mode and disappearing entirely once $H, A \to t \bar t$ becomes kinematically accessible.

In Type 2 2HDM the prospects are much greater due to the $\tan \beta$ enhacement of $\Gamma(H \to \tau^+ \tau^-)$. Features at low $\tan \beta$ are due to the interplay of the production modes. As $\tan \beta$ increases, gluon fusion is suppressed due to the falling $y_{Htt}, y_{Att}$ couplings, while $b\bar bH$ and $b\bar bA$ associated production is enhanced. The crossover takes place around $\tan \beta \sim 5$. The features around $m_H \sim 2 m_t$ are a result of the interplay between the production mode crossover and the emergence of the two-top threshold. 

Note that the parametric dependence for inclusive production of $H$ or $A$ with $H,A \to \mu^+ \mu^-$ is identical; the rate may be obtained from that of $H,A \to \tau^+ \tau^-$  by simply rescaling with a factor of $m_\mu^2 / m_\tau^2 \sim 0.0035$. Needless to say, this results in a vanishingly small rate for Type 1 2HDM, but a multi-fb rate for Type 2 2HDM. Searches in this channel are attractive due to the considerable mass resolution available in the di-muon final state. Given the important role of leptonic decays of $H$ and $A$ in the alignment limit, it is important to perform searches for resonant di-muon production out to high  masses. 

\subsection{Inclusive production with $H \to hh$}

Perhaps the most striking signal for the discovery of a heavy CP-even neutral Higgs is inclusive production of $H$ followed by $H\to hh$. As discussed earlier, this process often dominates the total width when kinematically available, and may even remain the largest branching ratio for $m_H > 2 m_t$ in Type 1 2HDM at large $\tan \beta$. The process leads to a variety of distinctive final states and the production cross section times branching ratio greatly exceeds the SM expectation of $\sim 8$ fb at 8 TeV. 

\begin{figure}[htbp] 
   \centering
   \includegraphics[width=3in]{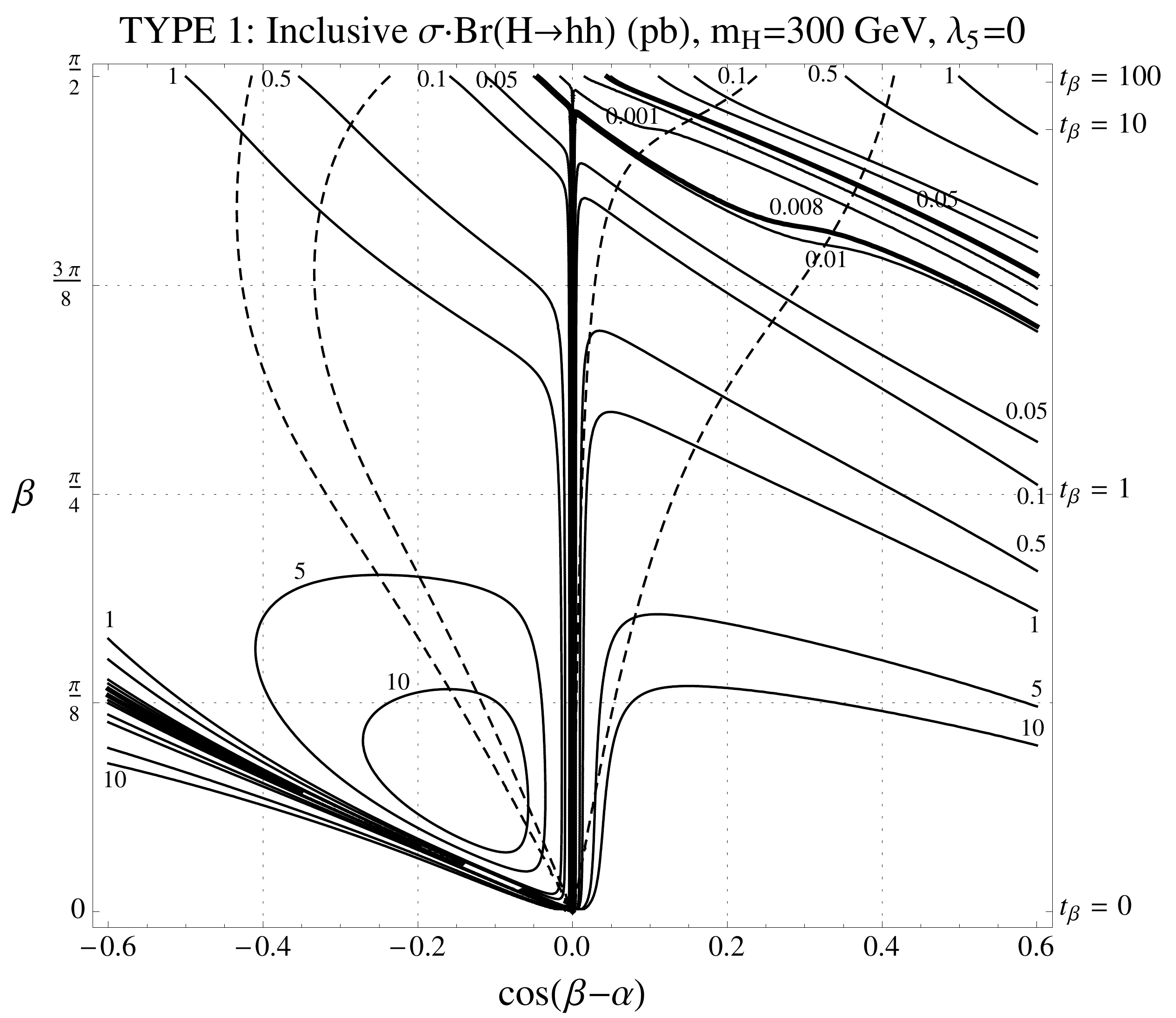} 
      \includegraphics[width=3in]{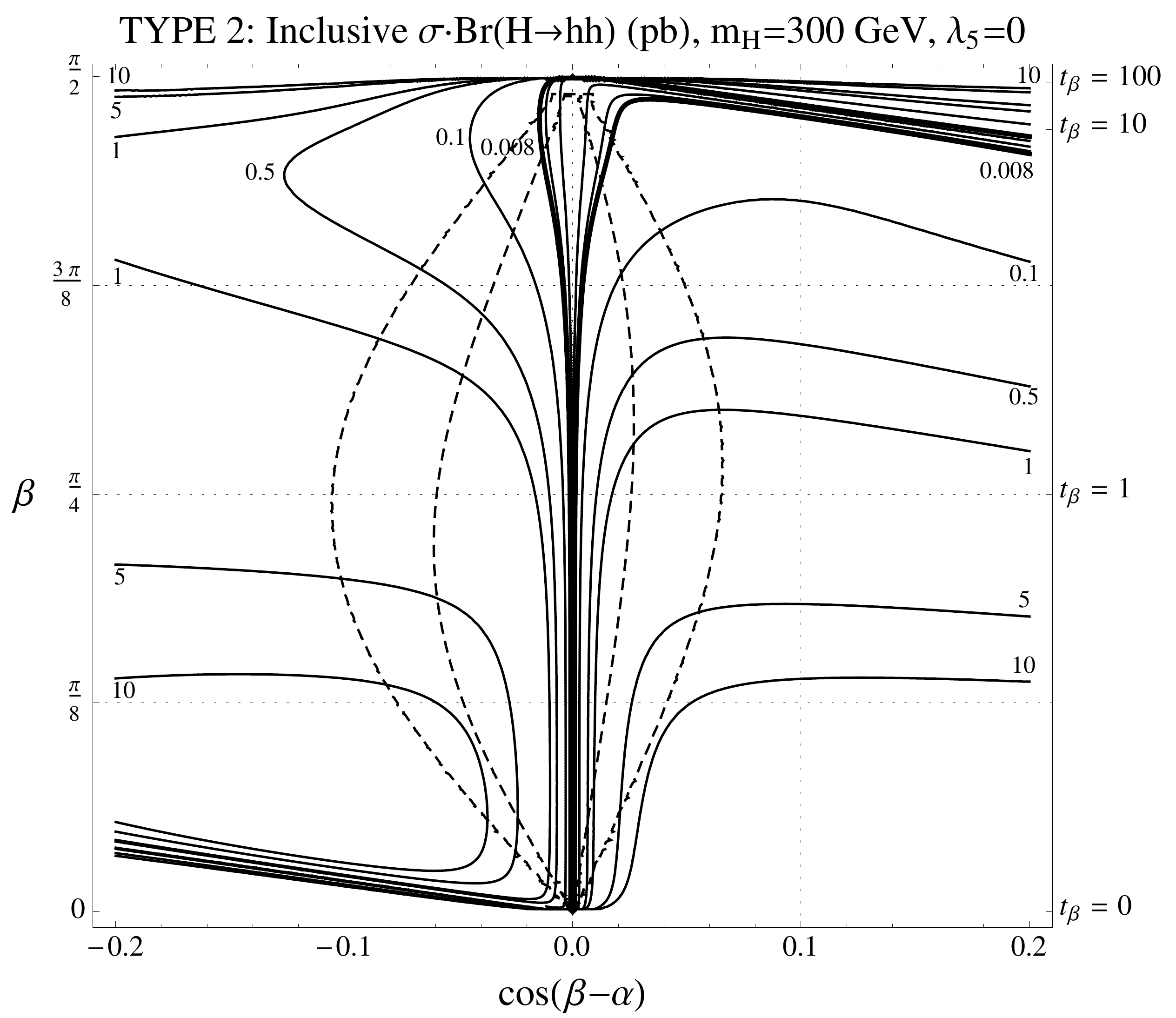} 
   \caption{Contours of the inclusive $\sigma \cdot {\rm Br}(H \to hh)$ in units of pb for 8 TeV $pp$ collisions for the non-SM-like scalar Higgs boson with $m_H = 300$ GeV, shown as a function of $\cos(\beta - \alpha)$ and $\beta$ for Type 1 (left) and Type 2 (right) 2HDM. Here we have chosen $\lambda_{5,6,7} = 0$ and $m_A = m_H$. The inner (outer) dashed contour denotes the 68\% (95\%) CL best fit to the signals of the SM-like Higgs.}
   \label{fig:Hhh}
\end{figure}

Contours of the inclusive $\sigma \cdot {\rm Br}(H \to hh)$ are shown in Fig.~\ref{fig:Hhh} for $\lambda_{5,6,7} = 0$. For Type 1 2HDM, the production mode is dominated by gluon fusion and the decay mode is parametrically identical to the total width except where $g_{Hhh} \to 0$. Thus the parametrics are governed by three features: (1) the falling production rate at large $\tan \beta$ due to diminishing $y_{Htt}$; (2) the vanishing of $g_{Hhh}$ in the alignment limit; and (3) the zeroes of $g_{Hhh}$ associated with $\cos(2 \beta - 2 \alpha) - \frac{\sin(2 \beta - 2 \alpha)}{\tan(2 \beta)} \approx \frac{1}{2} \frac{1}{1 - m_h^2/m_H^2}$. The zero for $\cos(\beta - \alpha) > 0$ approximately coincides with the zero in $y_{Htt}$, leading to the broad region of diminution around $\alpha \approx 0$. 

For Type 2 2HDM, the production mode is dominated by gluon fusion except at large $\tan \beta$ where $b\bar bH$ associated production takes over. The decay mode is parametrically identical to the total width at low $\tan \beta$ except in the exact alignment limit, while at high $\tan \beta$ the total width scales with $y_{Hbb}^2 \propto \tan^2 \beta$. Thus the features are again governed by the falling gluon fusion production rate at large $\tan \beta$ -- exacerbated by the growing width -- and the zeros of $g_{Hhh}$. 

In both cases, it bears emphasizing that $\sigma \cdot {\rm Br}(H \to hh)$ may be quite large -- more than a picobarn, two orders of magnitude larger than the SM rate -- while remaining consistent with current coupling fits to the signals of $h$. Even when $\lambda_{5,6,7} = 0$ the rate may remain high in light of direct limits on $H \to VV$, and nonzero $\lambda_{5,6,7}$ have the dual effect of increasing  $\sigma \cdot {\rm Br}(H \to hh)$ and weakening direct limits on $H \to VV$ by lowering $\sigma \cdot {\rm Br}(H \to VV)$.
 
Note in particular that for Type 1 2HDM, the region with enhanced VBF production of $h$ with $h \to \gamma \gamma, VV^*$ corresponds to $\sigma\cdot{\rm Br}(H \to hh) \gtrsim 1$ pb. This implies that if VBF production remains high due to Type 1 couplings, there is a large and readily discoverable rate for $\sigma\cdot{\rm Br}(H \to hh)$ if $H$ is not too heavy.

\begin{figure}[htbp] 
   \centering
   \includegraphics[width=3in]{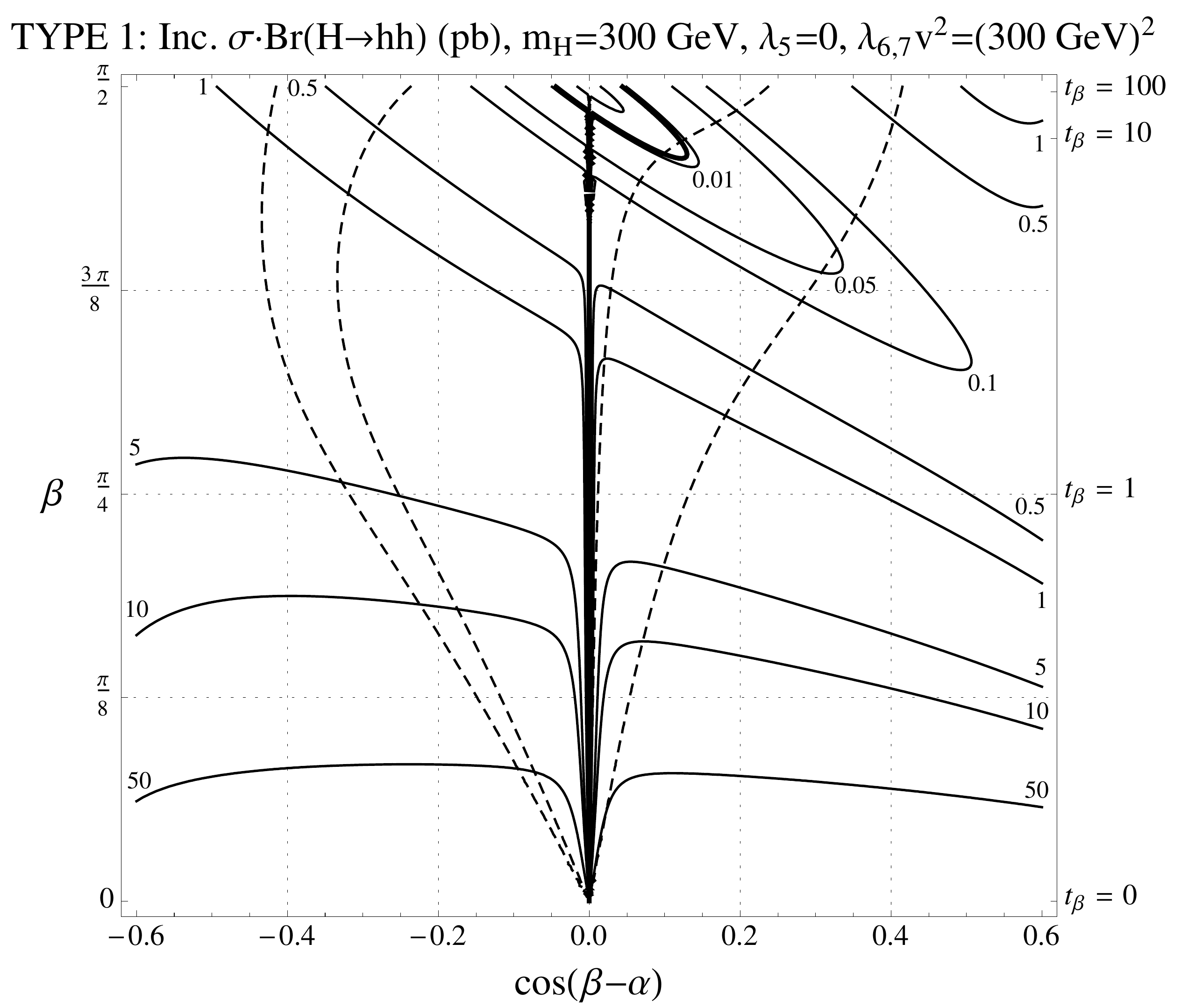} 
      \includegraphics[width=3in]{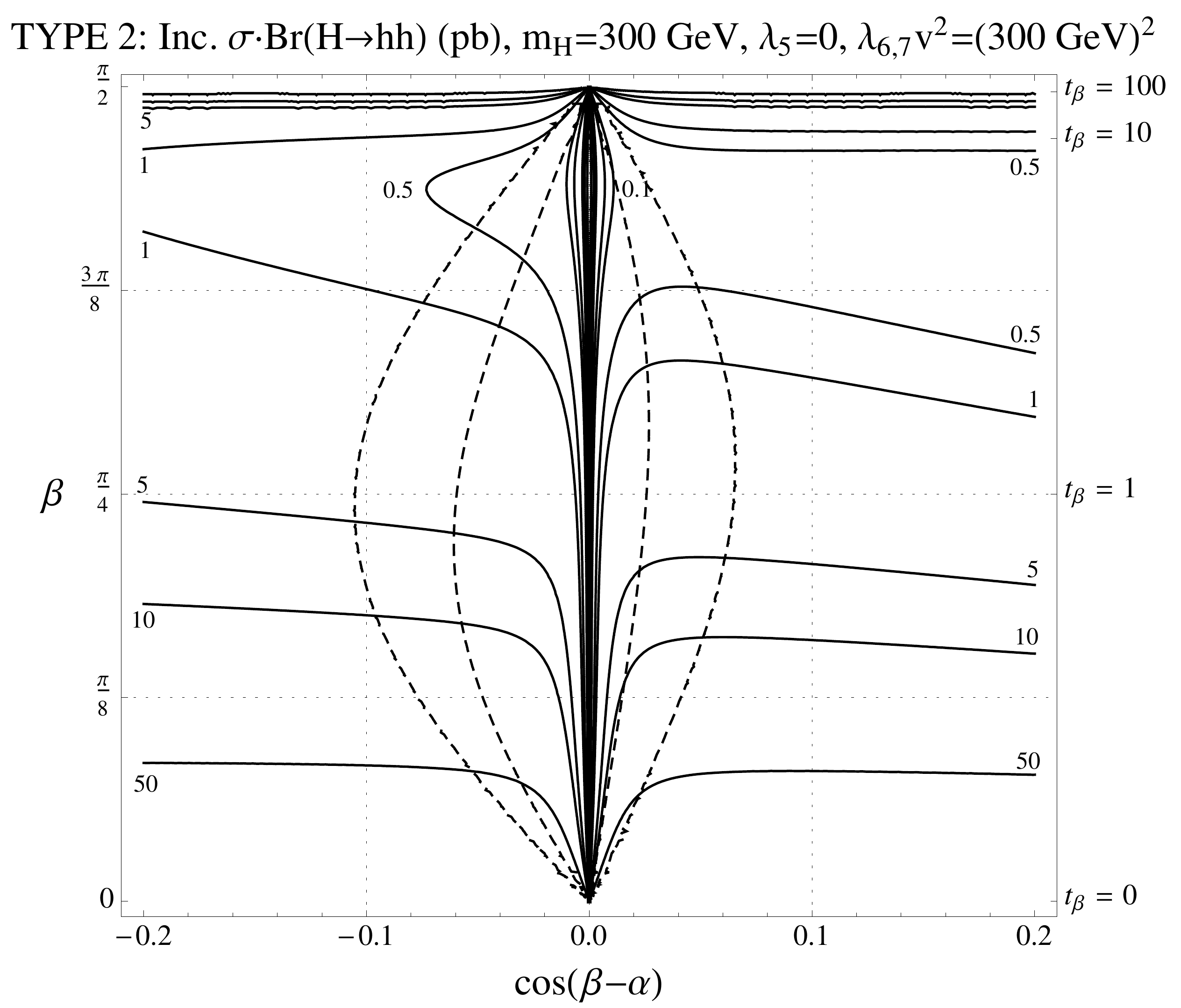} 
   \caption{Contours of the inclusive $\sigma \cdot {\rm Br}(H \to hh)$ in units of pb for 8 TeV $pp$ collisions for the non-SM-like scalar Higgs boson with $m_H = 300$ GeV, shown as a function of $\cos(\beta - \alpha)$ and $\beta$ for Type 1 (left) and Type 2 (right) 2HDM. Here we have chosen $\lambda_{5} = 0,$ $\lambda_{6,7} v^2 = (300 ; {\rm GeV})^2,$ and $m_A = m_H$. The inner (outer) dashed contour denotes the 68\% (95\%) CL best fit to the signals of the SM-like Higgs.}
   \label{fig:Hhh67}
\end{figure}

Contours of the inclusive $\sigma \cdot {\rm Br}(H \to hh)$ are shown in Fig.~\ref{fig:Hhh67} for $\lambda_{5} = 0$ and $\lambda_{6,7} v^2 = (300 \; {\rm GeV})^2$. When $\lambda_{6,7}$ are nonzero, the $\tan \beta$-enhanced contributions to $\Gamma(H \to hh)$ ensure that it comprises an even larger component of the total width and dominates even at large $\tan \beta$ in Type 2 2HDM.

\subsection{Inclusive production with $A \to Zh$}

Inclusive production of the pseudoscalar $A$ with $A \to Zh$ plays a role quite analogous to inclusive production of $H$ with $H \to hh$, often governing the total width when kinematically available and potentially providing one of the most promising modes for discovery of additional scalars at the LHC. Although there are $Zh$ associated production searches at both ATLAS and CMS, which place a constraint on excessive enhancement of the total $Zh$ cross section, sensitivity may be improved significantly by searching for resonant production of the $Zh$ final state.

\begin{figure}[htbp] 
   \centering
   \includegraphics[width=3in]{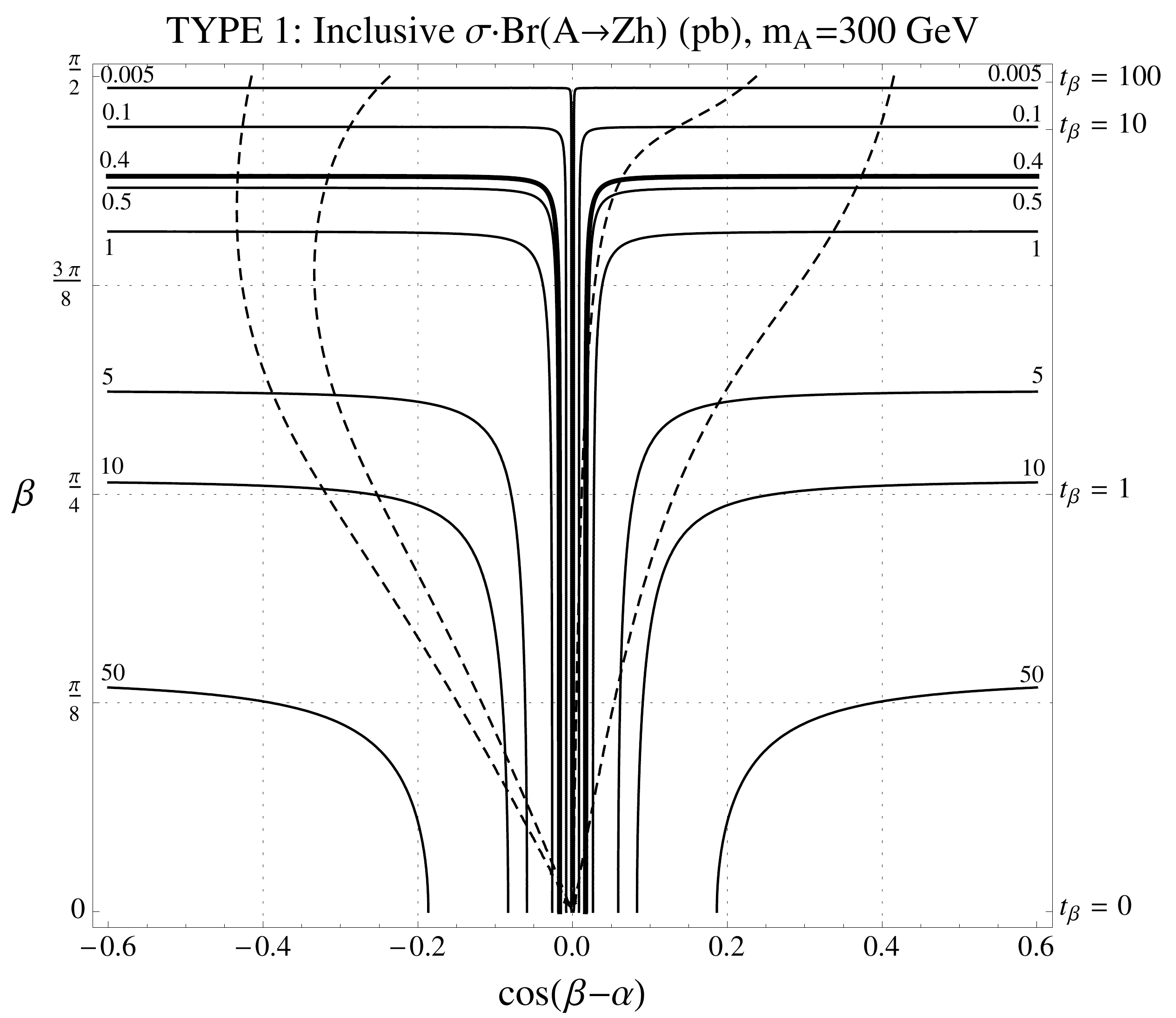} 
      \includegraphics[width=3in]{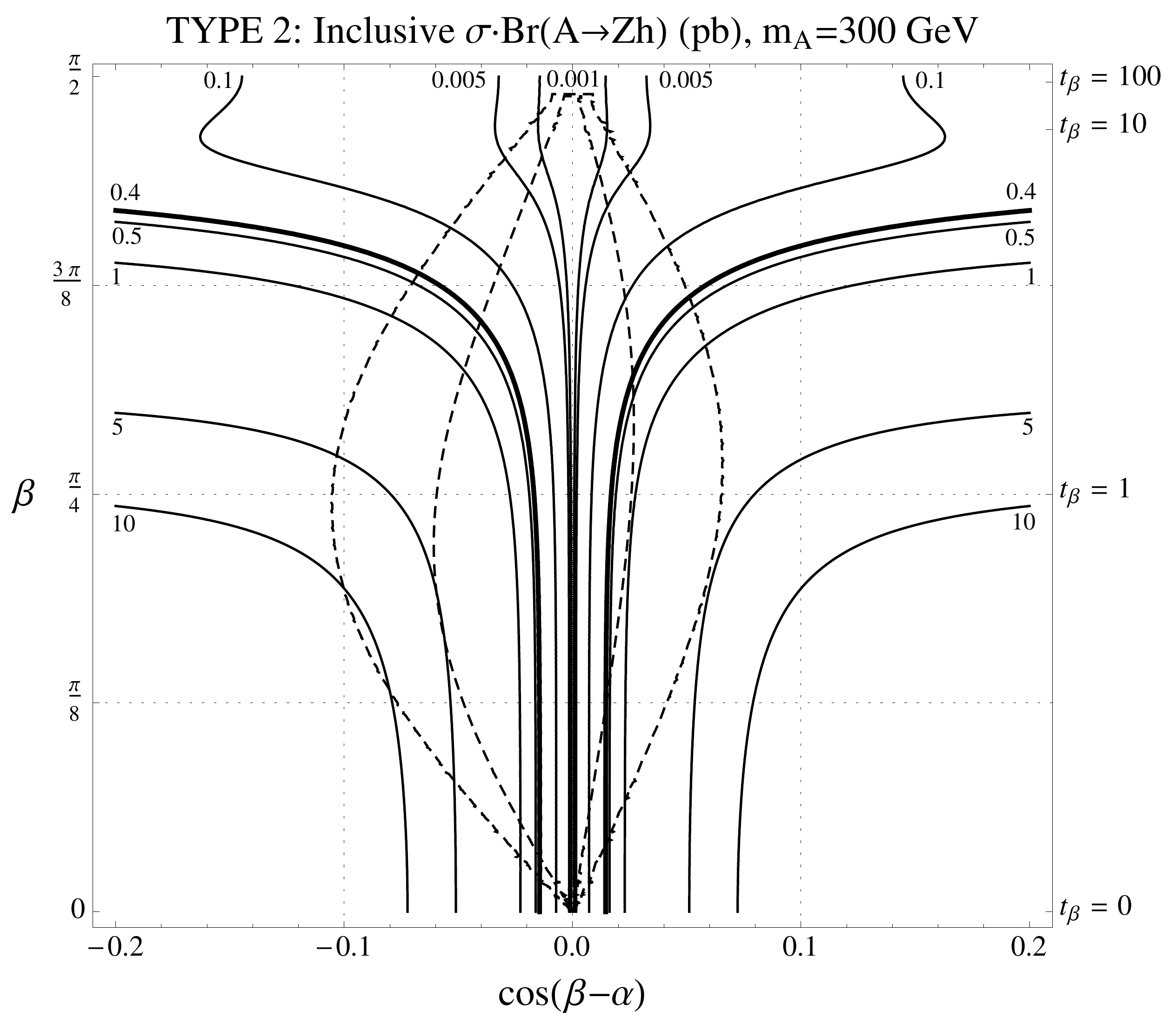} 
   \caption{Contours of the inclusive $\sigma \cdot {\rm Br}(A \to Zh)$ in units of pb for 8 TeV $pp$ collisions for the pseudoscalar Higgs boson with $m_A = 300$ GeV, shown as a function of $\cos(\beta - \alpha)$ and $\beta$ for Type 1 (left) and Type 2 (right) 2HDM. The inner (outer) dashed contour denotes the 68\% (95\%) CL best fit to the signals of the SM-like Higgs.}
   \label{fig:AZh}
\end{figure}

Contours of the inclusive $\sigma \cdot {\rm Br}(A \to Zh)$ are shown in Fig.~\ref{fig:AZh}. The parametric scaling is relatively straightforward. In Type 1 2HDM the production mode is primarily gluon fusion, and so falls with $y_{Att}$ at large $\tan \beta$, while the decay mode vanishes only in the exact alignment limit and otherwise controls the total width. Near the alignment limit, it is particularly dominant at large $\tan \beta$ where the fermion partial widths are $\tan \beta$-suppressed. This combination of features entirely explains the distinctive shape of the $\sigma \cdot {\rm Br}(A \to Zh)$ contours in Type 1 2HDM.

In Type 2 2HDM, the story is entirely analogous at low $\tan \beta$. At high $\tan \beta$, $b\bar bA$ associated production increases the production rate, but the total width is increasingly controlled by $\Gamma(A \to b \bar b)$. This explains the broadening of contours at large $\tan \beta$ relative to the Type 1 case.

Much as with $H \to hh$, for Type 1 2HDM, the region with enhanced VBF production of $h$ with $h \to \gamma \gamma, VV^*$ corresponds to $\sigma\cdot{\rm Br}(A \to Zh) \gtrsim 1$ pb. Again, if VBF production remains high due to Type 1 couplings, there is a large and discoverable rate for $\sigma\cdot{\rm Br}(A \to Zh)$ if $A$ is not too heavy.

\subsection{$t \bar t$ production with $t \to H^\pm \bar b$ and $H^\pm \to \tau^\pm \nu$}

The inclusive production cross section for $H^\pm$ is generally quite small, coming primarily from $t \bar b H^\pm$ associated production, but the rate may be appreciable when $m_{H^\pm} < m_t$ and $H^\pm$ appears as a rare decay mode in $t \bar t$ pair production. As with di-tau production, searches for $t \bar t$ production with $t \to H^\pm \bar b$ and $H^\pm \to \tau^\pm \nu$ are already carried out in the context of MSSM-like 2HDM, but we reproduce the parametrics here for both Type 1 and Type 2 2HDM for completeness. 

\begin{figure}[htbp] 
   \centering
         \includegraphics[width=3in]{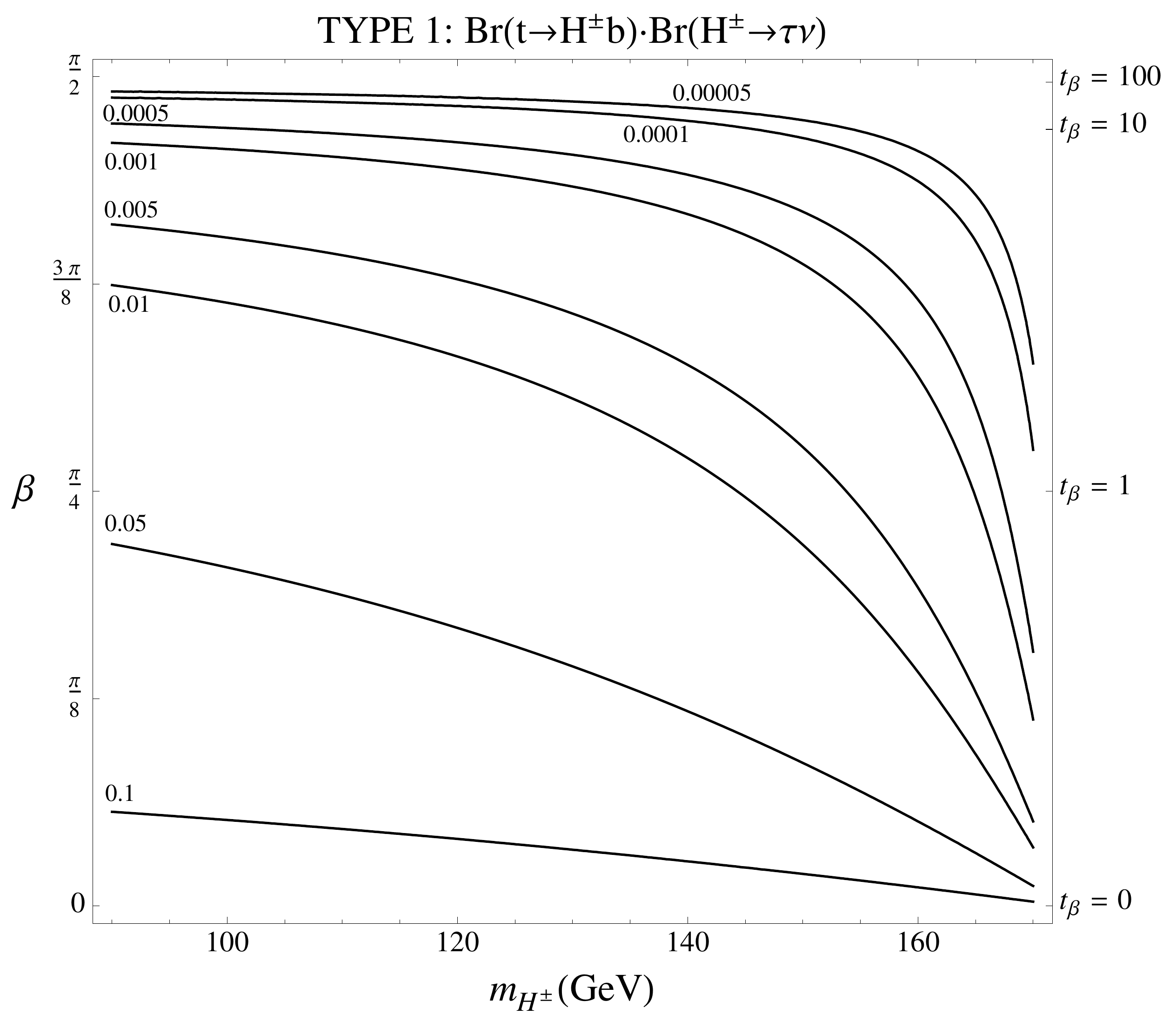} 
      \includegraphics[width=3in]{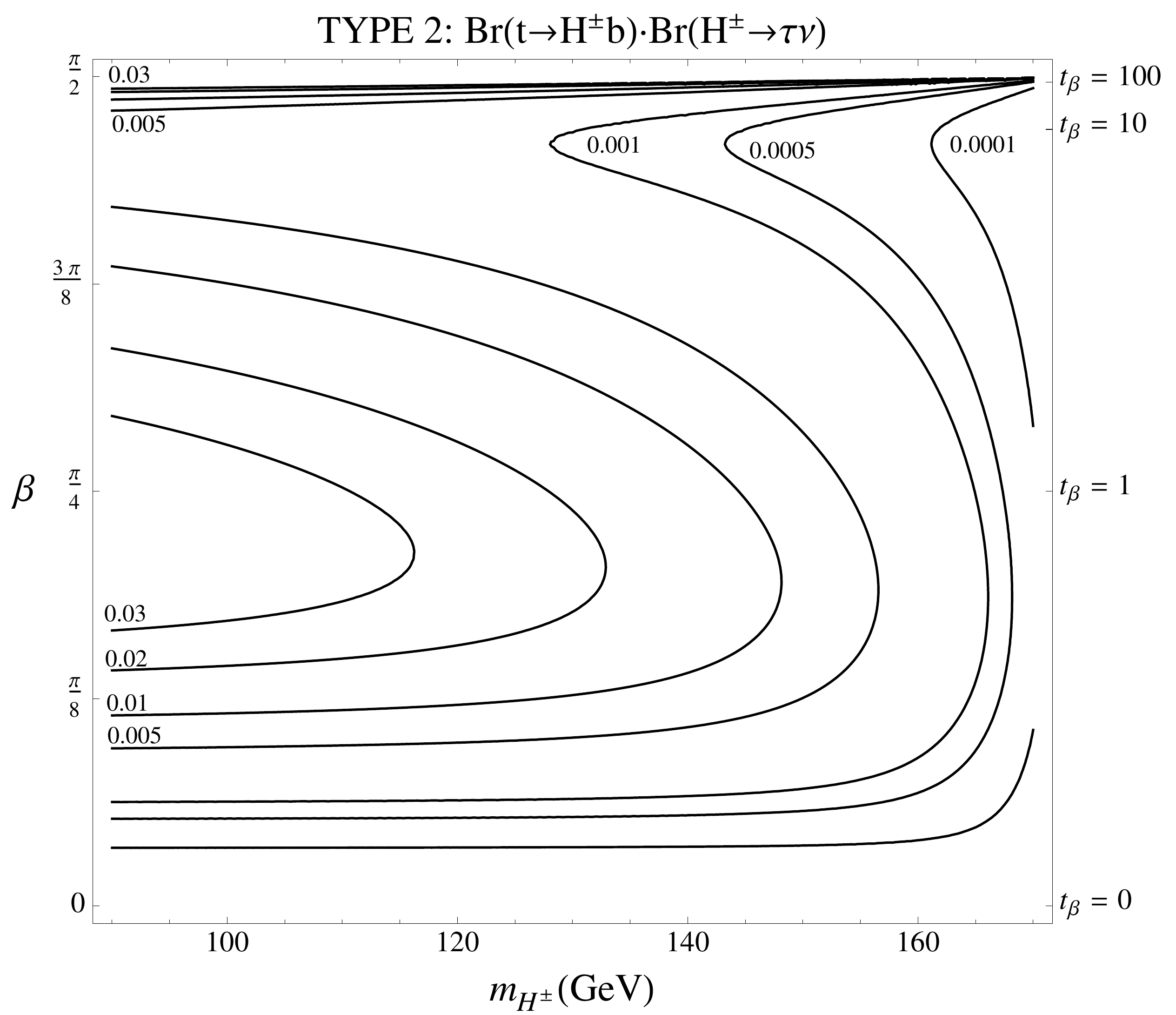}    
      \caption{Contours of the combined branching ratio 
      ${\rm Br}(t \to H^\pm \bar b) \cdot {\rm Br}(H^\pm \to \tau^\pm \nu)$, 
      shown as a function of $m_H$ and $\tan \beta$  for Type 1 (left) and Type 2 (right) 2HDM. }
   \label{fig:Hctaunu}
\end{figure}

Contours of the combined branching ratio ${\rm Br}(t \to H^\pm \bar b) \cdot {\rm Br}(H^\pm \to \tau^\pm \nu)$ are shown in Fig.~\ref{fig:Hctaunu}. For $m_{H^\pm} < m_t$, $\Gamma(H^\pm \to \tau \nu)$ uniformly dominates the total width. The contours are governed largely by $\Gamma(t \to H^\pm \bar{b})$, which accumulates contributions proportional to both $y_{Att}$ and $y_{Abb}$. Thus in Type 1 2HDM the branching ratio falls with both $\tan \beta$ and mass as the fermion couplings and phase space are suppressed. In Type 2 2HDM,  contributions due to $y_{Att}$ are important at low $\tan \beta$, and transition to contributions from $y_{Abb}$ at large $\tan \beta$, all modulated by the decreasing phase space as $m_{H^\pm}$ is increased.

\subsection{Signals of degenerate Higgses}

\begin{figure}[htbp] 
   \centering
   \includegraphics[width=3in]{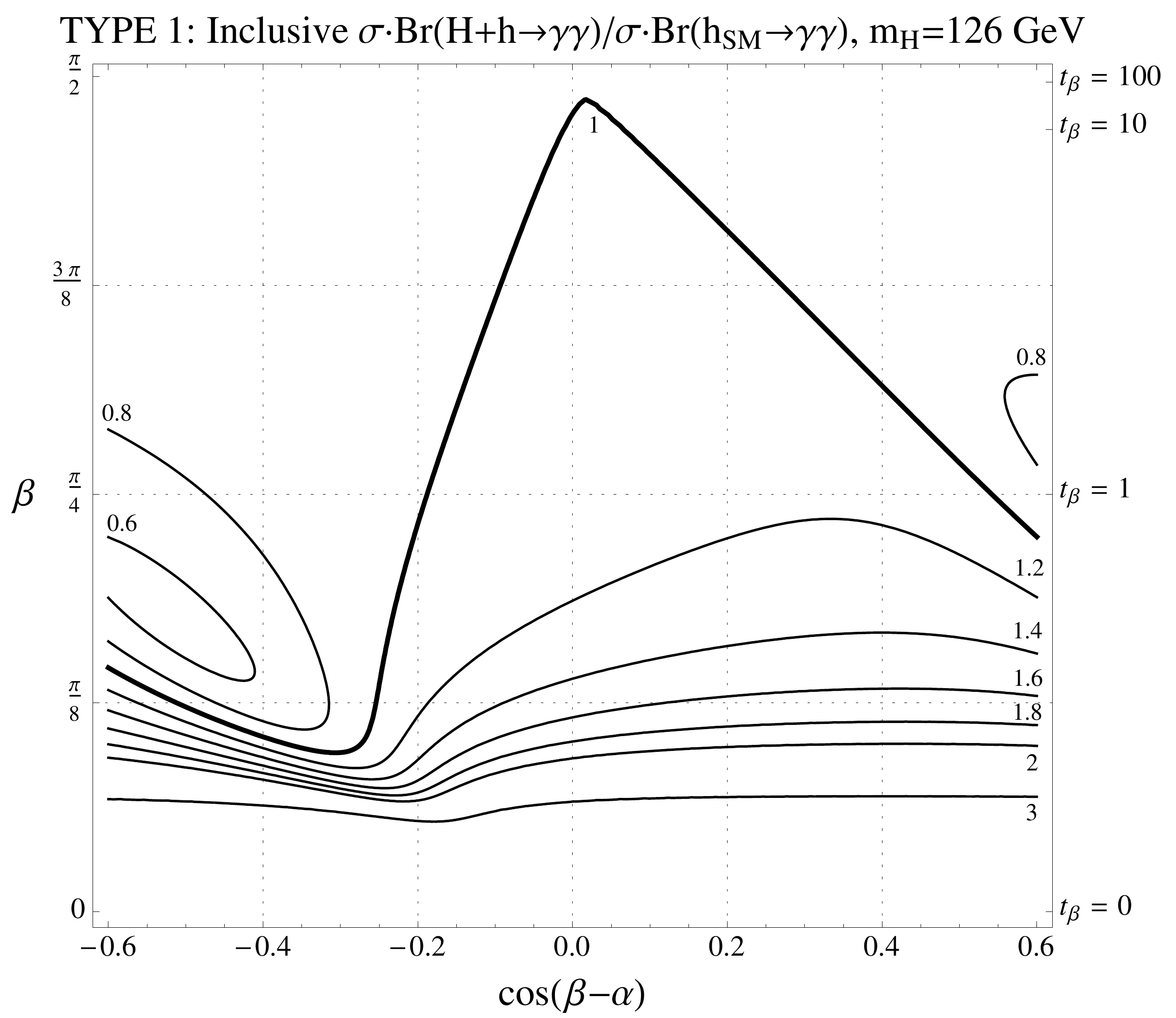} 
      \includegraphics[width=3in]{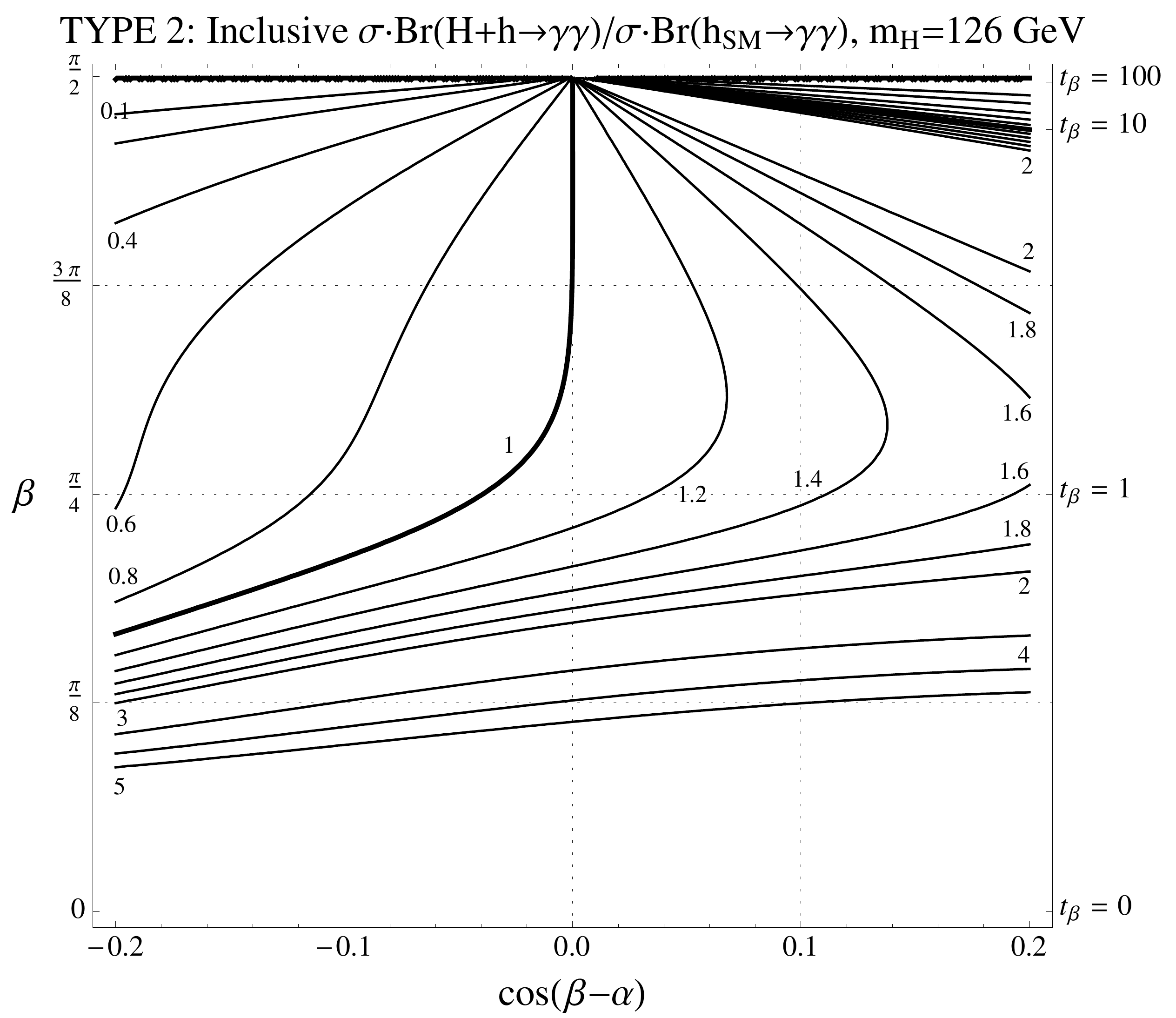} 
   \caption{Contours of the inclusive $\sigma \cdot {\rm Br}(H+h \to \gamma \gamma)/\sigma \cdot {\rm Br}(h_{SM} \to \gamma \gamma)$  for 8 TeV $pp$ collisions for the sum of contributions from the SM-like and non-SM-like scalar Higgs bosons with $m_H = m_h = 126$ GeV, shown as a function of $\cos(\beta - \alpha)$ and $\beta$ for Type 1 (left) and Type 2 (right) 2HDM. }
   \label{fig:Hhgammagamma}
\end{figure}

\begin{figure}[htbp] 
   \centering
   \includegraphics[width=3in]{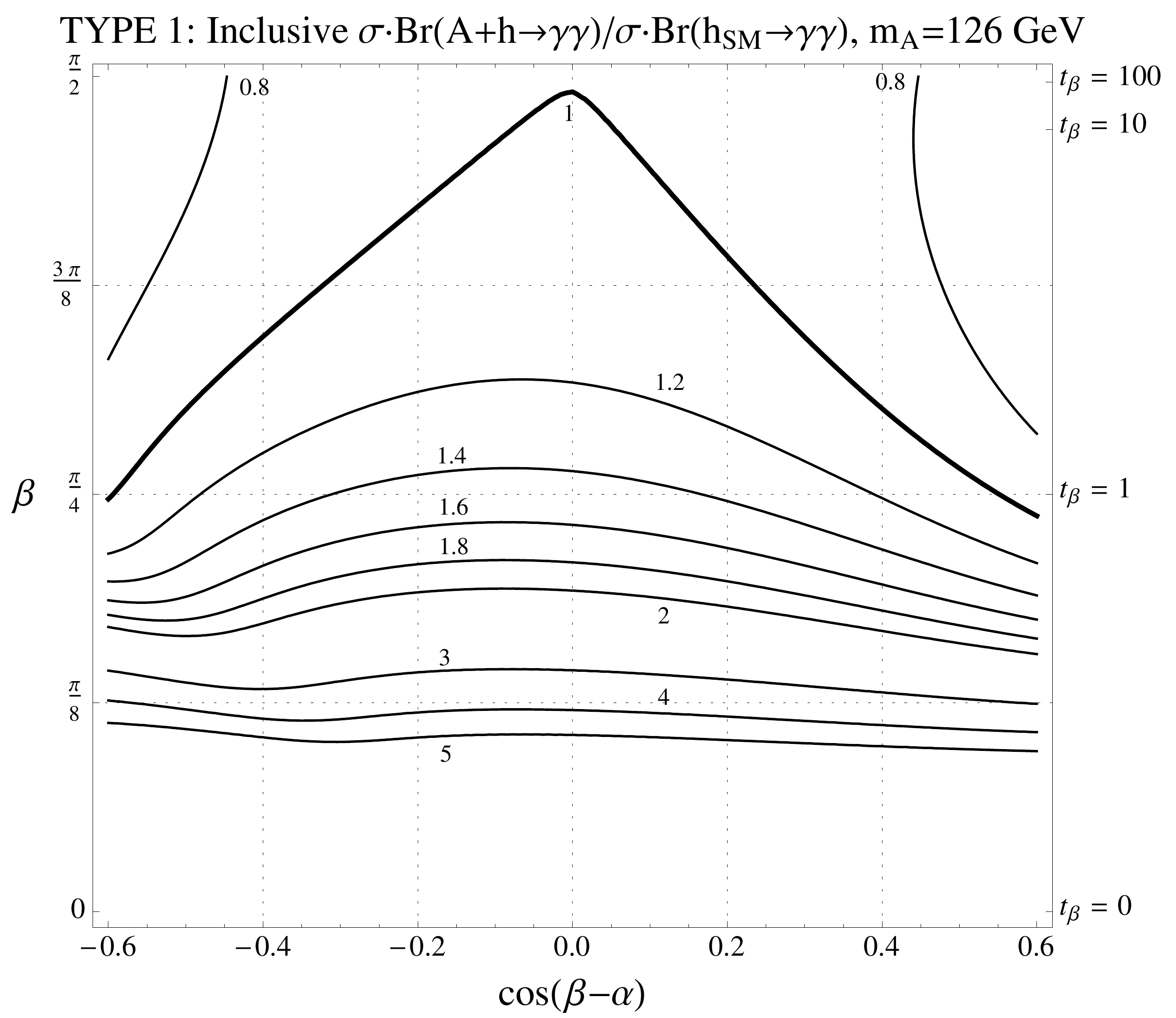} 
      \includegraphics[width=3in]{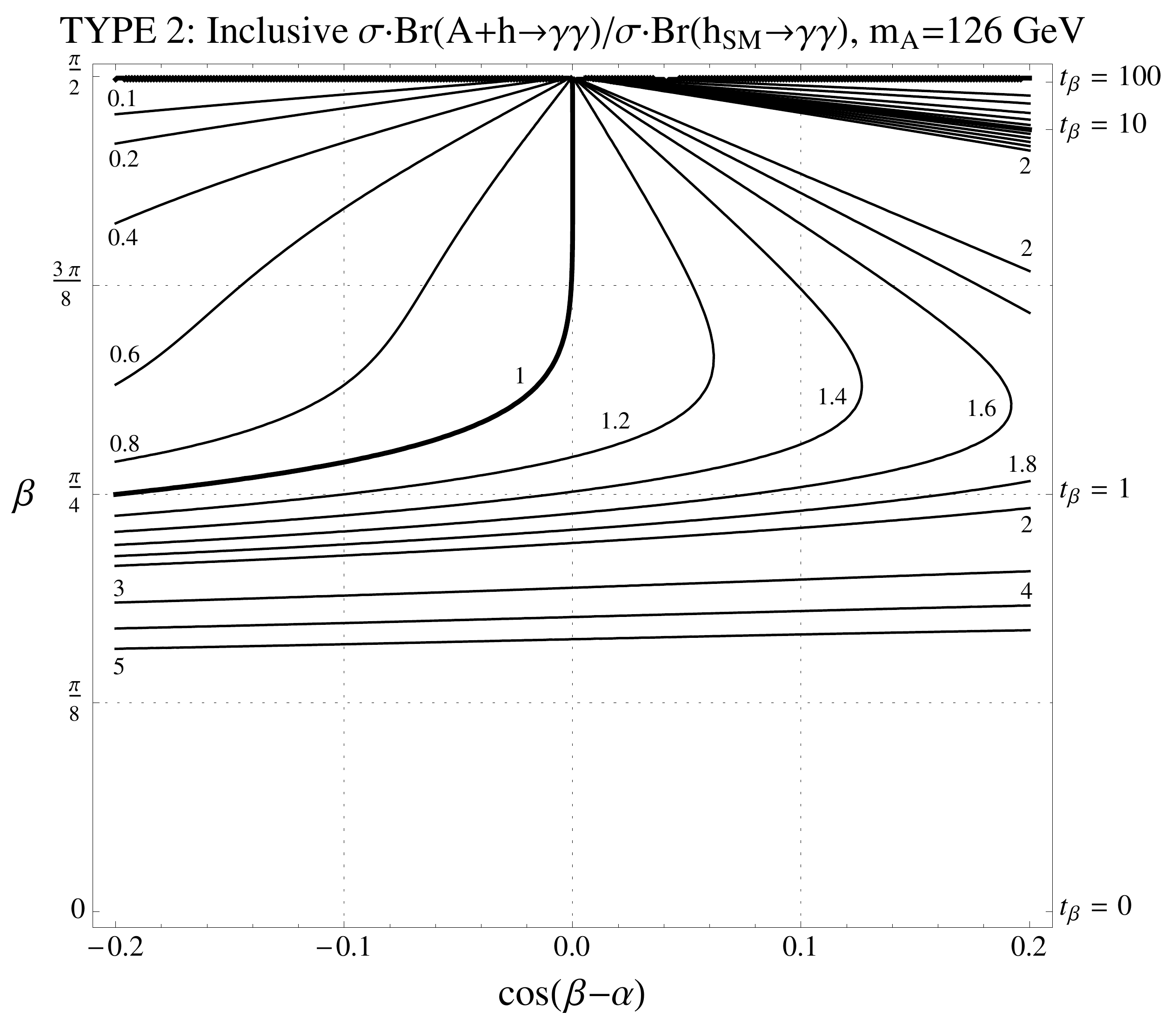} 
   \caption{Contours of the inclusive $\sigma \cdot {\rm Br}(A+h \to \gamma \gamma)/\sigma \cdot {\rm Br}(h_{SM} \to \gamma \gamma)$  for 8 TeV $pp$ collisions for the sum of contributions from the SM-like and pseudoscalar Higgs bosons with $m_A = m_h = 126$ GeV, shown as a function of $\cos(\beta - \alpha)$ and $\beta$ for Type 1 (left) and Type 2 (right) 2HDM. }
   \label{fig:Ahgammagamma}
\end{figure}

\begin{figure}[htbp] 
   \centering
   \includegraphics[width=3in]{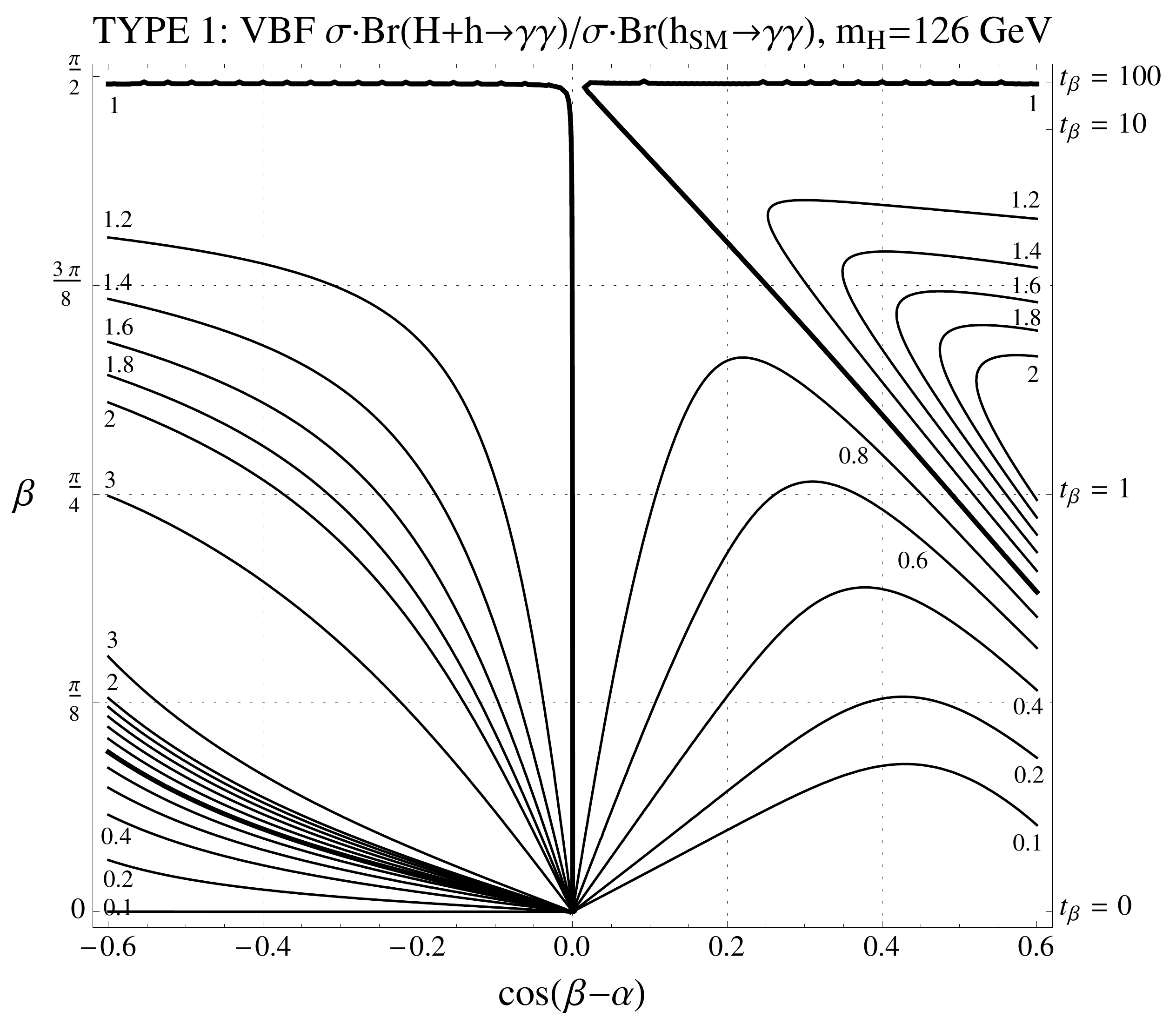} 
      \includegraphics[width=3in]{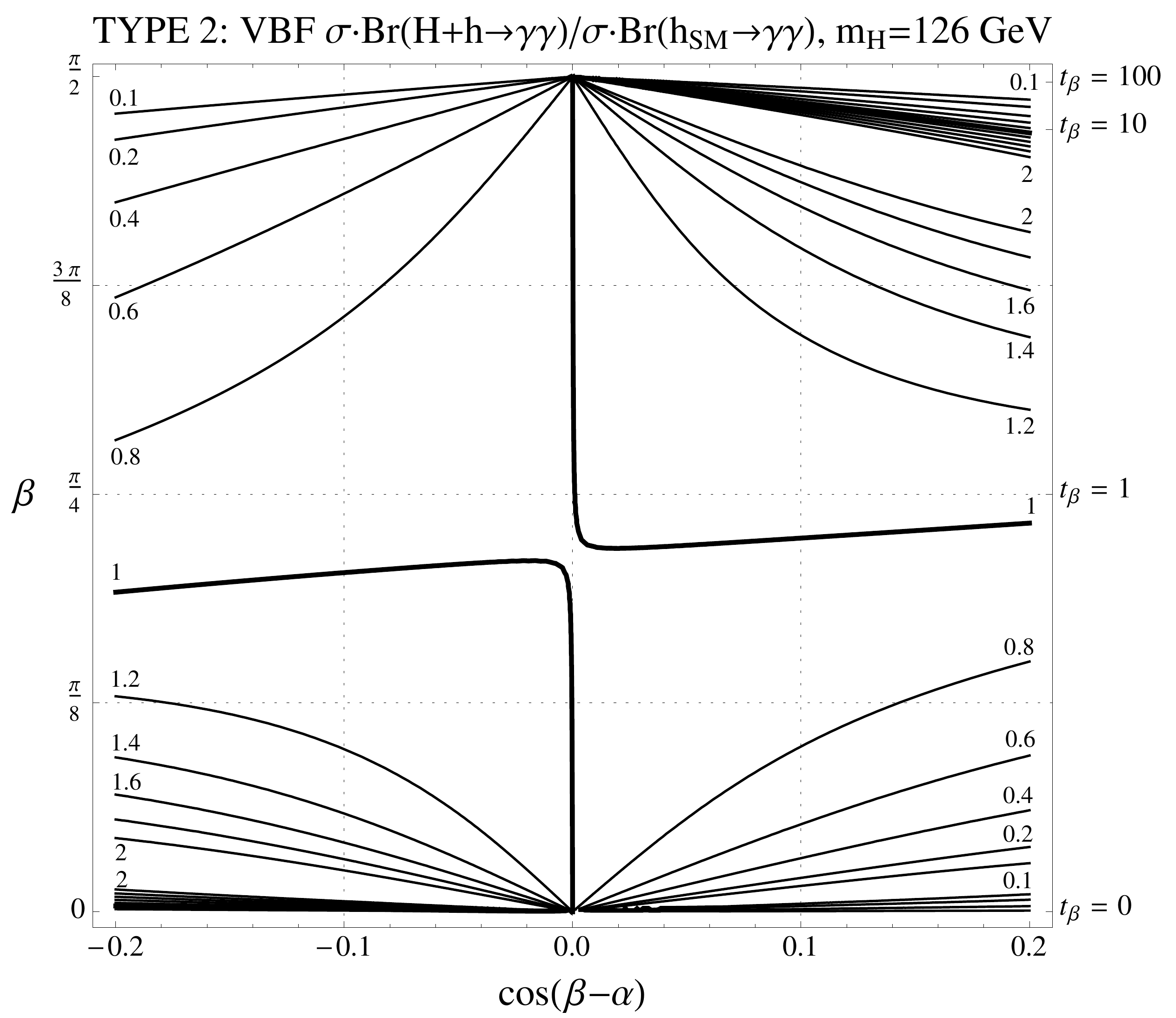} 
   \caption{Contours of the VBF $\sigma \cdot {\rm Br}(H+h \to \gamma \gamma)/\sigma \cdot {\rm Br}(h_{SM} \to \gamma \gamma)$  for 8 TeV $pp$ collisions for the sum of contributions from the SM-like and non-SM-like scalar Higgs bosons with $m_H = m_h = 126$ GeV, shown as a function of $\cos(\beta - \alpha)$ and $\beta$ for Type 1 (left) and Type 2 (right) 2HDM. }
   \label{fig:HhVBFgammagamma}
\end{figure}

Finally, it is interesting to consider the possibility that the observed signals at 126 GeV could arise from two near-degenerate scalars, which if sufficiently degenerate could not be distinguished with current experimental resolution. The inclusive contributions from $H$ or $A$ degenerate with $h$ are shown in Figs.~\ref{fig:Hhgammagamma} and \ref{fig:Ahgammagamma}, respectively. We also show the VBF contribution from $H$ degenerate with $h$ in Fig.~\ref{fig:HhVBFgammagamma}. We do not overlay fits to the current Higgs signal, since the fits were performed under the hypothesis that only $h$ contributes.\footnote{Performing a fit to signals originating from degenerate $h$ and $H/A$ would in any event require accurately accounting for the possibility of large production in association with $b$ quarks due to $\beta$-dependent couplings that can enhance this mode.  This is however not a production mode for which experimental efficiencies are presently known, so assuming such a signal would necessarily jeopardize the credibility of such fits.} Unsurprisingly, near the alignment limit, the signal typically exceeds that of the SM Higgs, since the additional contributions from $H$ or $A$ are maximized in the alignment limit. In practice, distinguishing such degenerate Higgses is experimentally challenging given the limited mass resolution at the LHC, but there is some potential to resolve degenerate states using appropriate cross-ratios of branching ratios \cite{Gunion:2012he}. 

\section{Conclusions \label{sec7}}

The search for the Higgs at the LHC is entering a new post-discovery phase in which measurements of the SM-like Higgs couplings and direct searches for additional scalars begin to explore the parameter space of extended electroweak symmetry breaking sectors. In this work we have constructed a map between current signal fits to the SM-like Higgs, potential deviations in future measurements of the SM-like Higgs, and possible signals of additional Higgs scalars in the context of theories with two Higgs doublets. Our work highlights a number of important observations that should be incorporated into LHC searches for additional Higgs scalars. Coupling measurements of the SM-like Higgs suggest that, in the context of 2HDM, the most promising channels for discrepancies in future measurements include VBF and $Vh$ production of $h$ with $h \to \gamma \gamma$ or $h \to VV^*$ as well as inclusive production of $h$ with $h \to \tau^+ \tau^-$, all of which are correlated and may be enhanced by as much as 80\% (20\%) above SM predictions consistent with current fits for case of a Type 1 (Type 2) 2HDM. 

Coupling measurements of the SM-like Higgs also constrain the production modes for heavier Higgses. Away from the alignment limit, the gluon fusion production of $H$ with decay to $hh$ (when kinematically available) may exceed the SM rate for Higgs pair production by more than two orders of magnitude consistent with current coupling fits. This process may dominate decays of $H$ even when $m_H > 2 m_t$, providing a promising search channel for high-mass Higgs bosons. It may also reduce the branching ratio of $H$ to vector bosons, weakening limits from current direct searches for $H \to VV$. Similarly, gluon fusion production of $A$ with decay to $Zh$ may be more than twice the SM rate for $Zh$ associated production. These rates remain appreciable close to the alignment limit since both $\Gamma(H \to hh)$ and $\Gamma(A \to Zh)$ scale as $\propto m^3 / v^2$, compensating somewhat for alignment limit-suppression. Very close to the alignment limit these modes become subdominant, but gluon fusion production of $H, A$ with decay to $\gamma \gamma, \tau^+ \tau^-,$ and $\mu^+ \mu^-$ may all be appreciable and provide promising avenues for discovering additional scalars even if the light Higgs is completely SM-like. 

Our results suggest that $H \to hh$ and $A \to Zh$ should become high priorities in searching for additional Higgs bosons. The possibility of observable signals in $\gamma \gamma, \tau^+ \tau^-,$ and $\mu^+ \mu^-$ final states in the exact alignment limit suggest these searches should be extended to higher Higgs masses. \\

\noindent {\bf Note Added:} While this manuscript was being completed, \cite{Djouadi:2013vqa,Coleppa:2013dya,Chen:2013rba} appeared, addressing aspects of the interplay between coupling fits and additional signals in certain 2HDM types.

\acknowledgments{We thank Kyle Cranmer, Sally Dawson, Aleandro Nisati, and Marc Sher 
for useful discussions. NC and ST are supported in part by  the DOE under grant 
DE-FG02-96ER40959. NC is also supported by the NSF under grant PHY-0907744 and the Institute for Advanced Study. The work of J.G. is supported by the ERC Advanced Grant No.~267985
{\it Electroweak Symmetry Breaking, Flavour and ÊDark Matter: One Solution for Three Mysteries (DaMeSyFla)}.}

\appendix


\section{$Hhh$ Coupling} \label{app:a}

The magnitudes of 2HDM couplings that involve more than two Higgs boson scalars 
depend in 
detail on the underlying interactions. 
For definiteness we consider here renormalizable models with general 
CP-conserving tree-level potential 
\begin{eqnarray} 
\nonumber
V &=& m_{11}^2 \Phi_1^\dag \Phi_1 + m_{22}^2 \Phi_2^\dag \Phi_2 - \left[m_{12}^2 \Phi_1^\dag \Phi_2 + {\rm h.c.} \right] \\ \label{eq:potential}
&+& \frac{1}{2} \lambda_1 (\Phi_1^\dag \Phi_1)^2 + \frac{1}{2} \lambda_2 (\Phi_2^\dag \Phi_2) + \lambda_3 (\Phi_1^\dag \Phi_1)(\Phi_2^\dag \Phi_2) + \lambda_4 (\Phi_1^\dag \Phi_2) (\Phi_2^\dag \Phi_1) \\ \nonumber
&+& \left[ \frac{1}{2} \lambda_5 (\Phi_1^\dag \Phi_2)^2 + \lambda_6 (\Phi_1^\dag \Phi_1) (\Phi_1^\dag \Phi_2) + \lambda_7 (\Phi_2^\dag \Phi_2) \Phi_1^\dag \Phi_2 + {\rm h.c.} \right]
\label{tree_pot}
\end{eqnarray} 
The Feynman diagram coupling for $Hhh$ 
with $SU(2)_L \times U(1)_Y \to U(1)_Q$ electroweak symmetry breaking 
may be obtained from 
the quartic interactions in (\ref{tree_pot}) by 
eliminating the fields $\Phi_1^0$ and $\Phi_2^0$
in favor of the mass eigenstates $h$ and $H$ 
and expectation values 
$v_1$ and $v_2$ using the mixing relations 
(\ref{hHmixing})
with the result 
\begin{eqnarray} 
g_{Hhh} = 
\frac{v}{4}  & \Big[& 
 \cos \alpha \big( - (3 \lambda_1 + \lambda_3 + \lambda_4 + \lambda_5) \cos \beta 
                         -3(\lambda_6 + \lambda_7 ) \sin \beta    \big)  \nonumber \\
& + & 
             \sin \alpha \big(  -(3 \lambda_2 + \lambda_3 + \lambda_4 + \lambda_5) \sin \beta 
                   -3(\lambda_6 + \lambda_7 ) \cos \beta    \big)   \nonumber \\
& + & 
         3 \cos (3 \alpha) \big( ( \lambda_1 - \lambda_3 - \lambda_4 - \lambda_5) \cos \beta
               + (\lambda_6 - 3 \lambda_7 ) \sin \beta    \big)    \nonumber \\
& + & 
         3 \sin (3 \alpha) \big( ( -\lambda_2 + \lambda_3 + \lambda_4 + \lambda_5) \sin \beta
               + (3 \lambda_6 -  \lambda_7 ) \cos \beta    \big) 
                          ~\Big]
                          \label{Hhh_coupling_lam}
\end{eqnarray}               
The 
combinations of quartic interactions 
$\lambda_1$, $\lambda_2$, and $\lambda_3 + \lambda_4$, 
appearing in the $Hhh$ coupling 
may be written in terms of the physical masses
$m_h, m_H, m_A$, the expectation value $v$, 
the mixing angles $\alpha$ and $\beta$, and the quartic interactions  
$\lambda_5$, $\lambda_6$, and $\lambda_7$
\cite{Gunion:2002zf}
\begin{eqnarray} 
\lambda_1 & = &   { m_H^2 \cos^2 \alpha  + m_h^2 \sin^2 \alpha - m_A^2 \sin^2 \beta \over 
   v^2 \cos^2 \beta}    - \lambda_5 \tan^2 \beta - 2 \lambda_6 \tan \beta 
      \\
\lambda_2 & = &   { m_H^2 \sin^2 \alpha  + m_h^2 \cos^2 \alpha - m_A^2 \cos^2 \beta \over 
   v^2 \sin^2 \beta}    - \lambda_5 \cot^2 \beta - 2 \lambda_7 \cot \beta 
      \\
\lambda_3 + \lambda_4 & = & 
  { (m_H^2 - m_h^2) \sin\alpha \cos \alpha \over v^2 \sin \beta \cos \beta}    
  + { m_A^2 \over v^2} -   \lambda_6 \cot \beta - \lambda_7 \tan \beta 
\end{eqnarray}
With these relations, the $Hhh$ coupling for the 
tree-level potential (\ref{tree_pot}) may be written
\begin{eqnarray} 
\label{gHhheq} 
g_{Hhh}  =   
\frac{\cos(\beta \! - \! \alpha)}{v} && \hspace{-0.5cm}
\left[ \left(3 m_A^2 + 3 \lambda_5 v^2 - 2 m_h^2 - m_H^2 \right) 
\left( \cos(2 \beta - 2 \alpha) - \frac{ \sin(2 \beta - 2 \alpha)}{\tan(2 \beta)} \right)  \right. \nonumber \\
& -&   m_A^2 - \lambda_5 v^2 + \frac{\lambda_6 v^2}{2}  
\left( - \cot \beta + 3 \sin(2\beta - 2 \alpha) + 3 \cot \beta \cos (2 \beta - 2 \alpha) \right) \nonumber \\ 
& +&  \left. \frac{\lambda_7 v^2}{2}  
 \left( - \tan \beta - 3 \sin(2 \beta - 2 \alpha) + 3 \tan \beta \cos(2 \beta - 2 \alpha) \right) \right] 
\end{eqnarray}
The $Hhh$ self coupling is homogenous in $\cos(\beta - \alpha)$,
and vanishes in the alignment limit.  
Near the alignment limit 
it may be expanded in a power series about $\cos (\beta - \alpha) =0$,
with the result 
\begin{eqnarray}
g_{Hhh} \simeq - \frac{\cos(\beta \! - \! \alpha)}{v} & \bigg[ &  4 m_A^2 - 2m_h^2 - m_H^2  + 4 \lambda_5 v^2
\nonumber \\
& & 
+ \frac{2 v^2}{\tan(2 \beta)} (\lambda_{6} - \lambda_7)  + \frac{2 v^2}{\sin(2 \beta)} (\lambda_{6} +\lambda_{7}  ) 
 + {\cal O}(\cos(\beta \! - \! \alpha)) ~ \bigg] 
\end{eqnarray}

The general expression (\ref{gHhheq}) for the $Hhh$ coupling simplifies 
in certain models.  
In the MSSM $\lambda_{5}=\lambda_6=\lambda_7  =0$. 
In this case the
$Hhh$ coupling reduces to a function of the physical masses $m_h, m_H, m_A$ 
and the mixing angles $\alpha$ and $\beta$
\begin{eqnarray} \label{mssmlike}
g_{Hhh}^{\rm MSSM} = \frac{\cos(\beta - \alpha)}{v} 
\left[ \left(3 m_A^2  - 2 m_h^2 - m_H^2 \right) 
\left( \cos(2 \beta - 2 \alpha) - \frac{ \sin(2 \beta - 2 \alpha)}{\tan(2 \beta)} \right) - m_A^2  \right] 
\hspace{0.5cm}
\end{eqnarray}    
This expression may be simplified using the 
the tree-level MSSM Higgs mass sum rule
$m_h^2 + m_H^2 = m_Z^2 + m_A^2$, 
as well as the MSSM tree-level relations
$m_A^2 = m_Z^2 {\sin(2\alpha + 2 \beta)}/ {\sin(2 \alpha - 2 \beta)}$ and 
$m_h^2 = m_Z^2 \cos(2 \beta) {\sin(\alpha + \beta)}/ {\sin(\alpha - \beta)}$, 
with the result 
\begin{eqnarray} \label{mssmlimit}
g_{Hhh}^{\rm MSSM} &=& - \frac{m_Z^2}{v} 
 \Big[2 \sin(2\alpha) \sin(\beta + \alpha) - \cos(2 \alpha) \cos(\beta + \alpha) \Big]
\end{eqnarray}    
As another limit, 
2HDMs with a $Z_2$ exchange symmetry $\Phi_1 \leftrightarrow \Phi_2$
have $m_{12}^2=0$ and 
$\lambda_{6}=\lambda_7 = 0$. 
In this case $m_A^2 = -\lambda_5 v^2$, 
and the $Hhh$ couplings reduces to a function of the masses $m_h$ and $m_H$ 
and the mixing angles $\alpha$ and $\beta$
\begin{eqnarray}
g_{Hhh}^{Z_2} &=&- \frac{\cos(\beta - \alpha)}{v} \left[ \left( 2 m_h^2 + m_H^2 \right) \left( \cos(2 \beta - 2 \alpha) - \frac{ \sin(2 \beta - 2 \alpha)}{\tan(2 \beta)} \right)  \right]
\end{eqnarray}
This form of the $Hhh$ coupling was used in a previous study of  
multi-lepton signatures of 2HDMs
\cite{Craig:2012pu}.


\section{Standard-Model Like Higgs Fit Data \label{app}}

For reference we collect here the data that is used in constructing fits of the SM-like 
Higgs couplings.  
In Tables~\ref{tab:ATLAS}, \ref{tab:CMS}, \ref{tab:Tevatron} we show results from ATLAS, CMS, and the Tevatron respectively.

\begin{table}[hhh]
\footnotesize
\centering
\renewcommand{\arraystretch}{1.1}
\begin{tabular}{| l | c | c | c | c | c |}
\hline
Channel & $\hat \mu$ (7 TeV) & $\zeta_i^{\rm (G, V, T)}$ (\%)&  $\hat \mu$ (8 TeV) & $\zeta_i^{\rm (G,V,T)}$ (\%) & Refs. \\
\hline\hline 
$b \bar b$ & comb. w/8 & --- & ${-0.42\pm 1.05}$ & $(0,100,0)$ & \cite{ATLASbb, ATLASbbT} \\
$b \bar b$ ($ttH$)& $3.81 \pm 5.78$ & $(0,30,70)$ & ---  & --- &\\
\cline{1-6}
$\tau \tau$ & comb. w/8 & --- & ${0.7\pm 0.7}$ & $(20,80,0)$& \cite{ATLAStautau} \\
\cline{1-6}
$WW\, (0j)$ & $0.06 \pm 0.60$ & inclusive & $0.92^{+0.63}_{-0.49}$ & inclusive &    \\
$WW\, (1j)$ & $2.04^{+1.88}_{-1.30}$ & inclusive & $1.11^{+1.20}_{-0.82}$ & inclusive &   \cite{ATLASww} \\
$WW\, (2 j)$ & --- & --- & $1.79^{+0.94}_{-0.75}$ & $(20,80,0)$ & \\
\cline{1-6}
$ZZ$ & comb. w/8  & --- & $1.7^{+0.5}_{-0.4}$ & inclusive& \cite{ATLASzz} \\
\cline{1-6}
$\gamma \gamma_{\rm (L)} $ (uc$|$ct) & $0.53^{+1.37}_{-1.44}$ & $(93,7,0)$& $0.86\pm 0.67$ & $(93.7,6.2,0.2)$& \\
$\gamma \gamma_{\rm (H)} $ (uc$|$ct) &$0.17^{+1.94}_{-1.91}$ & $(67,31,2)$ & $0.92^{+1.1}_{-0.89}$ & $(79.3,19.2,1.4)$ &  \\
$\gamma \gamma_{\rm (L)} $ (uc$|$ec) & $2.51^{+1.66}_{-1.69}$& $(93,7,0)$ &$2.51^{+0.84}_{-0.75}$  &  $(93.2,6.6,0.1)$ & \\
$\gamma \gamma_{\rm (H)} $ (uc$|$ec) & $10.39^{+3.67}_{-3.67}$ & $(65,33,2)$ & $2.69^{+1.31}_{-1.08}$ & $(78.1,20.8,1.1)$& \\
$\gamma \gamma_{\rm (L)} $ (c$|$ct) & $6.08^{+2.59}_{-2.63}$ & $(93,7,0)$ &  $1.37^{+1.02}_{-0.88}$ &$(93.6,6.2,0.2)$&\\
$\gamma \gamma_{\rm (H)} $ (c$|$ct) & $-4.40^{+1.80}_{-1.76}$ & $(67,31,2)$ &$1.99^{+1.50}_{-1.22}$  &$(78.9,19.6,1.5)$ & \\
$\gamma \gamma_{\rm (L)} $ (c$|$ec) & $2.73^{+1.91}_{-2.02}$ & $(93,7,0)$ & $2.21^{+1.13}_{-0.95}$ & $(93.2,6.7,0.1)$&  \\
$\gamma \gamma_{\rm (H)} $ (c$|$ec) & $-1.63^{+2.88}_{-2.88}$ & $(65,33,2)$ & $1.26^{+1.31}_{-1.22}$ & $(77.7,21.2,1.1)$& \cite{ATLASgaga,ATLASgaga1212}\\
$\gamma \gamma $ (c$|$trans.) & $0.35^{+3.56}_{-3.60}$ & $(89,11,0)$ & $2.80^{+1.64}_{-1.55}$ & $(90.7,9.0,0.2)$& \\
$\gamma \gamma $ (dijet) & $2.69^{+1.87}_{-1.84}$ & $(23,77,0)$ & --- & --- & \\
$\gamma \gamma $ (loose high mass $jj$) & --- & --- & $2.76^{+1.73}_{-1.35}$ & $(45,54.9,0.1)$& \\
$\gamma \gamma $ (tight high mass $jj$) & --- & --- & $1.59^{+0.84}_{-0.62}$ & $(23.8,76.2,0)$& \\
$\gamma \gamma $ (low mass $jj$) & --- & --- & $0.33^{+1.68}_{-1.46}$ & $(48.1,49.9,1.9)$& \\
$\gamma \gamma $ ($E_{\rm T}^{\rm miss}$ significance) & --- & --- & $2.98^{+2.70}_{-2.15}$ & $(4.1,83.8,12.1)$& \\
$\gamma \gamma $ (lepton tag) & --- & --- & $2.69^{+1.95}_{-1.66}$ & $(2.2,79.2,18.6)$& \\
\cline{2-4}
\hline
\end{tabular}
\caption{\small Light Higgs fit data from ATLAS.  We denote best fits on signal strength modifier as $\hat \mu$ and quote efficiencies $\zeta$ for production initiated by gluons (G), weak gauge bosons (V), and top quarks (T). For diphoton channels, `uc' (`c') corresponds to unconverted (converted) photons, `ct' indicates central photons, `ec' indicates one or more photon in the endcap, and subscripts (H, L) designate high and low $p_T$.}
\label{tab:ATLAS}
\end{table}
\begin{table}[hhh]
\footnotesize
\centering
\renewcommand{\arraystretch}{1.1}
\begin{tabular}{| l | c | c | c | c | c |}
\hline
Channel & $\hat \mu$ (7 TeV) & $\zeta_i^{\rm (G,V,T)}$ (\%) &  $\hat \mu$ (8 TeV) & $\zeta_i^{\rm (G,V,T)}$(\%) & Refs. \\
\hline\hline 
$b \bar b$ & comb. w/8 & --- & $1.30^{+0.68}_{-0.59}$ & $(0,100,0)$& \cite{CMSbb} \\
$b \bar b$ ($ttH$)& ${-0.81^{+2.05}_{-1.75} }$ & $(0,30,70)$ & ---  & --- & \cite{CMSbbT}\\
\cline{1-6}
$\tau \tau$ ($0/1j$)& comb. w/8 & --- & $0.74^{+0.49}_{-0.52}$ & inclusive &  \\
$\tau \tau$ (VBF) & comb. w/8 & ---& $1.38^{+0.61}_{-0.57}$ & $(0,100,0)$ & \cite{CMStautau} \\
$\tau \tau$ (VH) & comb. w/8 & --- & $0.76^{+1.48}_{-1.43}$ & $(0,100,0)$& \\
\cline{1-6}
$WW\, (0/1 j)$ & comb. w/8 & --- & $0.76\pm 0.21$ & inclusive & \\
$WW\, (2 j)$ & comb. w/8 & ---  & $-0.05^{+0.73}_{-0.56}$ & $(17,83,0)$ & \cite{CMSww} \\
$WW$ (VH)  & comb. w/8 & --- &$-0.31^{+2.24}_{-1.96}$ & $(0,100,0)$& \\
\cline{1-6}
$ZZ$ (untagged) & comb. w/8 & --- &  $0.84^{+0.32}_{-0.26}$ & $(95,5,0)$ & \cite{CMSzz} \\
$ZZ$ (dijet tag) & --- & --- &  $1.22^{+0.84}_{-0.57}$ & $(80,20,0)$ & \\
\cline{1-6}
$\gamma \gamma $ (untagged 0)  & $3.78^{+2.01}_{-1.62}$ & $(61.4, 35.5,3.1)$& $2.12^{+0.92}_{-0.78}$& $(72.9,24.6,2.6)$ & \\
$\gamma \gamma $ (untagged 1) & $0.15^{+0.99}_{-0.92}$ & $(87.6,11.8,0.5)$ & $-0.03^{+0.71}_{-0.64}$ & $(83.5,15.5,1.0)$ &  \\
$\gamma \gamma $ (untagged 2) & $-0.05 \pm 1.21$& $(91.3,8.3,0.3)$ & $0.22^{+0.46}_{-0.42}$ & $(91.7,7.9,0.4)$ &  \\
$\gamma \gamma $ (untagged 3) & $1.38^{+1.66}_{-1.55}$& $(91.3,8.5,0.2)$ & $-0.81^{+0.85}_{-0.42}$ & $(92.5,7.2,0.2)$ &\\
$\gamma \gamma $ (dijet) & $4.13^{+2.33}_{-1.76}$& $(26.8,73.1,0.0)$ & --- & ---  &  \cite{CMSgaga} \\
$\gamma \gamma $ (dijet loose) & --- & --- & $0.75^{+1.06}_{-0.99}$ & $(46.8,52.8,0.5)$ & \\
$\gamma \gamma $ (dijet tight) & --- & --- & $0.22^{+0.71}_{-0.57}$ & $(20.7,79.2,0.1)$ & \\
$\gamma \gamma $ (MET) & --- & --- & $1.84^{+2.65}_{-2.26}$ & $(0.0,79.3,20.8)$ & \\
$\gamma \gamma $ (Electron) & --- & --- & $-0.70^{+2.75}_{-1.94}$ &  $(1.1,79.3,19.7)$ & \\
$\gamma \gamma $ (Muon) & --- & --- & $0.36^{+1.84}_{-1.38}$ &  $(21.1,67.0,11.8)$ & \\
\hline
\end{tabular}
\caption{\small Light Higgs fit data from CMS.}
\label{tab:CMS}
\end{table}
\begin{table}[hhh]
\footnotesize
\centering
\renewcommand{\arraystretch}{1.1}
\begin{tabular}{| c | l | c | c | c |}
\hline
Exp. & Channel & $\hat \mu$ (2 TeV)  & $\zeta^{\rm (G,V,T)}$ (\%) & Ref. \\
\hline \hline
CDF & $b \bar b$ & $1.72^{+0.92}_{-0.87}$ & $(0,100,0)$ & \cite{CDFD0,TevUpdate} \\
& $\tau \tau$ & $0.00^{+8.44}_{-0.00}$ & $(50,50,0)$  & \\
& $WW$ & $0.00^{+1.78}_{-0.00}$ & inclusive & \\
& $\gamma \gamma $ &  $7.81^{+4.61}_{-4.42}$ & inclusive &\\ 
\hline
D\O & $b \bar b$ & $1.23^{+1.24}_{-1.17}$ & $(0,100,0)$ &  \cite{CDFD0,TevUpdate}  \\
& $\tau \tau$ & $3.94^{+4.11}_{-4.38}$ & $(50,50,0)$ &  \\
& $WW$ & $1.90^{+1.63}_{-1.52}$ & inclusive & \\
& $\gamma \gamma $ &  $4.20^{+4.60}_{-4.20}$ & inclusive &\\ 
\hline
\end{tabular}
\caption{\small Light Higgs fit data from CDF/D\O.  Production efficiencies in $\tau \tau$ channels are approximated from \cite{TevTau}.}
\label{tab:Tevatron}
\end{table}

\pagebreak

\bibliography{fitbib}
\bibliographystyle{jhep}

\end{document}